\pdfoutput=1
\documentclass[floatfix,aps,prc,reprint,groupedaddress,nofootinbib]{revtex4-1}

\usepackage{amssymb}
\usepackage{amsmath}
\usepackage[colorlinks=true,linkcolor=blue,citecolor=blue, urlcolor=blue]{hyperref}   
\usepackage{tikz}
\usepackage{graphicx}

\definecolor{uibred}{RGB}{167, 38, 47}
\def\Eq#1{Eq.~(\ref{#1})}
\def\Eqs#1{Eqs.~(\ref{#1})}
\def\eq#1{(\ref{#1})}
\def\app#1{Appendix~\ref{#1}}
\def\Fig#1{Fig.~\ref{#1}}
\def\Figs#1{Figs.~\ref{#1}}
\def\Sec#1{Sec.~\ref{#1}}
\def\App#1{Appendix~\ref{#1}}

\def\p{\mathbf{p}}
\def\x{\mathbf{x}}
\newcommand{\xt}{\mathbf{x}}
\def\r{\mathbf{r}}
\def\k{\mathbf{k}}

\newcommand{\kt}{\mathbf{k}}
\newcommand{\ktt}{|\mathbf{k}|}

\newcommand{\rt}{\mathbf{r}}
\newcommand{\rtt}{|\mathbf{r}|}

\newcommand{\vt}{\mathbf{v}}

\newcommand{\pt}{\mathbf{p}}
\newcommand{\ptt}{|\mathbf{p}|}
\newcommand{\kompost}{K{\o}MP{\o}ST} 

\newcommand{\TBg}{\overline{T}}
\newcommand{\TId}{T_\text{id.}}
\newcommand{\xSc}{\frac{\tau \TId}{\eta/s}}
\newcommand{\xScInv}{\frac{\eta/s}{\tau \TId}}
\newcommand{\tauekt}{\tau_{\scriptscriptstyle \text{EKT}}}
\newcommand{\tauhydro}{\tau_\text{hydro}}
\def\st{\begin{equation}}
\def\stp{\end{equation}}
\def\llangle{\left\langle}
\def\rrangle{\right\rangle}

\newcommand{\subfig}[2]{%
\begin{tikzpicture}%
\node[rectangle] (image) at (0,0) {#2};
\node[anchor=south west] (label) at (image.south west) {(#1)};
\end{tikzpicture}%
}
\begin{document}

\title{Effective kinetic description of event-by-event pre-equilibrium dynamics in high-energy heavy-ion collisions}

\author{Aleksi Kurkela}
\email[]{a.k@cern.ch}
\affiliation{Theoretical Physics Department, CERN, Geneva, Switzerland}
\affiliation{Faculty of Science and Technology, University of Stavanger, 
4036 
Stavanger, Norway}

\author{Aleksas Mazeliauskas}
\email[]{a.mazeliauskas@thphys.uni-heidelberg.de}
\affiliation{Institut f\"{u}r Theoretische Physik, Universit\"{a}t Heidelberg, 
69120 Heidelberg, Germany}
\affiliation{Department of Physics and Astronomy, Stony Brook University, Stony 
	Brook, NY 11794, USA}

\author{Jean-Fran\c cois Paquet}
\email[]{jeanfrancois.paquet@duke.edu}
\affiliation{Department of Physics, Duke University, Durham, NC 27708, USA}
\affiliation{Department of Physics and Astronomy, Stony Brook University, Stony 
Brook, NY 11794, USA}

\author{S\"{o}ren Schlichting}
\email[]{sslng@uw.edu}
\affiliation{Department of Physics, University of Washington, Seattle, WA 
98195-1560, USA}

\author{Derek Teaney}
\email[]{derek.teaney@stonybrook.edu}
\affiliation{Department of Physics and Astronomy, Stony Brook University, Stony 
	Brook, NY 11794, USA}

\date{\today}

\begin{abstract}

We develop a macroscopic description of the space-time evolution of the energy-momentum tensor during the pre-equilibrium stage of a high-energy heavy-ion collision. Based on a weak coupling effective kinetic description of the microscopic equilibration process (\`a la ``bottom-up"), we calculate the non-equilibrium evolution of the local background energy-momentum tensor as well as the non-equilibrium linear response to transverse energy and momentum perturbations for realistic boost-invariant initial conditions for heavy ion collisions. We demonstrate how this framework can be used on an event-by-event basis to propagate the energy momentum tensor from far-from-equilibrium initial state models, e.g.\ IP\nobreakdash-Glasma,
 to the time $\tauhydro$ when the system is well described by relativistic viscous hydrodynamics.
The subsequent hydrodynamic evolution becomes essentially independent of the hydrodynamic initialization time $\tau_\text{hydro}$ as long as $\tau_\text{hydro}$ is chosen in an appropriate range where both kinetic and hydrodynamic descriptions overlap. We find that for $\sqrt{s_{NN}}=2.76\,\text{TeV}$ central Pb-Pb collisions, the typical time scale when viscous hydrodynamics with shear viscosity over entropy ratio $\eta/s=0.16$ becomes applicable is $\tau_\text{hydro}\sim 1\,\text{fm/c}$ after the collision.

\end{abstract}

\maketitle

\tableofcontents

\section{Introduction\label{sec:introduction}}

High-energy heavy-ion collisions at the 
Relativistic Heavy Ion Collider (RHIC) and the Large Hadron Collider 
(LHC) probe nuclear matter at extreme densities and temperatures.
One of the primary goals of heavy-ion research is to study the properties of 
the new phase of deconfined matter created in such collisions: the Quark-Gluon Plasma (QGP).

Sophisticated multistage models, in which the QGP evolution is described by
viscous relativistic hydrodynamics, have been remarkably successful in
describing the soft hadronic observables measured in  heavy-ion collisions~\cite{Heinz:2013th,Teaney:2009qa,Luzum:2013yya,Gale:2013da,deSouza:2015ena}.
Comprehensive model-to-data comparisons have been made to quantify
systematically the constraints provided by measurements on the transport
coefficient of the QGP, and to understand the impact of different observables
on these constraints~\cite{Niemi:2015qia,Bernhard:2015hxa,Sangaline:2015isa}.

Considerable progress has been  made to increase the predictive power of hydrodynamic simulations and to fully understand the assumptions built into such models. 
These advances include an evolving understanding of the conditions necessary
for hydrodynamics to be
applicable~\cite{Denicol:2014tha, Heller:2015dha, Heller:2013fn,Strickland:2017kux,Romatschke:2017acs}, 
and a more consistent treatment of the transition  from hydrodynamics to hadronic kinetics at late times~\cite{Molnar:2014fva}.
Progress is also being made on understanding the kinetics of early stages of the collision
and the subsequent transition to hydrodynamics, which is the topic of this work.

The initial conditions of hydrodynamic models remain one of the major sources of uncertainty in phenomenological studies of heavy ion collisions.
We will provide a practical way to propagate the energy-momentum tensor in a far-from-equilibrium initial state to a time when viscous hydrodynamics becomes applicable. 
Our goal is to have consistent overlapping descriptions of the early time
dynamics, and to limit the dependence of the hydrodynamic model on ad hoc
parameters such as the hydrodynamic initialization
time $\tauhydro$~\cite{vanderSchee:2013pia,Kurkela:2016vts}.

One approach to the  initial conditions is simply to  
parameterize the initial 
energy density  and its fluctuations, sidestepping the thermalization process with additional parameters.  Glauber-based models are
commonly used for this purpose, and provide  the energy density at $\tauhydro$
~\cite{Miller:2007ri,Moreland:2014oya,Blaizot:2014nia,Gronqvist:2016hym}.
Besides the energy density, the remaining hydrodynamic fields (such as the
flow velocity and the shear and bulk tensors)  must also be parameterized,
leading  to an uncomfortable growth  in the number of free parameters.
Physically
motivated models such as free streaming~\cite{Broniowski:2008qk,Liu:2015nwa}
and the gradient
expansion~\cite{Vredevoogd:2008id,vanderSchee:2013pia,Romatschke:2015gxa} have
been used to relate the initial energy density to the full stress tensor which is ultimately needed to
start the hydrodynamic simulation.

At weak coupling significant progress has been made in constructing a
complete picture of the early time dynamics before $\tauhydro$. Since the relevant
degrees of freedom change as a function of time, a consistent theoretical
description of the pre-equilibrium stages requires a combination of different
weak-coupling methods. Based on the Color Glass Condensate  framework (CGC),
the initial state immediately after the collision is characterized by strong
color fields, whose dynamics is essentially non-perturbative and best-described
in terms of classical-statistical field
theory~\cite{Iancu:2002xk,Iancu:2003xm,Gelis:2010nm, Gelis:2007kn,
Lappi:2011ju}. After a short period of time $\sim1/Q_s$ the system becomes
increasingly dilute. Genuine quantum effects can then be no longer neglected,
and the subsequent dynamics is better described in terms of QCD effective
kinetic theory~\cite{Arnold:2002zm}.
Several studies (which include some of the present authors) have investigated
the various stages of the equilibration process in detail, including the early
time dynamics using classical-statistical real-time lattice
techniques~\cite{Berges:2013fga,Gelis:2013rba,Berges:2014yta,
Schenke:2015aqa}, as well as the subsequent approach towards
local thermal equilibrium using effective kinetic theory simulations~\cite{ Xu:2004mz,El:2007vg,Kurkela:2015qoa,Keegan:2016cpi}, i.e.\ the
``bottom-up" thermalization
scenario~\cite{Baier:2000sb}.  

Although the output of classical field simulations has been used 
to initialize hydrodynamic codes~\cite{Gale:2012rq},  
a consistent  treatment at weak coupling  would pass the classical
output  through the kinetic theory simulation to 
determine the initial conditions for  the subsequent hydrodynamic evolution.
In this paper we  provide a concrete realization of this set of steps,
allowing for an event-by-event description
of the early time dynamics of high-energy heavy-ion collisions  which
smoothly approaches hydrodynamics.

Elaborating on
the ideas formulated in Ref.~\cite{Keegan:2016cpi}, we describe the pre-equilibrium
dynamics macroscopically in terms of non-equilibrium response functions of the
energy-momentum tensor. Specifically, linearized energy and momentum
perturbations are propagated on top of a boost-invariant and locally
homogeneous background,  and the energy-momentum tensor is evolved
from a non-equilibrium initial state to a later time when viscous hydrodynamics becomes applicable. We demonstrate that the pre-equilibrium
evolution smoothly matches onto  hydrodynamics, and
the subsequent hydrodynamic evolution becomes essentially independent of the
matching time $\tau_\text{hydro}$.

In order to obtain a smooth transition from kinetic description to realistic
viscous hydrodynamic evolution with typical shear viscosity over entropy ratio
$\eta/s\sim0.16$, the coupling constant $\lambda=N_c g^2$ (the single parameter
of kinetic theory) has to be extrapolated to large values of
$\lambda=10\text{--}25$. For such values of $\lambda$, the entire
non-equilibrium kinetic evolution is very well described by universal functions
of scaled evolution time $\tau T/(\eta/s)$.   The scalability of background and
linear response functions greatly simplifies practical application, since the
kinetic pre-equilibrium evolution needs only to be calculated once and then can
 be applied to any event-by-event hydrodynamic simulations.  In order to
facilitate the use of our results in phenomenological description of
event-by-event high-energy heavy-ion collisions, we make public the linearized
kinetic theory response functions and our implementation of the linear
pre-equilibrium propagator \kompost~\cite{kompost_github}.

The paper is organized as follows.  We introduce a general macroscopic description of local pre-equilibrium evolution based on linear response theory out of equilibrium in \Sec{sec:macro}, and discuss how to obtain the relevant inputs from an underlying microscopic description in effective kinetic theory in \Sec{sec:ekt}. We further study the equilibration of a locally uniform background in \Sec{subsec:background} and derive local hydrodynamization time for realistic initial conditions in \Sec{sec:hydtime}. In \Sec{sec:generalresponse} we provide the general decomposition of energy-momentum tensor response functions to initial energy and momentum perturbations and in \Sec{subsec:ektresponse} we discuss their realization in effective kinetic theory. The implementation of the kinetic theory pre-equilibrium phase for hydrodynamic models of heavy-ion collisions is detailed in \Sec{sec:implementation}, and then applied to two types of initial conditions, MC-Glauber and IP-Glasma initial conditions, in Sections~\ref{sec:glauber} and \ref{sec:ipglasma} respectively. We conclude with a compact summary of our findings and discussion of future directions in \Sec{sec:summary}. Several appendices provide details on the background scaling functions (\App{app:parameters}), determination of the kinetic response functions (\App{app:Tmunu_decompose}), free streaming response functions (\app{sec:freestreaming}), hydrodynamic response functions (\app{sec:hydroresp}) and kinetic response in the low-$k$ limit (\app{app:lowk}).

\section{Pre-equilibrium evolution}

\subsection{Macroscopic description of equilibration\label{sec:macro}}
\begin{figure}
\centering
\includegraphics[width=0.9\linewidth]{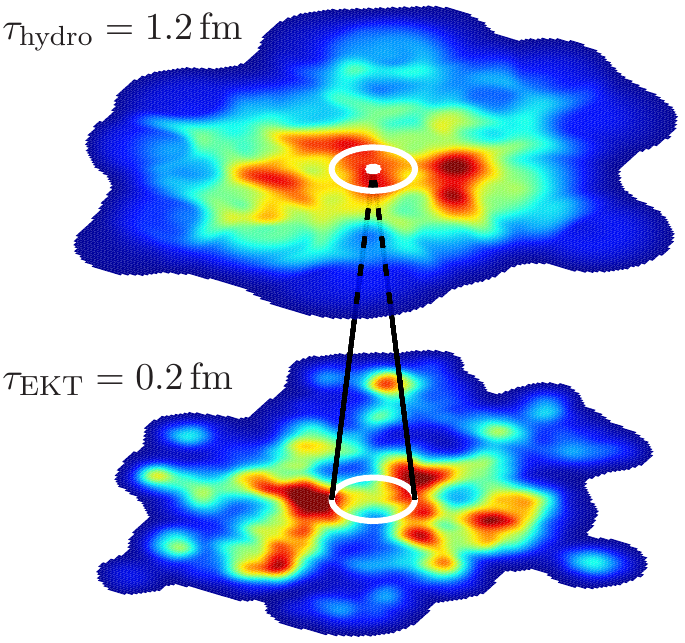}

\caption{The transverse energy density distribution for boost invariant IP-Glasma initial conditions   at the start of the kinetic theory pre-equilibrium evolution at $\tau_\text{EKT}=0.2\,\text{fm}$  and after the linearized kinetic evolution  given by \Eq{one}  at the hydrodynamic initialization time 
$\tauhydro=1.2\,\text{fm}$.  
The white circle indicates the size of the causal neighbourhood  in the transverse plane, \Eq{eq:causalnb}.\label{fig:glasma3d}}
\end{figure}

Although the initial state
shortly after the collision 
of two heavy
nuclei 
is presumably very complicated, many of the microscopic details 
wash out 
 over the first $\sim
1\,\text{fm/c}$ 
as the Quark Gluon Plasma (QGP)
equilibrates and becomes a hydrodynamically expanding fluid.
Since the late time hydrodynamic behavior is fully characterized by the
energy and momentum densities, it is
conceivable that the most important features of the pre-equilibrium evolution
also can  be characterized by the energy-momentum tensor $T^{\mu\nu}$.
Following the computational strategy
of Ref.~\cite{Keegan:2016cpi},
 we will use QCD  kinetics as a microscopic
theory to determine the non-equilibrium
evolution of $T^{\mu\nu}$.  
This evolution propagates the non-equilibrium stress
from an initial  time $\tauekt\, {\sim}\, 0.1\,\text{fm}$
to a time  when
hydrodynamics becomes applicable $\tauhydro\,{\sim}\,1\,\text{fm}$,
 and provides a map of the form
\begin{equation}
\left.T^{\mu\nu}(\tauekt,\x)\right|_\text{out-of-equilibrium} \longrightarrow 
T^{\mu\nu}(\tauhydro,\x),
\end{equation} 
as illustrated by \Fig{fig:glasma3d}.

We first note that, by virtue of causality, all contributions to the energy-momentum tensor at a given space-time point $(\tauhydro,\x)$ are fully determined by the initial conditions at earlier time $\tauekt$ in the causal neighborhood of point $\x$
\begin{equation}
|\x'-\x|<c(\tauhydro-\tauekt),\label{eq:causalnb}
\end{equation}
which is illustrated by a circle in \Fig{fig:glasma3d}. The energy-momentum tensor can always be split into a locally homogeneous background and perturbations around it\footnote{We will consider a boost-invariant form of the energy-momentum tensor throughout this work and bold spatial vectors $\x$ lie entirely in the transverse plane. Note that in principle this framework could be extended to include an inhomogeneous background and variations of the energy-momentum tensor in rapidity.}
\begin{equation}
T^{\mu\nu}(\tauekt,\x')=\TBg^{\mu\nu}_\x(\tauekt)+\delta 
T^{\mu\nu}_\x(\tauekt,\x').
\end{equation}
Since we anticipate the time scale of the equilibration process $(\tauhydro-\tauekt)\lesssim 1\,\text{fm}$ to be small compared to the system size $R_\text{Pb}\sim 5\,\text{fm}$, the long wavelength variations of the energy-momentum tensor within the causal circle are small 
$\sim (\tauhydro-\tauekt)/R_\text{Pb} \ll 1$. Short wavelength fluctuations (of order nucleon size or less) are $1/\sqrt{N}$  suppressed by the number of participant sources.
Small perturbations around the background $\TBg_\x^{\mu\nu}(\tauekt)$ can be described (to first approximation) by linear response theory. Based on this approach, the energy-momentum 
tensor at $(\tauhydro,\x)$ can be expressed as a sum of the evolved background $\TBg^{\mu\nu}_\x(\tauhydro)$ and the response to the initial out-of-equilibrium 
perturbations $\delta T_\x^{\alpha\beta}$:
\begin{align}
&T^{\mu\nu}(\tauhydro,\x) =\TBg^{\mu\nu}_\x(\tauhydro)
+\frac{\TBg^{\tau \tau}_\x(\tauhydro)}{\TBg^{\tau\tau}_\x(\tauekt)}\times\label{one}\\
&\times\int 
d^2\xt'~G^{\mu\nu}_{\alpha 
\beta}\left(\xt,\xt',\tauhydro,\tauekt\right)\delta 
T_\x^{\alpha\beta}(\tauekt,\xt').\nonumber
\end{align}
The first term represents the non-linear equilibration of the boost
invariant homogeneous background.  This contribution, discussed in detail in
\Sec{subsec:background}, has no transverse flow, but constitutes the major part of
the final energy density and pressure. 
The second term in \Eq{one} is a convolution of initial perturbations with the response function
$G^{\mu\nu}_{\alpha \beta}\left(\xt,\xt_0,\tauhydro,\tauekt\right)$, which in our work is the sole contributor to the off-diagonal components of $T^{\mu\nu}(\tauhydro,\x)$. The  multicomponent structure of 
$G^{\mu\nu}_{\alpha \beta}$ and its realization in kinetic theory is examined
in \Sec{sec:response}. Finally, note that the longitudinal expansion
has been factored out from the response functions by 
normalizing perturbations
with the background energy density. 

By applying the propagation formula in \Eq{one} to all points
$\x$ in the transverse plane, we construct initial conditions
$T^{\mu\nu}(\tauhydro,\x)$, which can be used as input to the subsequent
hydrodynamic description of high-energy heavy-ion collisions. Since the out-of-equilibrium evolution of the
background and response functions smoothly approaches
hydrodynamics, 
 the 
initialization time $\tauhydro$ can be changed without affecting physical
observables. Most importantly, this evolution 
consistently describes the
pre-equilibrium production of entropy  and
transverse flow.

\Eq{one} provides a general framework  to study the macroscopic features of the
pre-equilibrium evolution. However, its 
inputs, the  non-equilibrium evolution of the background stress $\TBg_\x^{\mu\nu}(\tau)$
and  response functions   $G^{\mu\nu}_{\alpha
\beta}\left(\xt,\xt',\tauhydro,\tauekt\right)$,
have to be computed based on an underlying microscopic description. In this work we use a weakly coupled  
kinetic theory to compute these inputs, and extrapolate our results to realistic values of the coupling constant. However, we note that in future applications, these could be also based on different microscopic descriptions e.g.\ strongly coupled
holographic models. 

While the details of the
calculation are described in the following sections, 
we point out an important feature of the dynamics 
that greatly simplifies the practical use of
\Eq{one} for realistic values of coupling constant. Specifically, we find that for the relevant range of couplings $\lambda$ (or
equivalently $\eta/s$), all of the $\eta/s$ dependence of
the background and 
response functions is described by $\eta/s$-independent functions
of a \emph{scaled time}, $\tau T/(\eta/s)$.
Thus, these universal functions can be tabulated for one value of $\eta/s$, 
and the results can be used in \Eq{one}
to initialize  hydrodynamic simulations over a significant $\eta/s$ range.
Interestingly, this scaling property also allows us to smoothly extrapolate the weakly coupled
results into the strongly coupled regime~\cite{Keegan:2015avk,Hong:2011bd}.

\subsection{Effective kinetic theory \& bottom-up thermalization\label{sec:ekt}}

\begin{figure*}
\centering
\begin{tikzpicture}
\node[anchor=south west] (image) at (0,0){%
\includegraphics[width=0.24\linewidth]{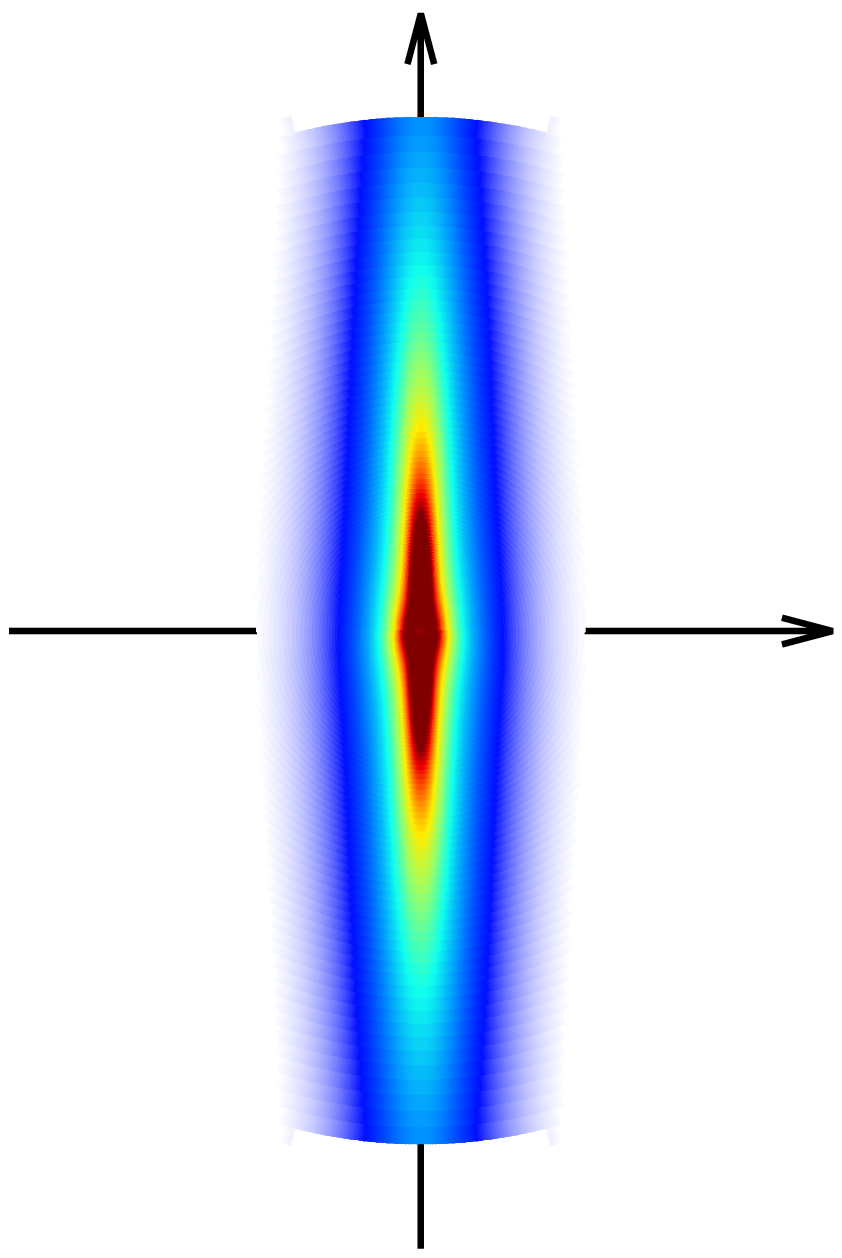}};
\node[anchor=north west, xshift=-0.5cm] at (image.north west) {$\tfrac{\tau \TId}{4\pi\eta/s}\sim 0.1$};
\node[anchor=north west, xshift=1.55cm] at (image.north west) {\Large $p_x$};
\node[anchor=north west, xshift=-0.7cm,yshift=-0.1cm] at (image.east) {\Large $p_z$};
\node[anchor=south west] at (image.south west) {(a)};
\end{tikzpicture}
\begin{tikzpicture}
\node[anchor=south west] (image) at (0,0){%
	\includegraphics[width=0.24\linewidth]{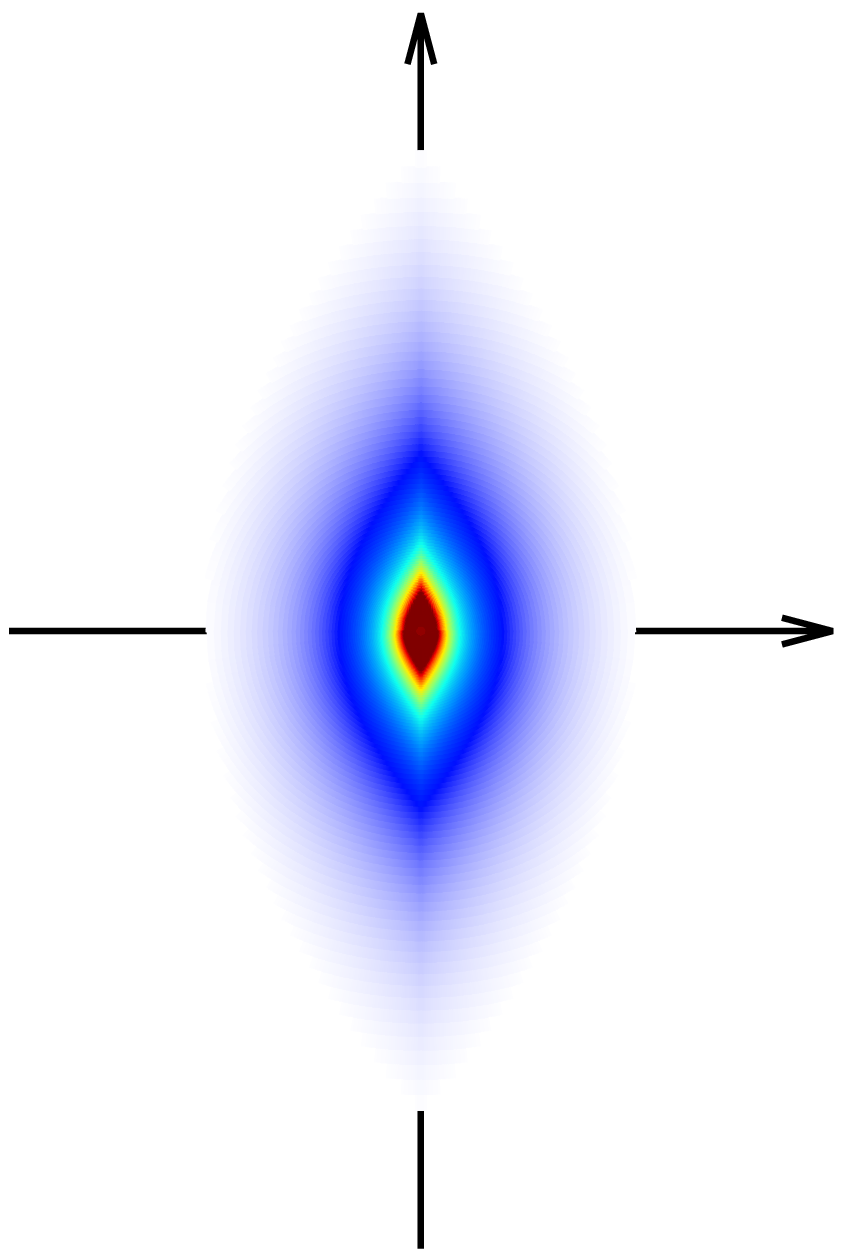}};
\node[anchor=north west, xshift=-0.5cm] at (image.north west) {$\tfrac{\tau \TId}{4\pi\eta/s}\sim 0.5$};
\node[anchor=north west, xshift=1.55cm] at (image.north west) {\Large $p_x$};
\node[anchor=north west, xshift=-0.7cm,yshift=-0.1cm] at (image.east) {\Large $p_z$};
\node[anchor=south west] at (image.south west) {(b)};
\end{tikzpicture}
\begin{tikzpicture}
\node[anchor=south west] (image) at (0,0){%
	\includegraphics[width=0.24\linewidth]{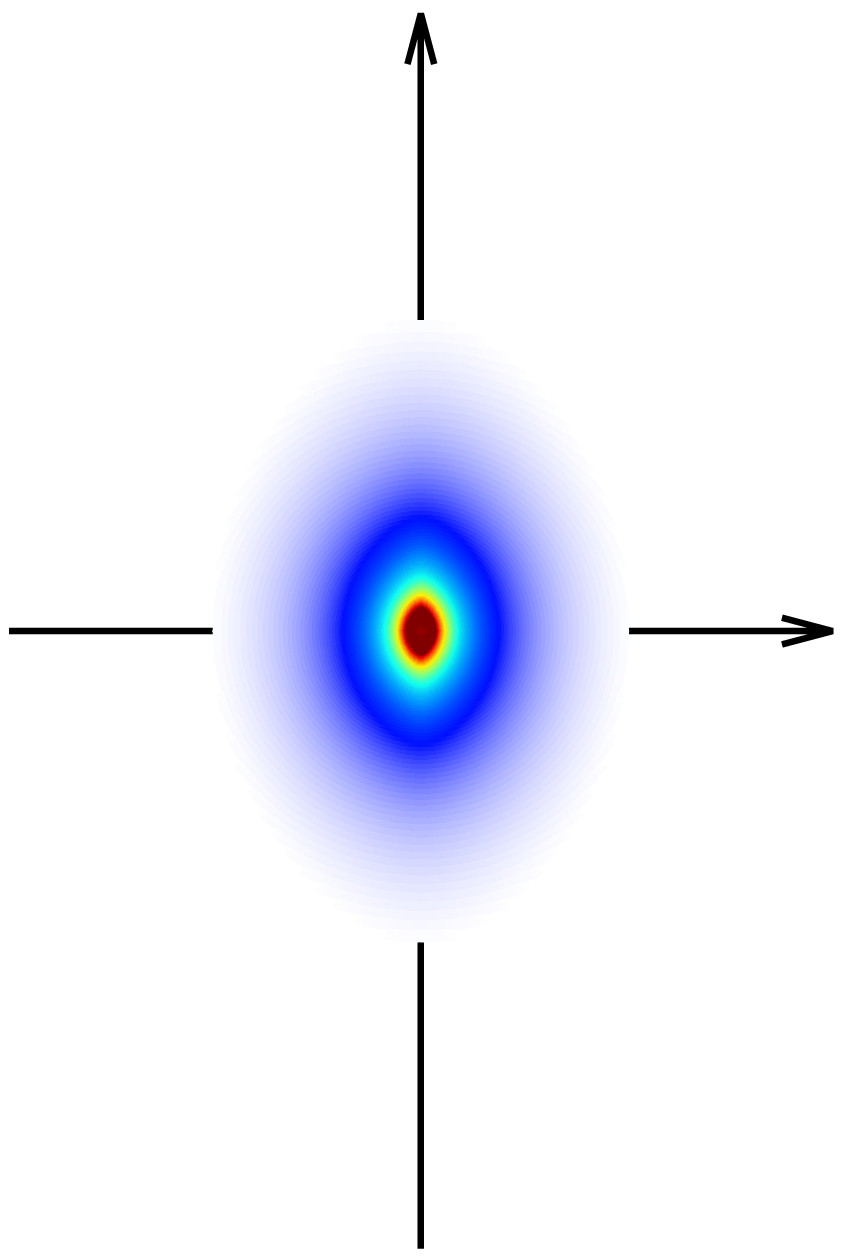}};
\node[anchor=north west, xshift=-0.5cm] at (image.north west) {$\tfrac{\tau \TId}{4\pi\eta/s}\sim 1.0$};
\node[anchor=north west, xshift=1.55cm] at (image.north west) {\Large $p_x$};
\node[anchor=north west, xshift=-0.7cm,yshift=-0.1cm] at (image.east) {\Large $p_z$};
\node[anchor=south west] at (image.south west) {(c)};
\end{tikzpicture}

\caption{Different phases of weak coupling equilibration (\`a la ``bottom-up") for a Bjorken
   expanding plasma.  Different panels show the gluon momentum distribution function
   $\bar f(\tau,\p)$  in the $(p^x,
   p^z)$ plane  as a function of scaled time $\tau \TId/(4\pi\eta/s)$: (a)~The initial distribution is broadened by elastic 
   $2\leftrightarrow 2$ scattering; (b)~inelastic $1\leftrightarrow 2$
   splitting and mini-jet quenching build up a low-momentum thermal
   bath; (c)~the distribution approaches local thermal equilibrium, but has significant
   corrections which are described by viscous hydrodynamics.  The initial conditions for $\bar f$ are given by \Eq{eq:init_cond}
   and the evolution is done for the  coupling constant $\lambda=10$ (with 
   $\eta/s\approx 0.62$). 
   The color scale represents the distribution function in the range
   $0.1 < \bar f < 5$.
   \label{fig:bottomup}
}
\label{fig:isotropization}
\end{figure*}
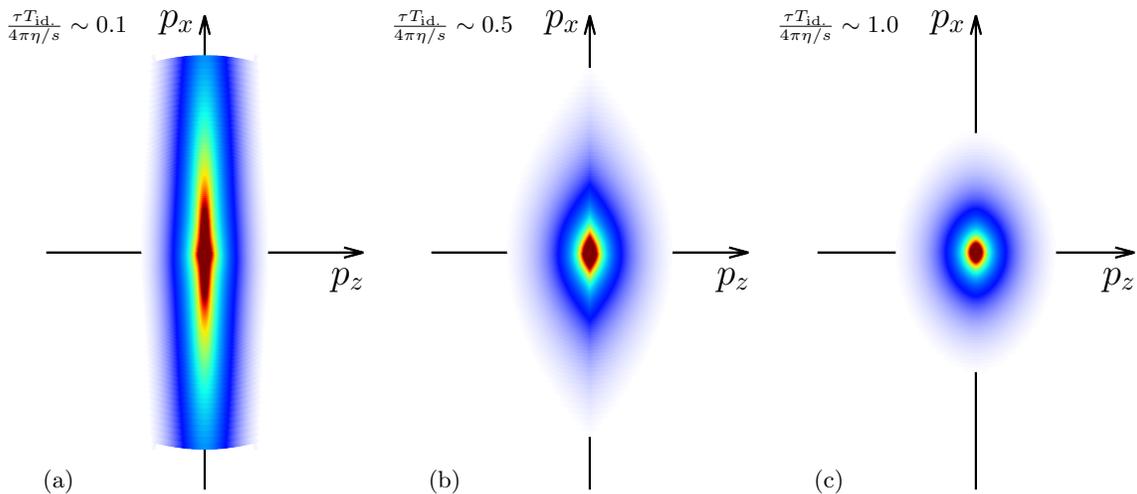

In this work the microscopic description of equilibration is provided by QCD effective kinetic theory~\cite{Arnold:2002zm,Keegan:2016cpi,Kurkela:2015qoa}.
Specifically, for the boost-invariant $\text{SU}(N_c)$ gluonic plasma 
considered here, the Boltzmann equation for the color and spin averaged gluon
distribution function $f_{\x,\p}$ takes the form\footnote{Here $\x\equiv(x,y)$ is a 2-dimensional transverse coordinate vector, while $\p\equiv(p^x,p^y,p^z)$ is a 
3-dimensional momentum vector.} 
\begin{align}
\partial_\tau f_{\x,\p}+ \frac{\bf p}{|\p|}\cdot \nabla_{\x} 
f_{\x,\p} - \frac{p^z}{\tau}\partial_{p^z} f_{\x,\p}= \nonumber\\
-\mathcal{C}_{2\leftrightarrow2}[f_{\x,\p}]-\mathcal{C}_{1\leftrightarrow2}[f_{\x,\p}].\label{bolz}
\end{align}
$\mathcal{C}_{2\leftrightarrow2}[f_{\x,\p}]$ is the collision integral for  elastic scatterings at leading order in the coupling constant, $\lambda=4\pi\alpha_s N_c$. The elastic scattering matrix element,
\begin{equation}
|\mathcal{M}|^2 = 2 \lambda^2 \nu_g \left( 9 + \frac{(s-t)^2}{u^2}+ 
\frac{(u-s)^2}{t^2}+ \frac{(t-u)^2}{s^2}\right), 
\end{equation}
diverges  at small momentum transfer, and is regulated by a
screening mass which is adjusted  so that 
the simulation
reproduces the leading order drag and
momentum diffusion coefficients for isotropic distributions~\cite{York:2014wja,Keegan:2016cpi}. 
Similarly,  $\mathcal{C}_{1\leftrightarrow2}[f_{\x,\p}]$ describes
 the leading order inelastic (particle number changing) bremsstrahlung processes. It includes the Landau-Pomeranchuk-Migdal suppression of
 collinear radiation, and also uses the isotropic screening approximation~\cite{Kurkela:2015qoa}.
Details of the numerical implementation can be found along with detailed expressions for the matrix elements in Appendix A of Ref.~\cite{Keegan:2016cpi}.

We determine the microscopic input to \Eq{one} by performing two independent
sets of calculations: one for the average background
$\TBg^{\mu\nu}$ described in \Sec{subsec:background},  and one for the response functions
$G^{\mu\nu}_{\alpha\beta}$ described in \Sec{subsec:ektresponse}. 
We calculate the evolution of the average energy-momentum
tensor $\TBg^{\mu\nu}$ by studying the equilibration of a spatially
homogeneous ``background" distribution $\bar f(\tau,\p)$, starting from an initial
condition
\begin{eqnarray}
\label{eq:init_cond}
\bar{f}_{0}\Big(|\p|,\theta\Big)=\frac{2 A}{\lambda} \frac{Q_0}{|\p|} \frac{e^{-\frac{2}{3} \frac{\p^2}{Q_0^2}[1 + (\xi^2-1) \cos^2(\theta)]}}{\sqrt{1 + (\xi^2-1) \cos^2(\theta)} }
\end{eqnarray}
where we denote $\cos(\theta)=p_z/|\p|$ and choose $Q_0 = 1.8\, Q_s$, $\xi=10$, and $A=5.24$ as in previous
publications~\cite{Keegan:2016cpi,Kurkela:2015qoa}.  This specific form is
motivated by classical simulations of the early time dynamics~\cite{Berges:2013eia},
where  the phase space distribution at $\tau \sim \tauekt$ approaches  a scaling form
characterized by a large initial momentum anisotropy  and 
a transverse momentum scale $\sim  Q_s$ as in the original
``bottom-up'' paper~\cite{Baier:2000sb}.

Before we analyze the evolution of the background energy-momentum tensor in
detail, we briefly review  how the phase
space distribution of gluons evolves in the ``bottom-up'' equilibration
scenario~\cite{Baier:2000sb},  which anticipated  the essential physics
at weak coupling.
Even at
moderate values of the coupling constant $(\lambda=10)$,  previous
numerical simulations have shown
that the system roughly follows
the three different stages of the bottom-up picture~\cite{Kurkela:2015qoa}. 
This is illustrated in \Fig{fig:bottomup}, where we present contour plots of the phase space distribution as a function of time. At early times, shown in \Fig{fig:bottomup}(a), the longitudinal expansion
competes with the elastic $2\leftrightarrow2$ broadening of the distribution
function. Subsequently, soft bremsstrahlung emissions from high momentum partons
begin to fill up the low momentum phase space as shown in \Fig{fig:bottomup}(b). These soft
particles can equilibrate via elastic $2\leftrightarrow2$ interactions, leading
to the formation of a soft and approximately thermal component at low momentum seen
in \Fig{fig:bottomup}(b); simultaneously the few remaining energetic partons (``mini-jets'') are
quenched and lose their energy to the soft thermal bath. Eventually, the
mini-jets are fully quenched, the system approaches local thermal
equilibrium, and the distribution function becomes well described by
Bose-Einstein distribution with viscous corrections, c.f. \Fig{fig:bottomup}(c).

\subsection{Approach to hydrodynamics for the homogeneous background \label{subsec:background}}

Based on the above picture of the underlying microscopic dynamics, we will now
discuss  how the boost invariant homogeneous background  
evolves towards viscous hydrodynamics.
As explained in \Sec{sec:macro}, our main interest is in the non-equilibrium
evolution of the energy-momentum tensor, which can be calculated directly from
the underlying particle distribution function $\bar f(\tau,\p)$
\begin{equation}
\TBg^{\mu\nu}(\tau) =\nu_g \int\! \frac{d^3\p}{(2\pi)^3} \frac{p^\mu p^\nu}{p}\,
\bar f(\tau,\p),
\end{equation}
where $\nu_g=2(N_c^2-1)=16$ is the number of degrees of freedom for a pure glue plasma. We
note that away from equilibrium, $\TBg^{\mu\nu}$ does not satisfy hydrodynamic
constitutive equations, and the energy-momentum tensor captures only the first
moments of the non-equilibrium distribution 
function. However,  the lowest moments of $\TBg^{\mu\nu}$ provide sufficient information for the late time hydrodynamic evolution, and we will therefore neglect the effect of higher-order 
moments throughout this work.

\begin{figure}
	\centering
	\includegraphics[width=\linewidth]{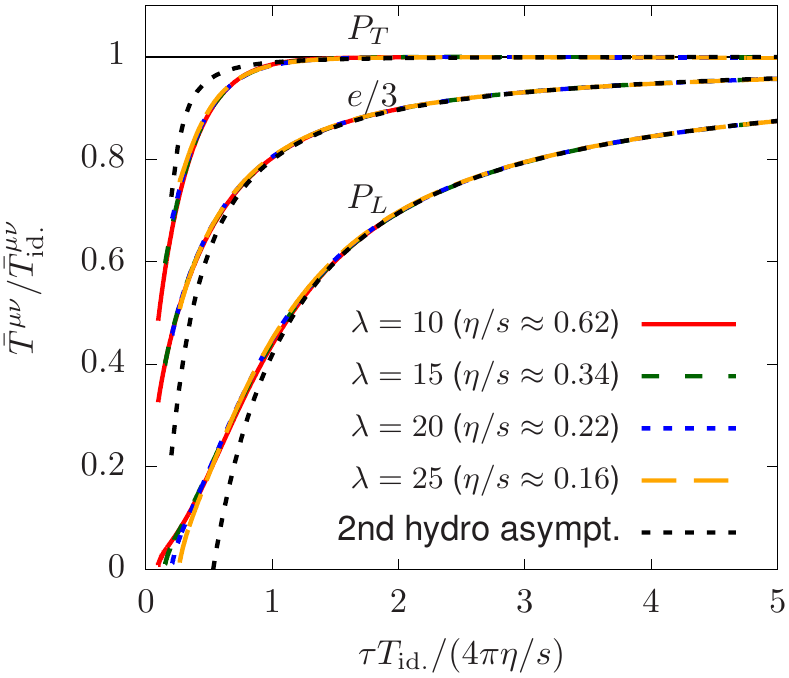}
	\caption{
		Equilibration of the different components of the background energy-momentum tensor.
		After normalizing the vertical axis by the asymptotic values,
		\Eq{eq:Tmunuideal}, and expressing time on the horizontal axis in terms of the scaled time variable, \Eq{eq:scaledtime},
		the entire evolution collapses onto a single curve for a
		range of  coupling constants $\lambda=10{-}25$ corresponding to 
		$\eta/s\approx 0.62{-}0.16$. 
	}
	\label{fig:paperplpt}
\end{figure}

Our results for the evolution of the diagonal elements of the background energy-momentum tensor\footnote{Off-diagonal elements of the background energy-momentum tensor vanish by the symmetries of the underlying distribution. Hence off-diagonal contributions of $T^{\mu\nu}$ that may exist at initial time $\tauekt$ will be treated as linearized perturbations.},
\begin{equation}\label{eq:defPLPT}
	\TBg^{\mu\nu}=\text{diag}\,( e, P_T, P_T, \tfrac{1}{\tau^2}P_L),
\end{equation}
are compactly summarized in \Fig{fig:paperplpt}, 
which shows the time evolution of the stress-energy tensor for different values of the 't Hooft coupling, $\lambda=10,15,20,25$. 
Motivated by previous work~\cite{Heller:2016rtz,Kurkela:2015qoa},
we have rescaled the time and stress axes in order to fairly compare the physics at different values of the coupling.
Specifically, at asymptotically late times the temperature approaches ideal hydrodynamics, parametrized as 
\begin{equation}
\TId(\tau;\Lambda_T)\equiv\frac{\Lambda_T}{(\tau \Lambda_T)^{1/3}},\label{eq:TId}
\end{equation}
where $\Lambda_T$ is a dimensionful integration constant 
\begin{equation}
\Lambda_{T}^2\equiv\lim_{\tau\rightarrow\infty} (\tau T^3)\,.
\end{equation}
In \Fig{fig:paperplpt} we first normalized the stress  on
the vertical axis by its asymptotic ideal hydrodynamics expectation
\begin{equation}
   \TBg^{\mu\nu}_\text{id.}(\tau)=\nu_g \frac{\pi^2}{30}\TId^4(\tau)\, \text{diag}\,( 1, \tfrac{1}{3},\tfrac{1}{3}, \tfrac{1}{\tau^2}\tfrac{1}{3}),\label{eq:Tmunuideal}
\end{equation}
and then rescaled the time on the horizontal axis by the equilibrium relaxation time $\tau_R(\tau)$, which for typical modes is determined by the shear viscosity 
and the ideal temperature $\TId(\tau;\Lambda_T)$ as
\begin{equation}
   \tau_\text{R}(\tau;\Lambda_T)
   \equiv\frac{\eta/s}{\TId(\tau;\Lambda_T)}.\label{eq:tauR}
\end{equation}
After these rescalings
the stress tensor follows a 
universal curve which is approximately independent of the coupling constant,
at least for the range of couplings considered in this work.

Such a scaling is guaranteed to work at late times where kinetic theory matches viscous hydrodynamics. 
Indeed,
in second-order conformal hydrodynamics
the energy density  
in a Bjorken expansion
 has the following  asymptotic form (see Ref.~\cite{Baier:2007ix} and \app{sec:transp}):
\begin{equation}
\frac{e(\tau)}{\nu_g \frac{\pi^2}{30}\TId^4(\tau)}=1-\frac{8}{3}\frac{\eta/s}{
\tau \TId}+\frac{8}{9}\big(3-C_2\big)\bigg(\frac{\eta/s}{\tau 
\TId}\bigg)^2,\label{eq:asymp} 
\end{equation}
where 
\begin{equation}
 C_2=\frac{\tau_\pi}{{\eta}/(sT)}\left(1-\frac{\lambda_1}{\tau_\pi
 \eta}\right)\label{eq:C2}
\end{equation}
is a dimensionless combination of second order transport
coefficients $\tau_\pi, \lambda_1$ and $\eta/s$.
In  leading-order  kinetic theory all transport coefficients are functions of the coupling 
constant $\lambda$---the only free parameter of the theory. Consequently
there is a one-to-one correspondence between the 
macroscopic parameter $\eta/s$ and the microscopic parameter $\lambda$ used in
the simulation. (The details of the matching between the macroscopic and microscopic
parameters are given in \App{app:parameters}.) Unlike  $\eta/s$, the ratio $C_2$ has a much weaker dependence on the coupling
constant~\cite{York:2008rr}, and thus  the coupling constant dependence
 in the hydrodynamic result (\Eq{eq:asymp}) is essentially 
completely contained in the \emph{scaled time} variable\footnote{In plots we 
include an additional factor of $1/4\pi$
so that scaled time $\tau \TId/(4\pi\eta/s)$ is typically of order unity.}
\begin{equation}
   \frac{\tau}{\tau_R(\tau)} \equiv	\frac{\tau \TId(\tau)}{\eta/s}.\label{eq:scaledtime}
\end{equation}
Returning to  \Fig{fig:paperplpt},
we  see that
for a substantial range of coupling constants (or $\eta/s$),
the non-equilibrium evolution of the energy-momentum tensor
follows universal functions of scaled time $\tau \TId/(\eta/s)$,
which are independent of the value of the coupling constant.  
Although such behavior is expected to emerge at late times, it is remarkable that the early time dynamics also exhibits such
a universal behavior\footnote{We note that at very weak coupling this near equilibrium scaling ansatz will fail. Indeed, in the bottom-up scenario the hydrodynamization time scales parametrically as $\tauhydro \sim \alpha_s^{-13/5}$, whereas the hydrodynamic scaling ansatz predicts a parametrically larger time scale $\tauhydro \sim \alpha_s^{-3}$ (c.f. Eqs.~(17),(18) with $\eta/s \sim \alpha_s^{-2}$), due to the fact that the energy density of the hard modes is assumed to decay $\sim \tau^{-4/3}$ rather than $\sim \tau^{-1}$. Of course, for moderate values of the coupling of practical interest the difference between $\alpha_s^{-13/5}$ and  $\alpha_s^{-3}$ is numerically insignificant.  }. Similar observations have also been reported in \cite{Heller:2016rtz,Romatschke:2017vte,Strickland:2017kux}, where this behavior is referred to as ``hydrodynamic attractors".

\begin{figure}
\centering
\includegraphics[width=\linewidth]{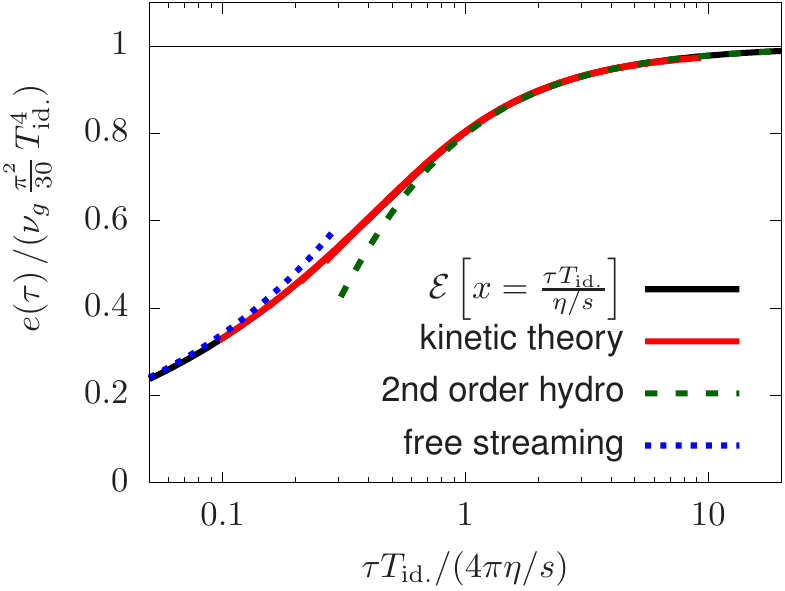}
   \caption{The universal scaling function, \Eq{eq:universalE}, fit to the kinetic theory evolution -- see \Fig{fig:paperplpt}. The early time behavior is approximately described by free streaming, while the late time behavior matches smoothly onto the second order hydrodynamic asymptotics, \Eq{eq:asymp}. 
   }
\label{fig:scaledTmunu}
\end{figure}

Exploiting the observed scaling property of the kinetic evolution, we will express the entire non-equilibrium energy-momentum tensor evolution in terms of a
universal function of the scaled time $\tau \TId/(\eta/s)$. Since for a Bjorken
expansion the longitudinal and transverse pressure can be readily determined
from the conservation law $P_L(\tau)=-\partial_\tau(\tau e(\tau))$ and the
tracelessness of the energy-momentum tensor $-e+2P_T+P_L=0$, there is in fact
only one independent function. Choosing the energy density $e(\tau)$ as the
independent variable, the entire pre-equilibrium evolution of the background
energy-momentum tensor can be parametrized as
\begin{align}
e&=\nu_g \frac{\pi^2}{30}\TId^4\,\mathcal{E}\left[x=\frac{\tau \TId}{\eta/s}\right].\label{eq:universalE}
\end{align}
with the universal scaling function $\mathcal{E}\left[x=\frac{\tau \TId }{\eta/s}\right]$. For asymptotically early times $x=\frac{\tau \TId }{\eta/s} \ll 1$ the effective kinetic theory evolution is approximately fitted onto the free-streaming behavior, while for late times $x\gg1$, $\mathcal{E}\left[x=\frac{\tau \TId }{\eta/s}\right]$ is very well described by the hydrodynamic asymptotics, \Eq{eq:asymp}, with $C_2\approx 1$ determined in \App{app:parameters}. It is straightforward to construct a parametrization of $\mathcal E(x)$ which interpolates between the two limits
 and the resulting fit curve is displayed in \Fig{fig:scaledTmunu}. The explicit parametrization of $\mathcal E(x)$ is provided in \App{app:parameters}. We emphasize once again that, thanks to the scaling of the background evolution with $\eta/s$, the same fitted kinetic theory curve can be used for different values of $\eta/s$ and different values of the initial energy density to map the early out-of-equilibrium energy density to the hydrodynamized energy-momentum tensor at later times.

\subsection{Hydrodynamization time and pre-equilibrium entropy production\label{sec:hydtime}}
\begin{figure*}
\centering%
\subfig{a}{\includegraphics[width=0.4\linewidth]{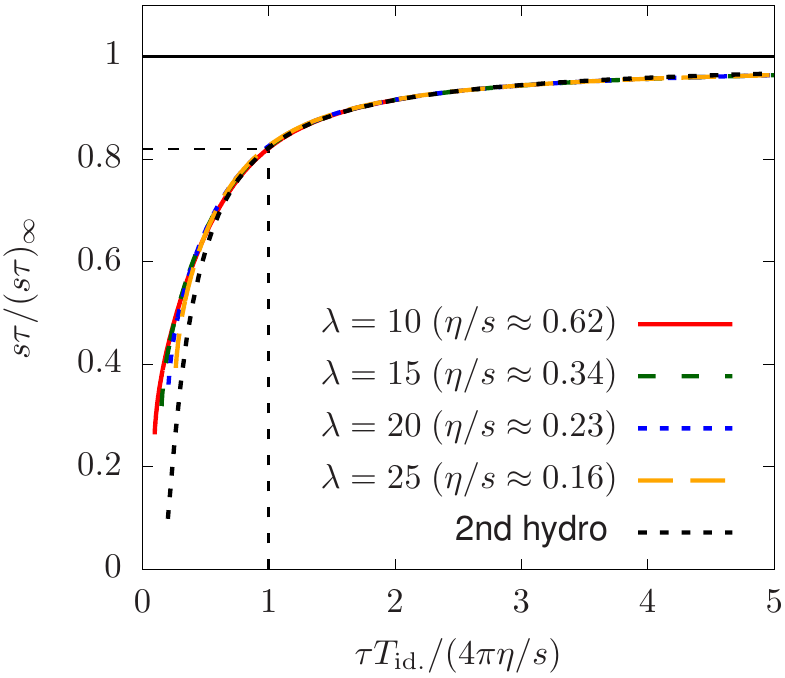}}\qquad
\subfig{b}{\includegraphics[width=0.4\linewidth]{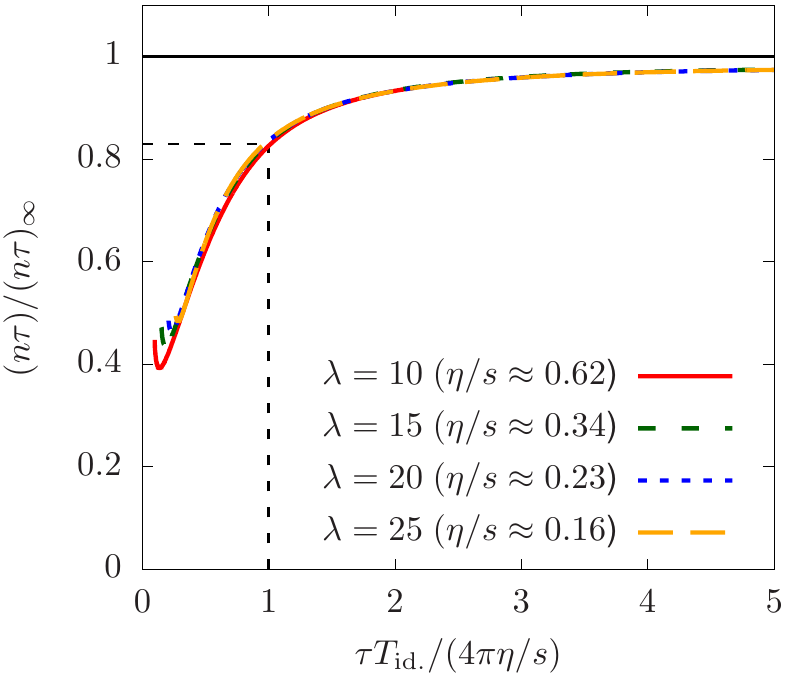}}
\caption{
   The entropy and gluon number density per rapidity as a function of scaled
   time $\xSc$. The  kinetic theory results for the entropy are compared 
   to viscous hydrodynamics (see text). 
}
\label{fig:entropya}
\end{figure*}

Based on the results in \Fig{fig:paperplpt}, we observe that, independently of the coupling constant, the kinetic description of equilibrating quark-gluon plasma overlaps well with hydrodynamics as soon as the reduced scaled time is larger than unity  $\tau \TId/(4\pi\eta/s)>1$. We therefore define
a hydrodynamization time, $\tauhydro$, as the boundary of 
applicability, i.e.\ $\tauhydro \TId/(4\pi \eta/s)=1$.
Substituting the definition of $\TId$, we can express 
$ \tauhydro$ in terms of $\eta/s$ 
\begin{equation}
   \tauhydro \equiv \left( \frac{4\pi\eta}{s}\right)^\frac{3}{2}  
   \frac{1}{\Lambda_T}\label{eq:tauhydro}\,,
\end{equation}
where $\Lambda_{T}$ is an energy scale defined 
through the asymptotic temperature \Eq{eq:TId}, or alternatively to the asymptotic
energy or entropy densities of the system.  
Specifically for a Bjorken expansion the temperature, energy density, and
entropy density have the asymptotic forms:
\begin{subequations}
	\begin{align}
   \lim_{\tau\rightarrow\infty} (\tau T^3) =& \Lambda_{T}^2\, , \label{eq:Tasymptotics}\\
   \lim_{\tau\rightarrow\infty} (\tau e^{3/4}) =& \Lambda_{E}^2 \, ,  \\
   \lim_{\tau\rightarrow\infty} (\tau s) =& \Lambda_{S}^2 \, .
\end{align}
\end{subequations}
where $\Lambda_{T}$, $\Lambda_E$, and $\Lambda_S$ are all related to each other by the equation of state. We will parametrize all equations of state with an effective number of degrees of freedom 
\begin{equation}
\label{eq:Landau}
e = \nu_{\rm eff}(T) \frac{\pi^2}{30} T^4  \, .
\end{equation}
where $\nu_{\rm eff}(T)=\nu_g=2(N_c^2-1)=16$ for the gluon gas used in simulations, while for a three flavor gas of quarks and gluons $\nu_{\rm eff}(T)= 47.5$. Finally,  
 at $T\sim 0.4\,{\rm GeV}$ lattice QCD simulations give $\nu_{\rm eff}\sim 40$~\cite{Bazavov:2014pvz,Borsanyi:2016ksw}. Using the definition \Eq{eq:Landau} and thermodynamic identities, it is straightforward to show that the relation between the integration constants is\footnote{In \Eq{eq:LambdaS} we used that $(e+p)/e=4/3$, which is true within 5\%  even for lattice equation of state at $T\sim 0.4\,{\rm GeV}$ ~\cite{Bazavov:2014pvz,Borsanyi:2016ksw}.}
 \begin{align}
 \Lambda_T^2 &\approx  0.3 \left( 
    \frac{16}{\nu_{\rm eff}} \right)^{3/4} \Lambda_E^2 \, ,  \\
    \Lambda_S^2 &\approx 2.0 \left( \frac{\nu_{\rm eff}}{16} \right)^{1/4} \Lambda_E^2 \, .\label{eq:LambdaS}
 \end{align}

Hydrodynamic simulations usually adjust the energy density $\Lambda_{E}$  (or entropy density $\Lambda_S$) at equilibrium time
to reproduce the multiplicity in the event. For central $\text{Pb}+\text{Pb}$ events at $\sqrt{s_{NN}}=2.76\,\text{TeV}$ discussed later in \Sec{sec:ipglasma} we estimated the average value of $\Lambda_E^2$ to be
\begin{equation}
   \llangle \tau e^{3/4} \rrangle \approx 1.6\,{\rm GeV}^2,
\end{equation}
which is consistent with other hydrodynamic simulations with a realistic equation of state (i.e. $\nu_{\rm eff}=40$) where $\llangle \tau s \rrangle \approx 4.1\, {\rm GeV}^2$~\cite{Keegan:2016cpi}.  Based on this estimate and using the hydrodynamization condition \Eq{eq:tauhydro}, we find that the hydrodynamization time of a boost invariant homogenous plasma is given by
\begin{equation}
\label{eq:hydrotime}
   \tauhydro\approx 0.8\,{\rm fm} \, \left( \frac{4\pi(\eta/s)}{2} 
   \right)^\frac{3}{2}  \left( \frac{ \llangle \tau e^{3/4}\rrangle }{1.6 \, {\rm  GeV}^2 } \right)^{-1/2} \left( \frac{\nu_{\rm eff}} {16}\right)^{3/8},
\end{equation}
which provides a realistic bound
 for the applicability of relativistic viscous hydrodynamics (for a constant value of $\eta/s=2/(4\pi)$). Changing the number of degrees of freedom from $\nu_{\rm eff}=16$ for a gluon gas to the more realistic $\nu_{\rm eff}=40$ increases the hydrodynamization time to $1.1\,{\rm fm}$.

One additional consequence of the pre-equilibrium evolution is a rather rapid entropy production associated with the increase of the gluon number density per rapidity. With the full gluon distribution at our disposal we can immediately calculate the Boltzmann entropy $s$ and the particle number $n$
\begin{align}
s(\tau) &= -\nu_g\int \frac{d^3\p}{(2\pi)^3}\big[ \bar f(\tau, \p) \ln \bar f(\tau, \p)\nonumber\\
&\qquad\qquad-(1+\bar f(\tau, \p))\ln (1+\bar f(\tau,\p))\big]\;, \\
n(\tau) &=~ \nu_g\int \frac{d^3\p}{(2\pi)^3} \bar f(\tau,\p) \;. 
\end{align}
Our results for the non-equilibrium entropy $(s)$ and particle number $(n)$ production 
  are summarized in \Fig{fig:entropya}. 
 We find that, independently of the coupling constant (or effective $\eta/s$), over $80\%$ of total entropy per rapidity is produced by the end of the pre-equilibrium stage $\tau \TId/(4\pi \eta/s)\sim 1$. One observes that the non-equilibrium production can be very well reproduced for later times $\tau \TId/(4\pi \eta/s) \gtrsim 1$ in second order hydrodynamics,  provided a non-equilibrium entropy definition $s_\text{non-eq}(\tau)=s_\text{eq}(\tau)\,(1-\frac{\tau_\pi}{4T\eta/s}\pi^{\mu\nu}\pi_{\mu\nu})$  is used~\cite{Baier:2007ix}.
Finally, the entropy and gluon number densities roughly doubles from the start of kinetic evolution to the time $\tauhydro$ (see \Fig{fig:entropya}). 
Clearly, the rapid generation of entropy during the approach to equilibrium is very important in relating  the initial state energy or gluon number density to the experimentally measured charged particle multiplicities (see Fig.~4 in Ref.~\cite{Aamodt:2010pb} and Fig.~3 in Ref.~\cite{Adam:2015ptt}, and  references therein).
Consequently, such large gluon multiplication factor needs to be taken into account in the correct estimation the properties of the initial state, e.g. the  saturation scale $Q_s$ in Color Glass Condensate picture, from the measured multiplicities.

\section{Response functions\label{sec:response}}

\subsection{General decomposition of macroscopic response functions}
\label{sec:generalresponse}
Continuing the discussion of \Sec{sec:macro}, we now look into the general properties of response functions for the linearized  energy-momentum perturbations evolving on top of the out-of-equilibrium background.
 We consider only boost 
invariant perturbations  in the transverse plane, and focus on the energy-momentum response to 
perturbations of the conserved charges---initial energy density $\delta 
T^{\tau\tau}$  and initial momentum density $\delta T^{\tau i}$.

 By normalizing 
the perturbations to the background energy density $\TBg^{\tau\tau}_\x(\tau)$, 
the evolution of energy-momentum perturbations can be compactly summarized as
\begin{equation}
\frac{\delta 
T^{\mu\nu}(\tau,\xt)}{\TBg^{\tau\tau}_\xt(\tau)}=\frac{1}{\TBg^{\tau\tau}_\x 
(\tau_0)}\int \!
d^2\xt_0 \,G^{\mu\nu}_{\alpha\beta}\Big(\xt,\xt_0,\tau,\tau_0\Big) {\delta 
T^{\alpha\beta}_\x(\tau_0,\xt_0)}\;.\label{eq:dTmunuconvolv}
\end{equation}
Since we consider perturbations on top of a (locally) homogenous and boost invariant background, translation invariance guarantees that the response functions depend only on the difference $\xt-\xt_0$ and it is often more convenient to work in Fourier space, where we define the Fourier transformed response function  $\tilde{G}^{\mu\nu}_{\alpha\beta}$ according to\footnote{Note that vectors $\x$ and $\k$ are both confined to the transverse ($x\text{-}y$) plane.}
\begin{align}
{G}^{\mu\nu}_{\alpha\beta}&\Big(\xt-\xt_0,\tau,\tau_0,\TBg_\x^{\tau\tau}(\tau_0)\Big)=\nonumber\\
&\int\!\frac{d^2\kt}{(2\pi)^2}~\tilde{G}^{\mu\nu}_{\alpha\beta}\Big(\kt,\tau,\tau_0,
\TBg_\x^{\tau\tau}(\tau_0)\Big)~e^{i\kt\cdot (\xt-\xt_0)}\;.\label{eq:Fourier}
\end{align}
Based on rotational symmetry in the transverse plane, one can further decompose 
the response functions into a tensor basis. For (scalar) energy perturbations 
the response function  $\tilde{G}^{\mu\nu}_{\tau\tau}(\kt,\tau,\tau_0, 
\TBg^{\tau\tau}_\x(\tau_0))$ has only four independent structures
\begin{align}
\tilde{G}^{\tau\tau}_{\tau\tau}(\kt)&=\tilde{G}_s^s(\ktt)\;, \quad  
\tilde{G}^{\tau i}_{\tau\tau}(\kt)=-i\tilde{G}_s^v(\ktt)  
\frac{\kt^i}{\ktt}\;,  \nonumber \\
\tilde{G}^{ij}_{\tau\tau}(\kt)&= \tilde{G}_s^{t,\delta}(\ktt) \delta^{ij} 
+\tilde{G}_s^{t,k}(\ktt) \frac{~\kt^i \kt^j}{\ktt^2}\;, \nonumber \\
\tilde{G}^{\tau\eta}_{\tau\tau}(\kt)&=0\;,  \qquad 
\tilde{G}^{i\eta}_{\tau\tau}(\kt)=0\;, \label{eq:G_energy_decomp}
\end{align}
while for (vector) momentum perturbations the decomposition reads
\begin{align}
 \tilde{G}^{\tau\tau}_{\tau k}(\kt)&=-i\tilde{G}_{v}^{s}(\ktt) 
 \frac{\kt^k}{\ktt}\;,\nonumber \\
 \tilde{G}^{\tau i}_{\tau k}(\kt)&=\tilde{G}_v^{v,\delta}(\ktt)\delta^{ik}  
 +\tilde{G}_v^{v,k}(\ktt) \frac{~\kt^i \kt^k}{\ktt^2} \;,\nonumber \\
 \tilde{G}^{ij}_{\tau k}(\kt)&= -i\tilde{G}_v^{t,\delta}(\ktt) \delta^{ij} 
 \frac{~\kt^k}{\ktt} -i \tilde{G}_{v}^{t,m}(\ktt)  \frac{ 
 \delta^{ik}\kt^{j}+\delta^{jk}\kt^{i}}{2\ktt}  \nonumber \\
 &\qquad \qquad -i\tilde{G}_v^{t,k}(\ktt) \frac{~\kt^i \kt^j \kt^k}{\ktt^3}\;, 
 \nonumber \\
\qquad \tilde{G}^{\tau\eta}_{\tau k}(\kt)&=0\;,   \qquad 
\tilde{G}^{i\eta}_{\tau 
k}(\kt)=0\;. \label{eq:G_momentum_decomp}
\end{align}
Since the longitudinal pressure components $\tilde{G}^{\eta\eta}_{\alpha\beta}$ 
are uniquely determined by the tracelessness of the energy-momentum tensor, one 
is then left with a total of ten independent response functions, which need to be determined by a particular microscopic model.

Similarly to the discussion in $\k$-space, the coordinate space response $G^{\mu\nu}_{\alpha\beta}$ can be 
decomposed in tensors constructed from the radial vector
$\r=\x-\x_0$. In practice, we first compute the response in $\k$-space and then  do the reverse Fourier transform~\cite{Keegan:2016cpi} .  The relations between momentum and 
coordinate space Green's functions are detailed in \app{app:Tmunu_decompose}.

\subsection{Non-equilibrium response functions from effective kinetic 
theory\label{subsec:ektresponse}}

The  independent components of macroscopic response functions $G_{\alpha \beta}^{\mu\nu}$ in \Eqs{eq:G_energy_decomp} and \eq{eq:G_momentum_decomp} need to be calculated by a particular microscopic theory. In this section we discuss the numerical realization of linear response in QCD kinetic theory around the non-equilibrium background presented in \Sec{subsec:background}. At late times and close to thermal equilibrium, kinetic response functions are bound to approach the hydrodynamic limit, which is studied  in \App{sec:hydroresp}. Similarly, the early time dynamics can be profitably compared to the analytic results of collision-free evolution, discussed in \App{sec:freestreaming}. 

 We follow the methodology of Ref.~\cite{Keegan:2016cpi} and linearize the 
 phase-space distribution function $f_{\x,\p}$ around the background 
 $\bar{f}_\p$, which is spatially 
 homogeneous, but 
 anisotropic in momentum space $\p=(p^x,p^y,p^z)$. We consider only boost 
 invariant $\delta f$
 perturbations, which can be decomposed into a Fourier integral of plane wave 
 perturbation $\delta f_{\k,\p}$ labelled by the wavenumber $\kt$ 
 in 
 the transverse plane
\begin{equation}
f_{\x,\p} = \bar{f}_\p+\int \frac{d^2 \k}{(2\pi)^2}\,\delta f_{\k,\p}~e^{i\k\cdot\x} .
\end{equation}
To linear order in perturbations, $\delta f_{\k,\p}$ evolves according to coupled Boltzmann equations
\begin{align}
\label{eq:bolz1}
\left(\partial_\tau - \frac{p_z}{\tau} \partial_{p_z}\right)  \bar f_{\p} &= - 
\mathcal{C}[\bar f],\\
\left(\partial_\tau - \frac{p_z}{\tau} \partial_{p_z} + \frac{i \p\cdot 
\k}{p} \right)  \delta f_{\k,\p} &= - \delta\mathcal{C}[\bar f, \delta f] ,\label{eq:bolz2}
\end{align}
where the collision kernel $ \delta\mathcal{C}[\bar f, \delta f]+\mathcal{O}(\delta f^2)= \mathcal{C}[\bar f+ \delta f]-\mathcal{C}[\bar f]$ is linearized in $\delta f$. Since the evolution for different values of $\kt$ on top of a homogenous background $\bar{f}_\p$ decouples from each other~\cite{Keegan:2016cpi}, the linearized Boltzmann equation can be solved independently for each wavenumber $\k$ in the transverse plane.

Even though the effective kinetic description of the pre-equilibrium dynamics 
requires the knowledge of the phase-space distribution $\delta f_{\k,\p}$ 
at the initial time $\tau_0$, one naturally expects the 
occurrence of memory loss during the evolution, so that the details of the initial 
phase-space distribution  become irrelevant as the system approaches local 
thermal equilibrium. Since our ambition is merely to extract the 
energy-momentum tensor, a representative choice of $\delta f_{\k,\p}$ to 
characterize initial energy and momentum perturbations should be sufficient 
to describe the non-equilibrium evolution.

Based on a weak-coupling picture of the initial state, where the properties of 
the background distribution $\bar{f}(\tau_0,\p)=\bar{f}_{0}(|\p|/Q_s,\theta)$ are determined by a 
single dimensionful scale $Q_s$,
it is natural to associate perturbations of 
the energy-momentum tensor $\delta T^{\mu\nu}(\xt,\tau_0)$ with local 
fluctuations of the scale $Q_s(\xt)=\bar{Q}_s+\delta Q_s(\k)e^{i\k\cdot 
\x}$. Hence one can motivate the  initial phase-space distribution of 
scalar perturbations to be of the form
\begin{equation}
\delta f_{\k,\p}^{(\text{Energy})}=\frac{\delta Q_s(\kt)}{\bar Q_s} ~\delta f_{\k,\p}^{(s)}
\end{equation}
where $\delta Q_s(\kt)/\bar Q_s=\delta T^{\tau\tau}/\bar T^{\tau\tau}$ denotes the amplitude of the perturbation and the spectral shape
\begin{equation}
\delta f_{\k,\p}^{(s)}=\frac{1}{4} \bar Q_s \partial_{\bar Q_s} \bar{f}_{0}\left(\frac{\ptt}{\bar Q_s},\theta\right)\;\label{eq:dfe}
\end{equation}
is determined from the variation of the background distribution, \Eq{eq:init_cond}, with respect to the scale $\bar Q_s$.

Similarly, considering that gradients of $Q_s(\xt)$ will lead to an initial 
velocity perturbation, e.g.\ $v_{i}(\tau_0) \propto  -\tau_0 \frac{\partial_{i} 
Q_s(\xt)}{\bar Q_s}$, we can motivate vector perturbations of the form
\begin{equation}
	\delta f_{\k,\p}^{(\text{Momentum})} = v^i(\kt) \delta f_{\k,\p}^{(v),i}\;,
\end{equation}
where $v^i(\kt) = \delta T^{\tau i}_{\k, (v)}/\bar T^{\tau \tau}$ denotes the amplitude of the initial velocity perturbation and the spectral shape
\begin{equation}	
\delta f_{\k,\p}^{(v),i}=\frac{1}{2} \partial_{v^{i}}\!\left. \bar{f}_{0}\left(\frac{|\pt-\vt \ptt|}{\bar Q_s},\theta \right) \right|_{\vt=0}\label{eq:dfg}
\end{equation}
is determined by a linearized velocity boost of the background momentum distribution. We note that in order to compute all response functions in the tensor decomposition in \Eq{eq:G_momentum_decomp} it is important that we keep track of the independent components (labeled by the index $i$) of the momentum response.

Even though at the level of the linearized Boltzmann equation, the 
actual magnitude of the perturbations is irrelevant in computing the response 
functions (as long as it remains sufficiently small to justify the linearized 
approximation at the relevant momentum scale), we find it convenient to choose an appropriate normalization. Defining the moments of the distribution function
\begin{subequations}
	\label{eq:dTmunudef}
\begin{align}
\delta T^{\mu\nu}_{\kt,(s)}(\tau)&= \nu_{g} \int 
\frac{d^3\p}{(2\pi)^3}~\frac{p^{\mu}p^{\nu}}{p}~\delta f^{(s)}_{\k,\p}\;,  
\\
\delta T^{\mu\nu,i}_{\kt,(v)}(\tau)&= \nu_{g} \int 
\frac{d^3\p}{(2\pi)^3}~\frac{p^{\mu}p^{\nu}}{p}~\delta 
f^{(v),i}_{\k,\p}\;, 
\end{align}
\end{subequations}
the corresponding energy and momentum perturbations associated with $\delta f_{\k,\p}^{(s)}$ and respectively $\delta f_{\k,\p}^{(v),i}$ at initial time are normalized such that for a highly oblate distribution ($\xi \gg 1$)
\begin{subequations}
	\begin{align}
	\frac{\delta T^{\tau \tau}_{\kt,(s)}(\tau_0) }{\TBg^{\tau 
	\tau}(\tau_0)}&= 1\,, 
		\\
\frac{\delta T^{\tau k,i}_{\kt,(v)}(\tau_0)}{\TBg^{\tau 
\tau}(\tau_0)}&= \delta^{ki}\;.  
	\end{align}
\end{subequations}

Given the above explicit form of the initial perturbations one can then 
determine the response of the energy-momentum tensor at any later time by 
numerically solving the kinetic equations, \Eqs{eq:bolz1} and \eq{eq:bolz2}. Once the solution to the linearized 
Boltzmann equation is calculated numerically, the response functions  
$\tilde{G}^{\mu\nu}_{\tau\tau}(\kt,\tau,\tau_0, 
\TBg_\x^{\tau\tau}(\tau_0))$  can be directly constructed from the moments of the distribution function. For example, the energy response to 
energy or momentum perturbations is determined by the ratios
\begin{align}
\label{eq:Gss}
\tilde{G}_{s}^{s}(\tau,\tau_0,\ktt)&= \left. \frac{\delta 
T^{\tau\tau}_{\kt,(s)}(\tau)} {\TBg^{\tau\tau}(\tau)} \right/ \frac{\delta 
T^{\tau\tau}_{\kt,(s)}(\tau_0) }{ \TBg^{\tau\tau}(\tau_0)}\;, \\
\tilde{G}_{v}^{s}(\tau,\tau_0,\ktt)&= \left. \frac{  
\frac{i\kt_{i}}{|\kt|} \delta T^{\tau\tau, i }_{\kt,(v)}(\tau)} {\TBg^{\tau\tau}(\tau)} \right/ 
\frac{ \frac{1}{2} \delta_{jk} \delta T^{\tau j, k}_{\kt,(v)}(\tau_0) }{ \TBg^{\tau\tau}(\tau_0)}\;.\label{eq:Gvs}
\end{align}
Similarly, all the other components can be constructed by linear combinations 
of different components of $\delta T^{\mu\nu}$ as described in 
\app{app:Tmunu_decompose}

\subsubsection{Scaling of response functions\label{sec:Gscaling}}

\begin{figure*}%
\centering
\subfig{a}{\includegraphics[width=0.4\linewidth]{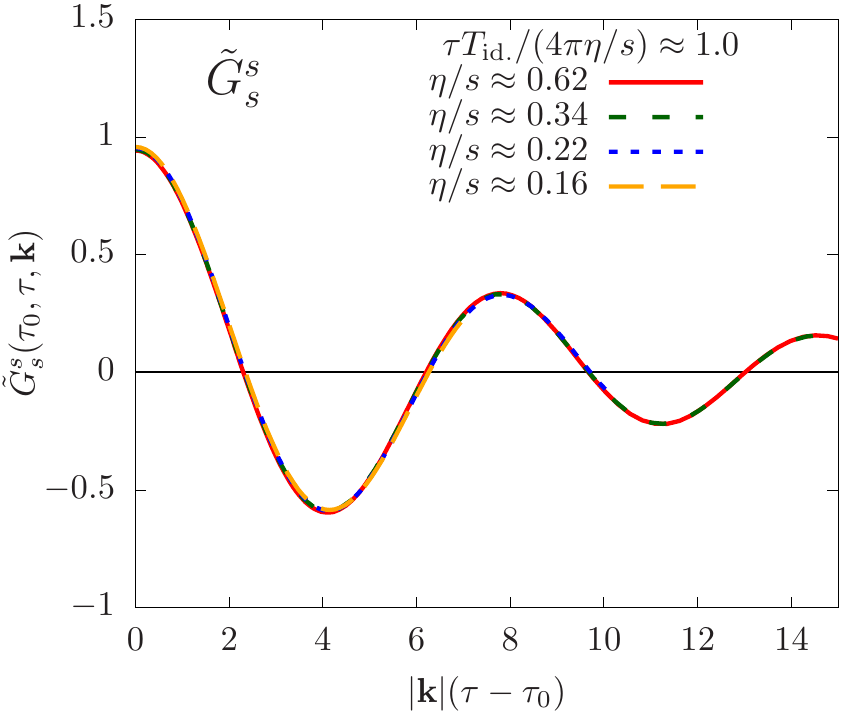}}\qquad
\subfig{b}{\includegraphics[width=0.4\linewidth]{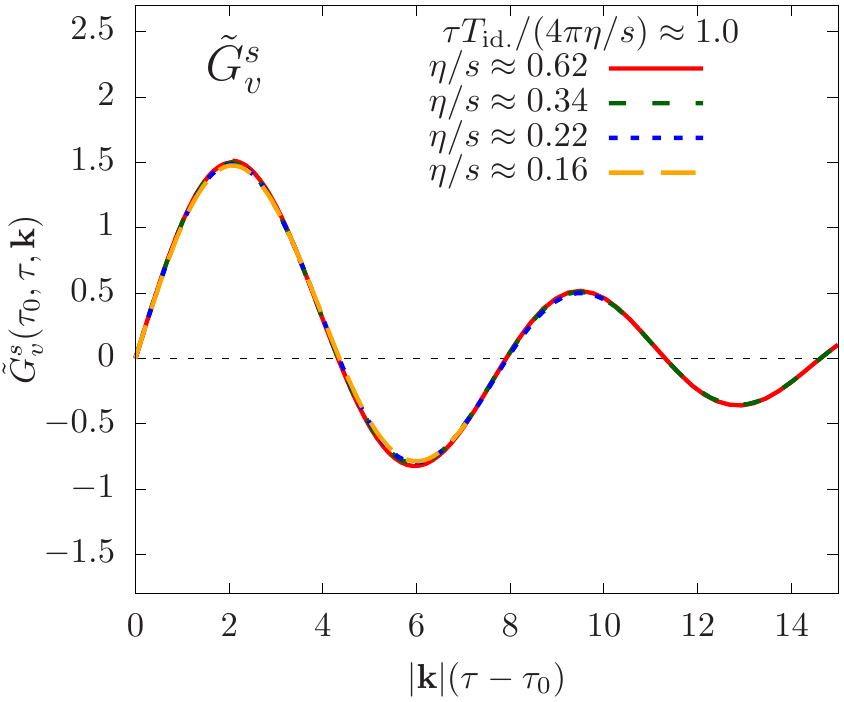}}
\caption{
   (a) The universal scaling function $\tilde{G}_s^s$ (see \Eq{eq:scalable})
   for the energy response to an initial energy perturbation as a function of
   the phase $\ktt(\tau -\tau_0)$ at a fixed scaling time, $\xSc$. The
   different curves correspond to different coupling strengths (or $\eta/s$)
   which collapse onto
   a universal curve.
    The response
      functions at different scaling times $\tau\TId/(4\pi\eta/s)=0.5,1.5,2.0$
      exhibit an equally good overlap (not shown). (b) The analogous plot for the
      energy response to an initial momentum perturbation, $\tilde{G}_s^v$.
}
\label{fig:rescaledresponse19de}
\end{figure*}

We  now present our numerical results for the non-equilibrium response 
functions calculated in effective kinetic theory. Even though generally 
the response functions 
$\tilde{G}^{\mu\nu}_{\alpha\beta}\big(\kt,\tau,\tau_0,\TBg_\x^{\tau\tau}(\tau_0)\big)$
 depend separately on the wavenumber $\kt$, the initial and final times $\tau$ 
and $\tau_0$,  the energy scale $\TBg_\x^{\tau\tau}(\tau_0)$, and the 
coupling constant $\lambda=g^2N_c$, we expect that in analogy to the evolution 
of the background the number of independent variables can be drastically 
reduced by identifying appropriate scaling variables. Based on our analysis of 
the background evolution in  \Sec{subsec:background}, the 
natural candidate variables are the scaled evolution time $\tau \TId / 
(\eta/s)$ and phase $\ktt (\tau-\tau_0)$.

Indeed  we find that in the relevant range 
of parameters the postulated scaling property holds and the response functions 
can be compactly expressed in terms of a universal function of the scaling 
variables such that, for example
\begin{align}
\label{eq:scalable}
\tilde{G}_{s}^{s}\Big(\ktt,\tau,\tau_0,\TBg_\x^{\tau\tau}(\tau_0),\lambda\Big)= \tilde{G}_{s}^{s,{\rm univ}}\left(\xSc,\ktt (\tau-\tau_0) 
\right)\;.
\end{align}
Even though this behavior is expected to emerge in the hydrodynamic limit\footnote{This hydrodynamic regime is discussed in more details in \app{sec:hydrolimit}.} of sufficiently 
late evolution times $\TId \tau / (\eta/s) \gg 1$ and for small wave-numbers $\ktt 
(\tau-\tau_0) \ll 1$, it is remarkable to observe that the scaling 
property holds across a much wider range of evolution times and wavelengths. 
As an illustrative example of the scaling, in \Fig{fig:rescaledresponse19de} we present our results for the response 
functions $\tilde{G}_{s}^{s}$ and $\tilde{G}_{v}^{s}$ given by \Eqs{eq:Gss} and \eq{eq:Gvs}  at scaled evolution time $\tau \TId/(4\pi \eta/s)\approx 1.0$.
 Different  curves  in \Fig{fig:rescaledresponse19de} 
correspond to simulations performed at different values of $\lambda$, which not only correspond to different effective $\eta/s$, but also amount to variations of the initial energy scale $\TBg_\x^{\tau\tau}(\tau_0)$, as seen from \Eq{eq:init_cond}.
 Other components of the response function in the decomposition given by \Eqs{eq:G_energy_decomp} and \eq{eq:G_momentum_decomp} also show a good scaling with $\tau \TId / 
(\eta/s)$ and phase $\ktt (\tau-\tau_0)$ (not shown).

Because of the scaling of the response functions with $\eta/s$---the same  kinetic theory response function computed for one set of initial conditions, can be used for different values of $\eta/s$ and different values of the initial energy density to map the early out-of-equilibrium energy and momentum perturbations to the hydrodynamized energy-momentum tensor perturbations at later times. This is the procedure adopted in the pre-equilibrium propagator \kompost{}.

Finally, the coordinate space response functions used in the  propagation formula \Eq{eq:dTmunuconvolv} are obtained by Fourier transforming $\ktt$-space components, e.g.\ \Fig{fig:rescaledresponse19de}, according to  \Eq{eq:Fourier}. The details of the procedure are given in \app{app:Tmunu_decompose} and also discussed in Ref.~\cite{Keegan:2016cpi}. Here we only point to the final result, i.e. the complete set of coordinate space response functions summarized in Figures \ref{fig:plot_grgss} and \ref{fig:plot_grgvs} in the \app{app:Tmunu_decompose}. Because the momentum space response functions are, to a good approximation, universal functions of scaling variables $\xSc$ and  $\ktt (\tau-\tau_0)$, the coordinate space response functions  are also universal when expressed in terms of the scaling variables  $\xSc$ and $|\x-\x_0| / (\tau-\tau_0)$.

\subsubsection{Hydrodynamic constitutive equations}
\begin{figure*}
	\centering
\subfig{a}{	\includegraphics[width=0.4\linewidth]{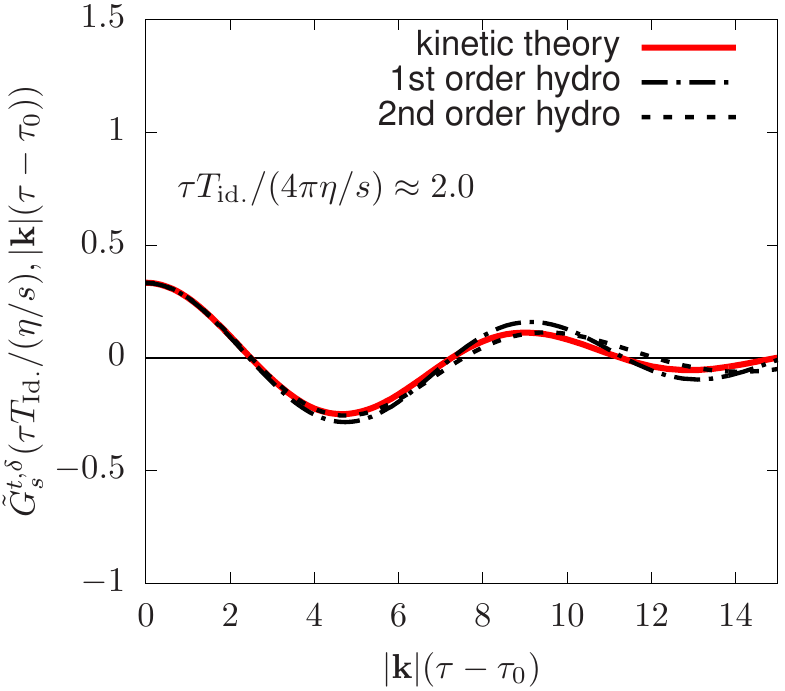}}
		\subfig{b}{\includegraphics[width=0.4\linewidth]{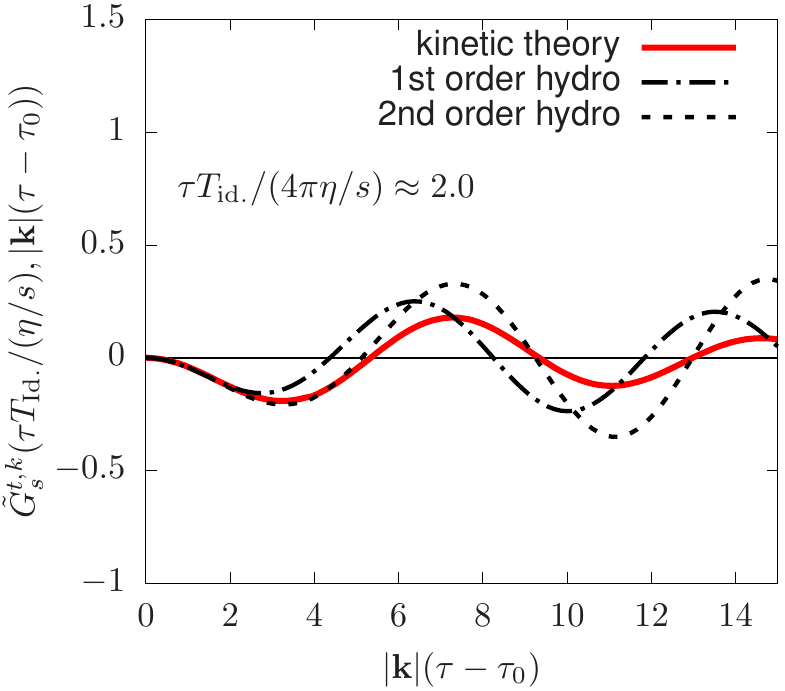}}
   \caption{Comparison of the response functions for $\delta T^{ij}$ from an initial energy perturbation (see $\tilde G_s^{t,\delta}$ and $\tilde G_s^{t,k}$ in \Eq{eq:G_energy_decomp})
     with the constitutive relations of first and second order hydrodynamics (\Eqs{eq:Gstk_const} and \eq{eq:Gtds_const}).
   }
	\label{fig:rescaledresponsegstdconst}
\end{figure*}

At late times when the system starts behaving hydrodynamically, the different 
components of response function decomposition \Eqs{eq:G_energy_decomp} and \eq{eq:G_momentum_decomp}, are no longer independent, but are related by hydrodynamic constitutive equations. In second order hydrodynamics and for small wavenumber perturbations, the spatial part of energy momentum tensor $\delta T^{ij}$ can be written as a  sum of energy $\delta T^{\tau\tau}$ and momentum $\delta T^{\tau i}$ density perturbations, with prefactors depending on first and second order hydrodynamic transport coefficients, i.e.\ $\eta,\tau_\pi$ and $\lambda_1$ ~\cite{Keegan:2016cpi}. By comparing constitutive equations with the 
decomposition in \Eq{eq:G_energy_decomp}, we find the following relation between response function components for initial energy perturbations
\begin{align}
&\tilde G_s^{t,k}(\tau,\tau_0,\ktt)=
\frac{3}{2}c_s^2\tau_\pi \eta \frac{\tilde G_s^s(\tau,\tau_0,\ktt) \ktt^2 
}{\TBg^{\tau\tau}}\nonumber\\
&+2(-\eta+\frac{1}{3}\eta \tau_\pi 
\frac{3c_s^2+1}{\tau}
-\frac{4}{3\tau}\lambda_1 
)\frac{\ktt\tilde 
	G_s^v(\tau,\tau_0,\ktt)}{\TBg^{\tau\tau}+\frac{1}{2}\TBg^k_k},\label{eq:Gstk_const}\\
&\tilde G_s^{t,\delta} (\tau,\tau_0,\ktt)=\nonumber\\ 
&\left(p+\frac{\eta}{2\tau}+\frac{1}{3}(1-c_s^2)\frac{\tau_\pi
	\eta-{\lambda_1}}{\tau^2}-\frac{1}{2}c_s^2 \tau_\pi \eta  k^2
	\right)\!\frac{\tilde 
	G_s^s 
(\tau,\tau_0,\ktt)}{\TBg^{\tau\tau}}\nonumber\\
&+\left(\frac{2}{3} 
\eta-\frac{2}{9} \eta 
\tau_\pi 
\frac{3c_s^2+2}{\tau}
+\frac{16}{9\tau}\lambda_1 \right)\frac{\ktt \tilde G_s^v(\tau,\tau_0,\ktt) 
}{\TBg^{\tau\tau}+\frac{1}{2}\TBg^k_k}\label{eq:Gtds_const}.
\end{align}
In \Fig{fig:rescaledresponsegstdconst} we explicitly test the first and second order constitutive relations, \Eqs{eq:Gstk_const} and \eq{eq:Gtds_const}, in the kinetic evolution at  scaled time $\tau \TId/(4\pi \eta/s)\approx 2$. Indeed for small wavenumbers $\ktt(\tau-\tau_0)< 6$ the second order constitutive equations are well satisfied, demonstrating  
that the low wavenumber perturbations approach hydrodynamic regime at 
sufficiently late times ${\tau \TId}/({4\pi\eta/s})>1$.

Similar constitutive relations can be derived for momentum response components and are given in \Eq{eq:G_momentum_const}.
The complete derivation is summarized in \App{sec:hydroresp}.

\section{Practical implementation: \kompost \label{sec:implementation}}

Based on the general formalism and results of linearized effective kinetic 
theory presented in the 
previous sections, we will now describe a practical implementation of the 
pre-equilibrium evolution for hydrodynamic modeling of heavy-ion collisions---\kompost. 
Starting from a given profile of the energy-momentum tensor 
$T^{\mu\nu}(\tauekt,\xt)$ at an early time $\tauekt$, for example from the 
IP-Glasma model, we follow the procedure outlined below to calculate the energy 
momentum tensor $T^{\mu\nu}(\tau_{\textrm{hydro}},\xt_{0})$ at a later time 
$\tauhydro>\tauekt$ when the system is sufficiently 
close to local thermal equilibrium for viscous hydrodynamics to become 
applicable\footnote{Equilibration is not necessarily achieved everywhere at 
the same time $\tau$, and the initial conditions could, in theory, be provided on a 
more complex $(\tau,x,y,\eta)$ hypersurface. However, it is the common practice 
to initialize hydrodynamic simulations on a constant $\tau$ hypersurface and we 
will follow this procedure in our present work.}.\\

\paragraph{ Decomposition in Background \& Perturbations}
We first split the initial energy-momentum tensor $T^{\mu\nu}$ at \kompost{} initialisation time $\tauekt$ in the background and perturbations
\begin{equation}
 T^{\mu\nu}(\tau,\x) = 
 \underbrace{\TBg^{\mu\nu}_{\x_0}(\tau)}_\text{background}+\underbrace{T^{\mu\nu}(\tau,\x)-\TBg^{\mu\nu}_{\x_0}(\tau)}_{\equiv\delta
  T_{\x_0}^{\mu\nu}(\tau,\x)}\;,\label{eq:split}
\end{equation}
where for linear evolution the decomposition into the background  
$\TBg^{\mu\nu}_{\x_0}$ and perturbations $\delta T^{\mu\nu}$ is arbitrary, as 
long as the perturbations are sufficiently small.
As discussed in \Sec{subsec:background} we consider locally homogeneous boost invariant background energy-momentum tensor $\TBg_{\x_0}^{\mu\nu}=\text{diag}\,(e, P_T,P_T,\tfrac{1}{\tau^2}P_L)$ with only one independent component $e(\tau)$. In order to obtain a smooth energy density profile from a discrete grid of input $T^{\tau\tau}$ values, we define the local background energy density as a 
Gaussian weighted average around the point $\x_0$ of interest 
\begin{equation}
\TBg^{\tau\tau}_{\x_0}(\tau)\equiv \int d^2\x'\, \frac{1}{2\pi \sigma^2} 
e^{\frac{-(\x'-\x_0)^2}{2\sigma^2}} T^{\tau\tau}(\tau ,\x')\;,\label{eq:avgBG}
\end{equation}
where the Gaussian width  is taken to be $\sigma=\Delta \tau/2$
\footnote{Note that changing the smearing 
width  $\sigma$ changes the decomposition into background and perturbations. We 
have 
checked by varying $\sigma$ by a factor of two that the sum of background 
and perturbations after the kinetic evolution remains remarkably invariant everywhere, 
except for edges of the fireball, where the linearized 
treatment breaks down.}. Here $\Delta \tau = (\tauhydro-\tauekt)$ is the duration of kinetic evolution, i.e.\ the causal circle radius discussed in \Sec{sec:macro}. 
Once the homogeneous diagonal background energy-momentum tensor $\TBg^{\mu\nu}_{\x_0}$ is determined in the neighbourhood of point $\x_0$,  the perturbation tensor $\delta T^{\mu\nu}_{\x_0}(\tau,\x)$ is obtained according to \Eq{eq:split}. In particular, the (small) initial off-diagonal components  of energy-momentum tensor $T^{\mu\nu}$ are completely absorbed in the perturbation tensor, e.g.\ $\delta T^{\tau i}_{\x_0}=T^{\tau i}$.
In accordance with the discussion in \Sec{sec:generalresponse} we only 
consider the response to initial energy $\delta T_{\x_0}^{\tau \tau}(\tauekt,\x)$ and 
momentum perturbations $\delta T^{\tau i}_{\x_0}(\tauekt,\x)$. The initial perturbations in the  shear-stress part of the energy-momentum tensor, i.e. $\delta T^{ij}_{\x_0}$ are not taken into account in this work\footnote{
Note that  in this implementation the diagonal components of the background energy-momentum tensor are entirely given by  $T^{\tau\tau}$.
Even though the initial diagonal components of energy-momentum tensor, $T^{xx}$, $T^{yy}$ and $T^{\eta\eta}$, are not used directly to determine the background energy-momentum tensor $T^{\mu\nu}_{\x_0}$, it is sufficient to assume that (on average) the system is highly anisotropic in the longitudinal direction, but approximately isotropic in the transverse plane, so that $T^{\tau\tau}$ can be used to estimate the diagonal components, i.e.\ $\tau^2 T^{\eta\eta}\ll T^{xx}\approx T^{yy}\approx T^{\tau\tau}/2$. This assumption is justifiable at early times in central heavy ion collisions.}.\\

\paragraph{ Background evolution \& scale parameter}
Once the decomposition in background and perturbations is determined at each 
point $\x_0$ in the transverse plane of the collision, we proceed to calculate the evolution of the background components
of the energy-momentum tensor. Since in the relevant range of parameters the 
effective kinetic theory evolution exhibits a universal behavior in terms of 
the scaling variable $\xSc$ (see \Sec{subsec:background}), we can 
immediately obtain the evolution for a specified values of $\eta/s$ by matching 
the point associated with the initial energy density 
$e(\tauekt)=\TBg^{\tau\tau}_{\x_0}(\tauekt)$ at time $\tauekt$ to 
the universal scaling curve. Specifically, we determine the scale parameter 
$\Lambda_T$ (which fixes the temperature function $\TId(\tau; \Lambda_T)$, see \Eq{eq:TId})  by solving the implicit 
equation
\begin{multline}
e(\tauekt)= 
\nu_{g}~\frac{\pi^2}{30}~\TId^4(\tauekt;\Lambda_T) \\ \times \mathcal{E}\left[x=\frac{\tauekt
 \TId(\tauekt;\Lambda_T) }{\eta/s}\right],\label{eq:implicit}
\end{multline}
where $\mathcal{E}(x)$ corresponds to the universal scaling curve for the 
evolution of the energy density (c.f.  \Sec{subsec:background} and \app{sec:paramtr}). 
\Eq{eq:implicit} is the requirement that the initial time and energy density, i.e.\ $\tauekt$ and
$e(\tauekt)$, lie somewhere on the scaling curve shown in \Fig{fig:scaledTmunu}.
Once the temperature 
function $\TId(\tau; \Lambda_T)$ is known, the energy density at the hydrodynamic 
initialization time $\tauhydro$ can be read off from the same universal curve as
\begin{multline}
e(\tauhydro) = 
\nu_{g}~\frac{\pi^2}{30}~\TId^4(\tauhydro;\Lambda_T) \\ \times \mathcal{E}\left[x=\frac{\tauhydro 
\TId(\tauhydro;\Lambda_T) }{\eta/s}\right]\;.
\end{multline}
Similarly, the background longitudinal and transverse pressure components,  $P_L=\tau^2 T^{\eta\eta}$ and  $P_T=T^{ii}$, can be determined from the scaling curve as 
detailed in \App{sec:paramtr}, and we obtain the background energy-momentum tensor $\TBg^{\mu\nu}_{\x_0}$  at time $\tauhydro$.\\

\paragraph{ Energy \& momentum perturbations}
Next we propagate the initial energy and momentum 
perturbations, $\delta T_{\x_0}^{\tau \tau}(\tauekt,\x)$ and $\delta T_{\x_0}^{\tau i}(\tauekt,\x)$, to calculate the contributions to all components of  the full energy-momentum tensor at hydrodynamic initialisation time $\tauhydro$. We use the pre-calculated  linear kinetic response functions $G^{\mu\nu}_{\alpha\beta}$ discussed in \Sec{sec:response}.  First, the tabulated Fourier-space
 functions $\tilde{G}^{\mu\nu}_{\alpha\beta}\big(\xSc, \ktt(\tau-\tau_0)\big)$ are 
 transformed to the coordinate space 
for the relevant values of scaled time $\xSc$ and radius
$|\x-\x_0|/(\tauhydro-\tauekt)$, see \app{app:Tmunu_decompose} for details\footnote{We note that in practice the high $\ktt$ 
tails are regulated by a small Gaussian damping and free-streaming extensions 
to stabilize numerical calculation of Fourier/Hankel 
transforms~\cite{Keegan:2016cpi}.}. Subsequently, the contributions to the 
energy-momentum tensor at each point $\x_0$ at the hydrodynamic initialisation surface 
$\tau_\text{hydro}$ are determined by convoluting the coordinate space 
response functions with the initial energy and momentum perturbation as in 
\Eq{eq:dTmunuconvolv}. The coordinate space response functions contributing to $\delta 
T^{\mu\nu}(\tauhydro,\x_0)$ have only limited support, namely the neigbourhood of points $\x$ in the causal past of $\x_0$
\begin{equation}
|\x-\x_0|<(\tauhydro-\tauekt),
\end{equation}
so in practice only a small number of spatial points contribute. 
We note 
that, according to the decompositions in Eqs.~\eqref{eq:G_energy_decomp}  and 
\eqref{eq:G_momentum_decomp}, we explicitly compute the contributions of 
energy and momentum perturbations to all components of energy-momentum tensor, 
without ever enforcing constitutive relations by hand. Adding the perturbations and the background produces the complete energy-momentum tensor at the end of \kompost{} evolution
\begin{equation}
T^{\mu\nu}(\tauhydro,\x_0)=\TBg^{\mu\nu}_{\x_0}(\tauhydro)+\delta 
T^{\mu\nu}_{\x_0}(\tauhydro,\x_0).
\end{equation}

\paragraph{Decomposition of $T^{\mu\nu}$ in hydrodynamic variables}
Once the full energy-momentum tensor is obtained at hydrodynamic initialization 
time $\tauhydro$, we perform a standard tensor decomposition into hydrodynamic 
variables~\cite{LandauFluids,Kovtun:2012rj}. Hydrodynamic simulations of heavy ion collisions describe the spacetime evolution of the 
energy-momentum tensor $T^{\mu\nu}$ in terms of the energy density $e$, the 
flow velocity $u^\mu$, the bulk pressure $\Pi$ and the shear-stress tensor 
$\pi^{\mu\nu}$. The evolution of these fields is given by the energy-momentum 
conservation equation $\nabla_\mu T^{\mu\nu}=0$ and, in second-order  Israel-Stewart  formulations of relativistic viscous hydrodynamics~\cite{Israel:1976tn}, relaxation-type 
equations for the viscous fields, $\Pi$ and $\pi^{\mu\nu}$. The bulk pressure $\Pi$ for conformal systems is exactly zero and the explicit expressions of $T^{\mu\nu}$  in terms of hydrodynamics fields is then given by
\begin{equation}
T^{\mu\nu}= e u^\mu u^\nu + p (e) \Delta^{\mu\nu}+\pi^{\mu\nu}\;,
\end{equation}
where $\Delta^{\mu\nu}\equiv g^{\mu\nu}+u^\mu u^\nu$ and the relation between energy and pressure is determined by the equation of state $p=p(e)$\footnote{In this paper we use mostly-plus metric convention $g^{\mu\nu}=\text{diag}(-1,1,1,\tfrac{1}{\tau^2})$.}.
The tensor decomposition is done by identifying the local rest-frame velocity $u^{\mu}$ and local energy density $e$  as the time-like eigenvector and eigenvalue of 
$T^{\mu}_{\phantom{\mu}\nu}u^\nu= -e u^{\mu}$.
Once $e$ and $u^{\mu}$ values are known,  $\pi^{\mu\nu}$ is obtained by tensor 
projection\footnote{$\pi^{\mu\nu}$ is the symmetric transverse traceless part of $T^{\mu\nu}$, i.e.\ $\pi^{\mu\nu}=\pi^{\nu\mu}$, $\pi^{\mu\nu}u_\nu=0$ and $\pi^{\mu\nu}g_{\mu\nu}=0$.}. 
The independent hydrodynamic fields $e$, $u^\mu$ and $\pi^{\mu\nu}$ are then passed to the subsequent hydrodynamic evolution.

Note that the linearized kinetic theory evolution does not guarantees the existence of a local fluid rest frame for arbitrary inputs; for sick cases, the procedure fails to find a meaningful rest frame. Such instances appear for the cases where the initial gradients are 
particularly steep (e.g.\ edges or peaks of the medium). 
Problems in extracting the flow velocity $u^{\mu}$ from $T^{\mu\nu}$ in certain 
spatial regions are thus indicative of the linear approximation of the kinetic 
theory being pushed too far.
 Although these points make only a small fraction of the total points in the transverse extent of the fireball (quantified below), they tend to introduce instabilities in hydrodynamics code.
Rather than attempting to address the problem  
in the hydrodynamic evolution, we developed a selective regulator of the kinetic  
theory output which we describe at the end of this section.

\paragraph{Hydrodynamic evolution\label{sec:implem_hydro}}

Once the hydrodynamic fields are initialized on a constant $\tau=\tauhydro$ 
hypersurface, their subsequent spacetime evolution is determined by second 
order relativistic hydrodynamics.  In this paper we use the publicly available viscous relativistic hydrodynamic  code MUSIC~\cite{Schenke:2010nt,Schenke:2010rr,Paquet:2015lta} to solve numerically the hydrodynamic equations and the subsequent particlization, which is described below. 
For the hydrodynamic phase we use a lattice-based QCD equation of state~\cite{Huovinen:2009yb}, except when comparing to the conformal equation of state $p=e/3$. As 
for the first order transport coefficients, a constant shear viscosity over 
entropy density ratio $\eta/s=2/4\pi\approx 0.16$ is used, while the bulk viscosity is 
neglected. 
The second order transport coefficients are determined by relating them to the first order ones in the relaxation-time 
approximation~\cite{Denicol:2012cn,Molnar:2013lta}. The complete list of second 
order transport coefficients and hydrodynamic equations can be found in Ref.~\cite{Ryu:2017qzn}.

\paragraph{Hadronization\label{sec:implem_hadr}}

At the end of hydrodynamic evolution the hadronic observables are computed from a constant temperature freeze-out surface with $T_\text{FO}=145\,\text{MeV}$ using the standard Cooper-Frye procedure~\cite{Cooper:1974mv}.  We note that for simplicity only thermal hadronic observables, i.e.\ without hadronic decays, are used in this work, but the viscous corrections to the hadronic momentum distribution are taken into account as described in Refs.~\cite{Paquet:2015lta,Ryu:2017qzn}.

\paragraph{Regulators}\label{par:regulators}

Below we document the regulator procedure for the limited instances when the energy-momentum tensor $T^{\mu\nu}$ obtained at the end of the linear kinetic evolution cannot be inverted to define a local fluid restframe.
The regulator we developed was motivated by free streaming, which is a meaningful
and robust initial stage model.
The regulator identifies regions of large gradients or low 
densities and drives the kinetic theory response towards free streaming in these
regions by selectively lowering the scaling variable $x=\tau \TId(\tau)/(\eta/s)$ 
to compute the kinetic response. We
emphasize that while the regulator is important for making the hydrodynamic simulations 
run, it produces only minimal modifications of the hydrodynamic input and does not affect the physical results discussed in \Sec{sec:results}. Thus, our pragmatic purpose here is to document the computer code.

Physically in heavy ion collisions the low density regions at the edges of the fireball can be never meaningfully described as a hydrodynamic medium. Hence there is no need to assume that 
$\eta/s$ is constant throughout. For the purposes of estimating 
the scaling variable we therefore define $(\eta/s)(T)$ which grows large in regions of low density, i.e.
\begin{equation}
 (\eta/s)(T)\equiv \left(\eta/s\right)_0 \left(1  + \frac{C_0^2}{T^2} \right) \, ,
\end{equation}
where  $C_0=100\,{\rm MeV}$, and $(\eta/s)_0$ is a constant physical input parameter, typically of order $1/4\pi$\footnote{$(\eta/s)_0$ can easily be replaced with a physical dependence on temperature for the interior of the fireball in future implementations of the code.}. The scaling variable 
is then replaced by the number of 
relaxation times between times $\tau_1$ and $\tau_2$
 \begin{align}
\label{eq:x0}
    x_{0}(\tau_2, \tau_1) \equiv& 
    \frac{3}{2} \int^{\tau_2}_{\tau_1}  d\tau \frac{\TId(\tau)}{(\eta/s)_0\left(1 + \frac{C_0^2}{\TId^2(\tau)}\right) }  \, .
 \end{align}
Taking  $C_0$ and $\tau_1$ to zero, returns $x_{0}$ to the canonical scaling variable
\begin{equation}
\label{xcanonical}
    x \equiv \frac{\tau \TId(\tau)}{(\eta/s)_0 } \, .
\end{equation}
For obtaining the background energy density from the scaling curve, \Eq{eq:universalE}, we use the scaling variable value $x_{0}(\tauhydro,0)$, while for propagating
the perturbations 
we use $x_{0}(\tauhydro,\tauekt)$, which provides a slightly better scaling parametrization
of the Green functions\footnote{Typically $\tauhydro\gg\tauekt$ and the difference between the two values of the scaling variable is small.}.
Using $x_{0}$ as opposed to the canonical value, \Eq{xcanonical}, removes a number of  instabilities near the edge of the grid, but does not regulate the occasional regions of very high gradients in the central region of the fireball. 

\begin{figure}
	\centering
	\includegraphics[width=0.9\linewidth]{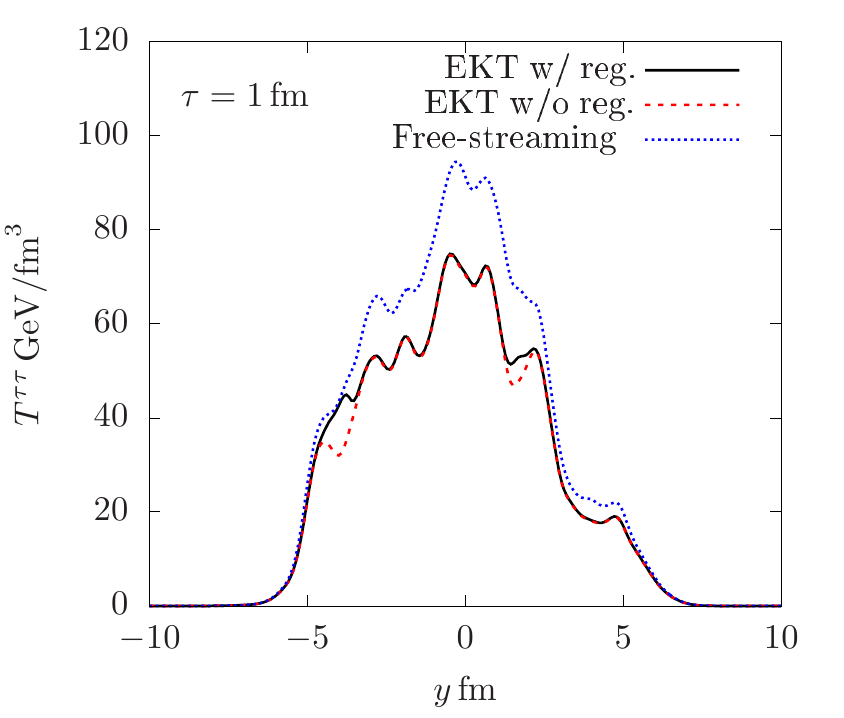}
	\caption{Profile of the energy density $T^{\tau\tau}$ along the $y$-direction for a kinetic theory evolution with and without a regulator, \Eq{eq:reg}. For points with large  gradients, the system response is driven towards free-streaming evolution. Note that only a small fraction of the  total number of grid points in the transverse plane are affected, see Table~\ref{table:reg}. Here  IP-Glasma initial conditions are used  with pre-equilibrium evolution from $\tauekt=0.2\,\text{fm}$ to $\tauhydro=1.0\,\text{fm}$.}
	\label{fig:regulator}
\end{figure}

The remaining instabilities arise when non-linearities become important. Examining 
when the $T^{\mu\nu}$ decomposition fails to find a rest frame, we determined (semi-empirically)
that for
\begin{equation}
     z  \equiv  \frac{ \sqrt{\delta T^{0x} \delta T^{0x} + \delta T^{0y} \delta T^{0y}} }{\tfrac{4}{3} \TBg^{00} }  - \frac{2}{3}  \frac{\delta T^{00}}{\TBg^{00}} > 0.5  \, , \label{eq:z}
\end{equation}
the $T^{\mu\nu}$ decomposition may fail to find a rest frame.  Certainly
when $z$ is of order $0.5$ the linearized kinetic theory has reached its limit of applicability.
For large $z$ we regulated $x_{0}$  according to
\begin{equation}
   x_{\rm reg} = x_{0} \, S(z; z_1, z_2) \, ,\label{eq:reg}
\end{equation}
where  $S(z; z_1, z_2)$ is a monotonic cubic spline interpolating between unity for $z< z_1$, and zero 
for $z > z_2$. In practice we take $z_1 = 0.4$ and $z_2=0.7$.
 When a regulated value $x_{\rm reg}$ is used as opposed to $x_{0}$ the dynamics is pushed  closer to free streaming limit in localised regions of steep gradients as shown in \Fig{fig:regulator}.

Technically the event is processed in two passes. In the first pass no regulator is used,
and the scaling variable $x_0$ is calculated according to \Eq{eq:x0}. From the $\TBg^{\mu\nu} + \delta T^{\mu\nu}$
of the first pass
we record the size of the unregulated perturbations as measured by the variable $z$ in \Eq{eq:z}, but we do
not attempt to find a local fluid rest frame yet. In the 
second pass we use $z$ from the first pass  to determine a regulated value of scaling variable $x_{\rm reg}$, \Eq{eq:reg}. Then $x_{\rm reg}$ is finally used
to propagate the background and perturbations from $\tauekt$ to $\tauhydro$. The second pass $T^{\mu\nu} = \TBg^{\mu\nu} + \delta T^{\mu\nu}$  has a rest frame decomposition which is passed on to the hydrodynamics code. We quantify the effect of the regulator by looking at the relative change in $T^{\tau\tau}$ component with and without a regulator
\begin{equation}
\delta \equiv \frac{\int d^2 x \left| T^{\tau\tau}_{\textrm{w/ reg}} - T^{\tau\tau}_{\textrm{w/o reg}} \right|}{\int d^2 x \left| T^{\tau\tau}_{\textrm{w/o reg}} \right|  }\label{eq:delta}.
\end{equation}
The relative change $\delta$ for MC-Glauber  and IP-Glasma initial conditions for different evolution times is recorded in Table~\ref{table:reg}. The transverse momentum flow grows with time and, according to criterion in \Eq{eq:z}, more points need to be regulated. However, from Table~\ref{table:reg} it is clear that only a small fraction of points in the transverse extend of the fireball are affected and only for longest evolution times the change is at a few percent level.

\begin{table}
	\centering
\begin{tabular}{|c|c|c|}
	\hline 
	& MC-Glauber & IP-Glasma \\ 
$\tauhydro$ (fm)	& $\delta$ & $\delta$\\ 
	\hline 
0.4	            &   0.002   & 0.001 \\ 
	\hline 
0.6	            &  0.004 & 0.005 \\ 
	\hline 
0.8	            &  0.006 & 0.01 \\ 
	\hline 
1.0	            & 0.01 & 0.03 \\ 
	\hline 
1.2	            & 0.02  & 0.05 \\ 
	\hline 
\end{tabular} 
\caption{Effect of the regulator as quantified by relative change in integrated energy density, \Eq{eq:delta}, for MC-Glauber and IP-Glasma initial conditions, and different kinetic evolution times.\label{table:reg}}
\end{table}

\section{Event-by-event pre-equilibrium dynamics \& matching to viscous hydrodynamics}
\label{sec:results}
We will now illustrate the applicability of our framework to perform event-by-event simulations of the pre-equilibrium dynamics of high-energy heavy-ion collisions. Since the kinetic theory equilibration scenario described in the previous sections provides a smooth crossover from the early stage of heavy ion collisions to the viscous hydrodynamics regime, we will demonstrate with the example of two initial state models how initial conditions for hydrodynamic simulations can be obtained within our framework. 

We first consider the Monte Carlo Glauber (MC-Glauber) model~\cite{Miller:2007ri}, which provides a phenomenological ansatz for the energy deposition in the transverse plane of heavy ion collisions, based on the location of binary nucleon collisions. Since  MC-Glauber is a not a dynamical  model, most phenomenological studies use an initialization time $\tauhydro\sim 0.5-1\,\text{fm}$ which is chosen empirically. However, we will show that the framework described in this work greatly reduces  the sensitivity of the hydrodynamic evolution to the initialization time $\tauhydro$. 

In the second part of this section, we will also consider the IP-Glasma model~\cite{Schenke:2012wb,Schenke:2012fw}, which provides a dynamical description of particle production and energy deposition. In this model color fields in each nucleus are sampled from a saturation model~\cite{Bartels:2002cj,Kowalski:2003hm} and subsequently evolved with classical Yang-Mills evolution to times $\tau\sim 1/Q_s \sim 0.1$~fm. Since the early time dynamics of IP-Glasma matches smoothly onto our effective kinetic description, this implementation amounts to a complete dynamical evolution within a weak coupling framework.

While the microscopic IP-Glasma model provides an initialization for the entire energy-momentum tensor of the collision in 2+1D, the MC-Glauber model is typically used as an ansatz only for the transverse energy density\footnote{It is also common to use Glauber model as an ansatz for the entropy density, which is then related to the energy density through the equation of state. In this work, the Glauber model is used as an ansatz for the energy density directly.}, without specifying the other components of the energy-momentum tensor. In the language of this paper, this means that IP-Glasma initial conditions provide both energy and momentum perturbations\footnote{We note that IP-Glasma also provides higher order fluctuations, e.g.\ of the different $T^{ij}$ components. However, as discussed in \Sec{sec:generalresponse} we limit ourselves to the energy-momentum response to the fluctuations of conserved quantities like energy and momentum.}, 
while the Glauber model only contains energy perturbations. 

We note that the hydrodynamic initialization time $\tauhydro$ is treated as a variable in this section. The values of $\tauhydro$ used are of the order of the \emph{background} hydrodynamization time given by Eq.~\ref{eq:hydrotime}, but not equal to it.

\subsection{Energy perturbations with MC-Glauber initial conditions\label{sec:glauber}}

We start our discussion with the Monte Carlo Glauber initial conditions, which provides an ansatz for the energy density distribution $e(\mathbf{x})$ at each point in the transverse plane of the collision\footnote{We used the publicly available T\raisebox{-0.5ex}{R}ENTo code~\cite{Moreland:2014oya} to generate the event. In our study, the reduced nuclear thickness parameter $p$ of the model was set to unity to obtain a typical participant Glauber model, while the negative binomial parameter for nucleon fluctuations was set to $k=1$, and the nucleon smearing width $w=0.5\,{\rm fm}$.}.  Energy is deposited at the location of the participant nucleons, and   the normalization of the energy distribution (i.e. total deposited energy) is adjusted so as to reproduce experimentally observed charged hadron multiplicities for typical central Pb+Pb events at LHC  energies $\sqrt{s_{NN}}=2.76\,\text{TeV}$. The energy-momentum tensor at time $\tauekt$ takes the form
\begin{equation}
	\label{eq:GlauberInitial}
	T^{\mu\nu}(\tauekt,\mathbf{x})=\begin{pmatrix} e(\x) &0 &0 &0\\
		0 & \frac{1}{2} e(\x)  & 0 & 0\\
		0 & 0 & \frac{1}{2} e(\x)  & 0\\
		0 & 0 & 0 & 0 
	\end{pmatrix}
\end{equation}
 Even though the Glauber model itself does not provide an intrinsic time scale $\tauekt$ for the dynamic evolution, it is clear that this time scale should be at least on the order of the formation time $\sim 1/Q$ required for semi-hard particles to go on-shell. Since we anticipate the typical momenta $Q\sim 2\,\text{GeV}$ for central Pb+Pb events, we will use $\tauekt=0.1\,\text{fm}$ in the following if not stated otherwise\footnote{We checked explicitly that the sensitivity of our results to this choice is relatively small, as long as the initial energy density is rescaled by an appropriate factor, which can be deduced from the relation $e\tau = e_0\tau_0+\int_{\tau_0}^{\tau}d\tau T^{zz}$ for Bjorken expansion. Since at very early times $T^{zz} \ll e$, such rescaling effectively mimics a free-streaming evolution.}.  Similarly, the specific form of the energy-momentum tensor $T^{\mu\nu} =e (\tauekt,\mathbf{x})  \times \text{diag}\left( 1,1/2,1/2,0 \right)$ in $(\tau,x,y,\eta)$ coordinates, can be motivated from the fact that due to the kinematics of high-energy collisions, the longitudinal momentum of each particle in the local rest frame is negligibly small compared to its transverse energy, such that the longitudinal pressure approximately vanishes  at very early times (c.f. \Sec{sec:ipglasma} for a microscopic description of the early time dynamics). Since the $T^{\tau i}$ momentum components of the energy-momentum tensor are also initialized as zero in the MC-Glauber model, the effective kinetic theory evolution of the energy-momentum tensor in \Eq{eq:GlauberInitial} involves only energy perturbations.

\begin{figure}
	\centering
	\includegraphics[width=0.9\linewidth]{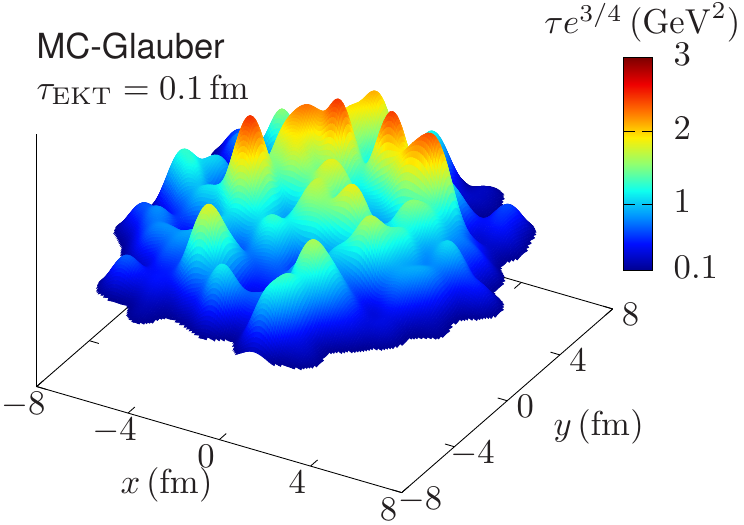}
	\caption{Transverse density of $\tau e^{3/4}\sim s\tau$ for MC-Glauber 
	event used in \Sec{sec:glauber} at initial time $\tau_\text{EKT}=0.1\,\text{fm}$ corresponding to a central PbPb event at center-of-mass energy $\sqrt{s_{NN}}=2.76\,\text{TeV}$. }
	\label{fig:trento3dtauekt}
\end{figure}

We used a single MC-Glauber event with $b=0.9\,{\rm fm}$ impact parameter, $N_{\rm part} =408$ participants and spatial eccentricity $\epsilon_2 = 0.064$.  The total transverse energy per unit rapidity in the event at $\tauekt=0.1\,{\rm fm}$ was set to 
$
\int d^2\x\, 
T^{\tau\tau} =  5.3 \times 10^4\, {\rm GeV/fm}
$
to normalize the energy density distribution.
 The transverse distribution of a proxy quantity $\tau e^{3/4}\sim s\tau$ for entropy per rapidity is shown in \Fig{fig:trento3dtauekt} for reference.
 Ultimately, after pre-equilibrium, hydrodynamic evolution, and freeze-out, this event corresponds to a midrapidity charged hadron multiplicity of $dN_{\textrm{ch}}/d\eta=1870$. The event is thus in line with a central LHC at $\sqrt{s_{NN}}=2.76\,\text{TeV}$, which, for reference, have $dN_{\textrm{ch}}/d\eta=1601\pm 60$ and $\langle N_{\textrm{part}} \rangle=383\pm 3$ for the 5\% most central events~\cite{Aamodt:2010cz}.

The energy momentum tensor in Eq.~(\ref{eq:GlauberInitial}) features a large pressure anisotropy indicating that the system at $\tauekt$ is still far from local equilibrium, 
and cannot be described properly by ordinary viscous hydrodynamics.  However, the use of \kompost{} to describe the subsequent pre-equilibrium evolution ($\tauekt<\tau<\tauhydro$) leads to the onset of hydrodynamic behavior that can be used as proper initial conditions for hydrodynamic evolution at $\tauhydro$.
The overlap in the range of validity of the pre-equilibrium and hydrodynamic phase ensures that the subsequent evolution is essentially independent on the switching time $\tauhydro$. In practice, the smoothness of the transition from the early stage of heavy ion collisions to hydrodynamics can be quantified in multiple manners. In what follows, we look at averages and profiles of the hydrodynamics fields  --- or equivalently the energy-momentum tensor -- as well as hadronic observables, and investigate their dependence on the hydrodynamic initialization time $\tau_{\textrm{hydro}}$.

\subsubsection{Average hydrodynamic fields}
\label{sec:Glauber_average}

\begin{figure*}
	\centering
(a-c) with lattice QCD equation of state\\
\subfig{a}{\includegraphics[width=0.3\linewidth]{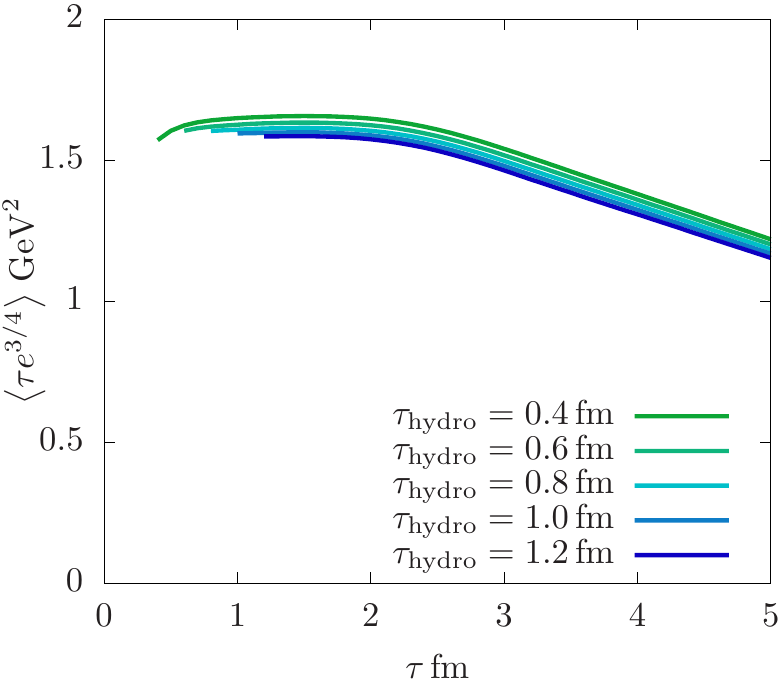}}\quad
\subfig{b}{\includegraphics[width=0.3\linewidth]{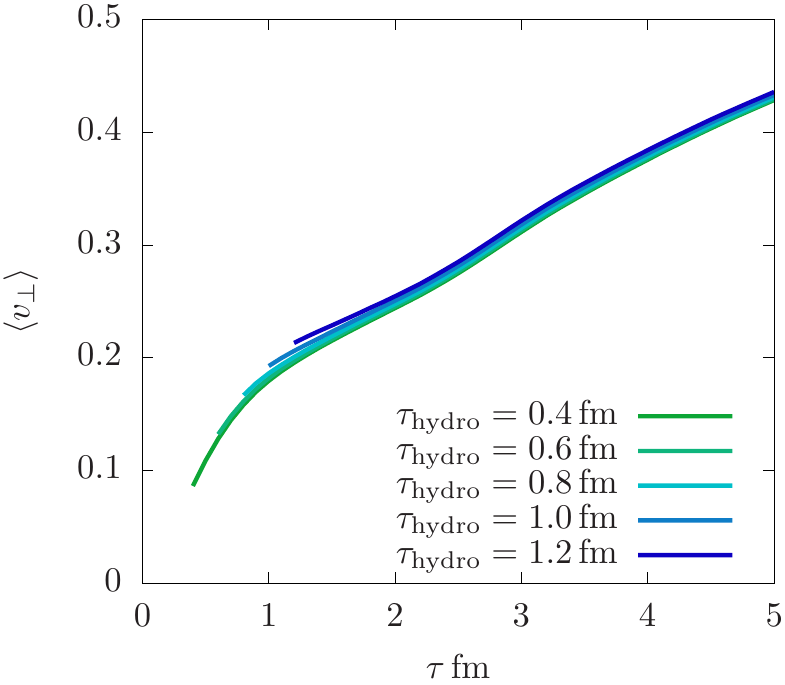}}\quad
\subfig{c}{\includegraphics[width=0.31\linewidth]{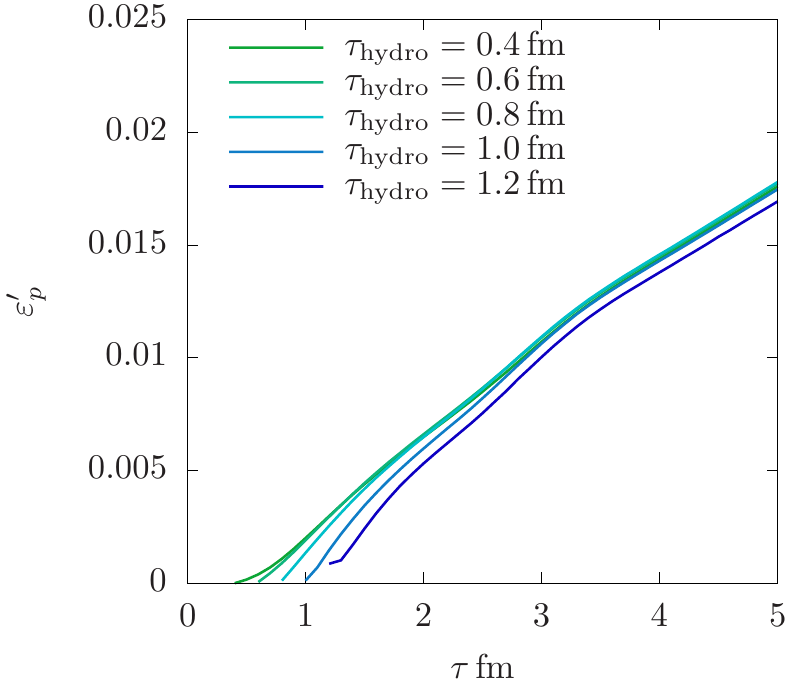}}
(d-f) with conformal equation of state\\
\subfig{d}{\includegraphics[width=0.3\linewidth]{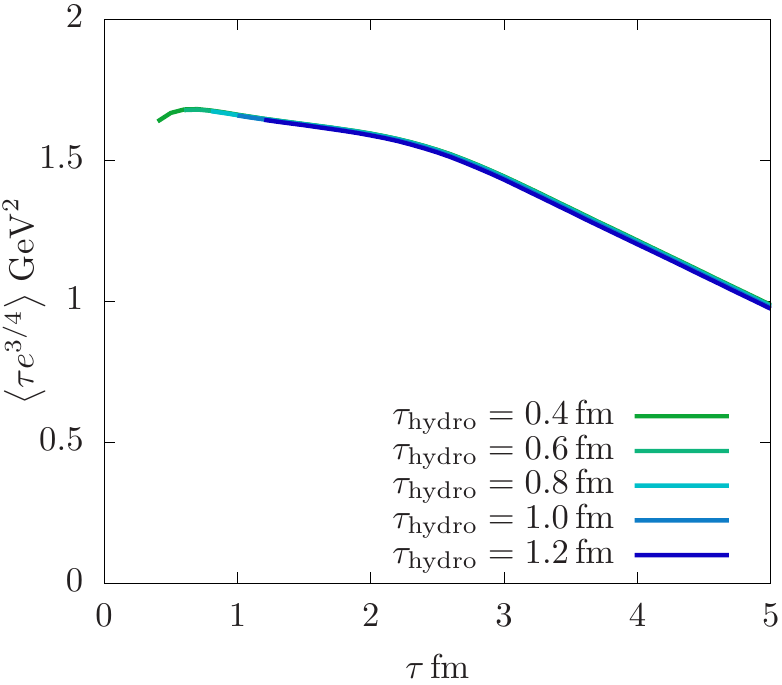}}\quad
\subfig{e}{\includegraphics[width=0.3\linewidth]{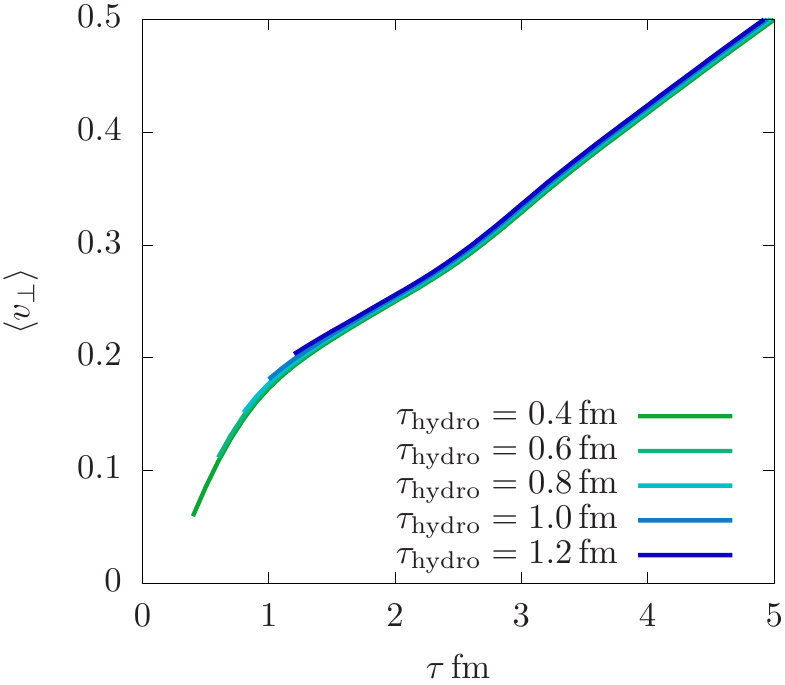}}\quad
\subfig{f}{\includegraphics[width=0.31\linewidth]{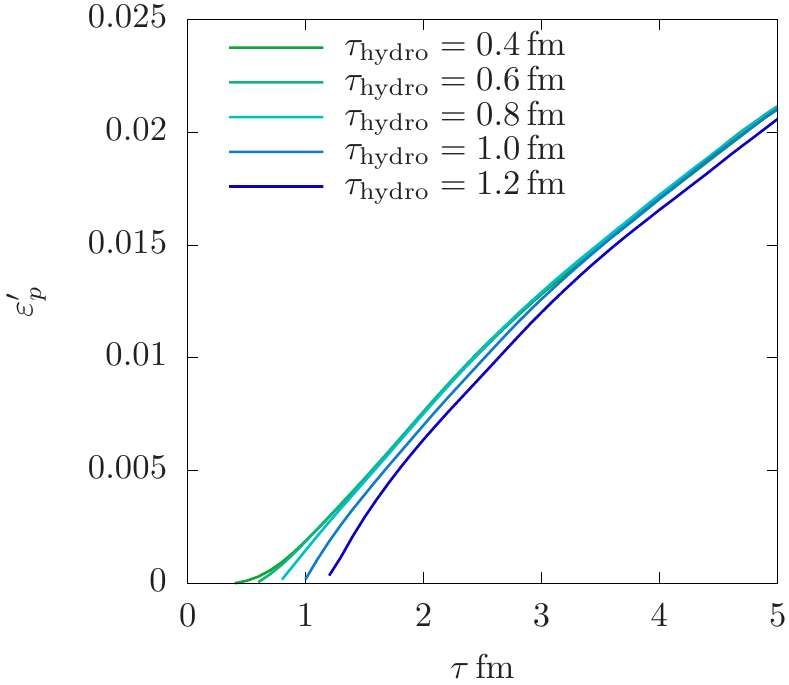}}
	\caption{(top row) The transverse averages (as defined in \Eq{eq:energy_average}) of (a) $\tau \epsilon^{3/4}$ and (b) transverse 
	velocity $v_\perp$, and (c) momentum 
	eccentricity (as defined by \Eq{eq:momentum_aniso}) in the hydrodynamic phase as a function of  
	time $\tau$. The different lines correspond to different 
	hydrodynamic initialization time $\tau_{\textrm{hydro}}$, i.e.\ different duration of kinetic pre-equilibrium evolution. The initial 
	condition of the effective kinetic theory at time 
	$\tauekt=0.1\,\text{fm}$ is a central participant MC-Glauber event 
	normalized to correspond to a $\sqrt{s_{NN}}=2.76$~TeV Pb-Pb collisions (see \Fig{fig:trento3dtauekt}). (bottom row) the same as (a-c), except a conformal equation of state is used in the hydrodynamics evolution instead of a lattice QCD one (see \Fig{fig:conformality}).}
	\label{fig:average_glauber_conformal}\label{fig:average_glauber_qcd}
\end{figure*}

Since realistic fluctuating initial conditions are used, hydrodynamic fields have a complicated profile in the transverse $xy$-plane. As a starting 
point, we look at transversely averaged values of hydrodynamic fields. We define averages $\left<\ldots\right>$ as
\begin{equation}
\left<\ldots\right>\equiv \frac{\int d^2\x u^\tau e \ldots}{\int d^2\x u^\tau e}
\label{eq:energy_average}
\end{equation}
where $\int d^2\x$ denotes an integral over the transverse coordinates. The factor $u^\tau$ is inserted for covariance, as would be obtained from the (covariant) surface flux $d\Sigma_\mu u^\mu$ with $d\Sigma_\mu$ in the (proper) time direction.

In \Fig{fig:average_glauber_qcd}(a) we present the evolution of $\left<\tau e^{3/4}\right>$, which is akin to the entropy per rapidity $\left<s\tau\right>$ of the boost-invariant system, as a function of physical time $\tau$. Initial conditions at $\tauekt=0.1$~fm are evolved with the kinetic theory in \kompost{} (c.f. Sec.~\ref{sec:implementation}), and subsequently passed to the hydrodynamic model at five values of $\tau_{\textrm{hydro}}$: $0.4, 0.6, 0.8, 1.0\text{ and }1.2\,\text{fm}$. 
We first note that \Fig{fig:average_glauber_qcd}(a) shows the expected transition from $\left<\tau e^{3/4}\right>$ being approximately constant at early time to subsequently dropping when the transverse expansion becomes important.
More importantly one observes from \Fig{fig:average_glauber_qcd}(a) that the evolution of $\langle\tau e^{3/4}\rangle$ depends very weakly on the 
hydrodynamics initialization time $\tauhydro$.

In \Fig{fig:average_glauber_conformal}(b) we show the rise of radial velocity $\left< v_\perp \right>$=$\left< \sqrt{v_x^2+v_y^2} \right>$ with time $\tau$. The kinetic theory pre-equilibrium captures well the rapid rise in the radial flow at early times, which levels off to a steady radial increase at later times. It is again remarkable that the spread between the calculations for different values of $\tauhydro$ between $0.4$~fm and $1.2$~fm is at a percent level. 

In order to follow the evolution of the azimuthal anisotropy, we computed
the integrated transverse stress tensor $[T^{ij}]_s = ([T^{xx}]_s, [T^{xy}]_s, [T^{yy}]_s)$,  where 
\begin{equation}
\left[\ldots\right]_s \equiv \int d^2\x u^\tau \ldots \, .
\end{equation}
 denotes an integral over the transverse plane without the energy weight, as $T^{\mu\nu}$ is already ``energy weighted''.
Examining the principal axes of $[T^{ij}]_s$,  we define the 
momentum ellipticity
\begin{equation}
\varepsilon_p^\prime(\tau) \equiv \frac{\sqrt{( \left[T^{xx}\right]_s-\left[T^{yy}\right]_s)^2+4\left[T^{xy}\right]_s^2}}{\left[T^{xx}\right]_s+\left[T^{yy}\right]_s} \, ,
\label{eq:momentum_aniso}
\end{equation}
which provides a measure of the elliptic flow as function of time.

Our results for momentum ellipticity are shown in
\Fig{fig:average_glauber_qcd}(c), and indicate that the dependence 
of the averaged hydrodynamic fields on $\tauhydro$ is modest even if
$\tauhydro$ is varied from $0.4$ to $1.2$~fm. Generally, the kinetic theory
slightly under-predicts the pressure anisotropy
$\varepsilon_p^\prime(\tau)$ in the hydrodynamic evolution. 
$\varepsilon_p^\prime(\tau)$ is inherently quadratic in the flow velocity, 
suggesting that including the first nonlinear couplings between
the background radial flow and the generated elliptic flow could improve 
the agreement here.

 \begin{figure}
    \centering
        \includegraphics[width=0.95\linewidth]{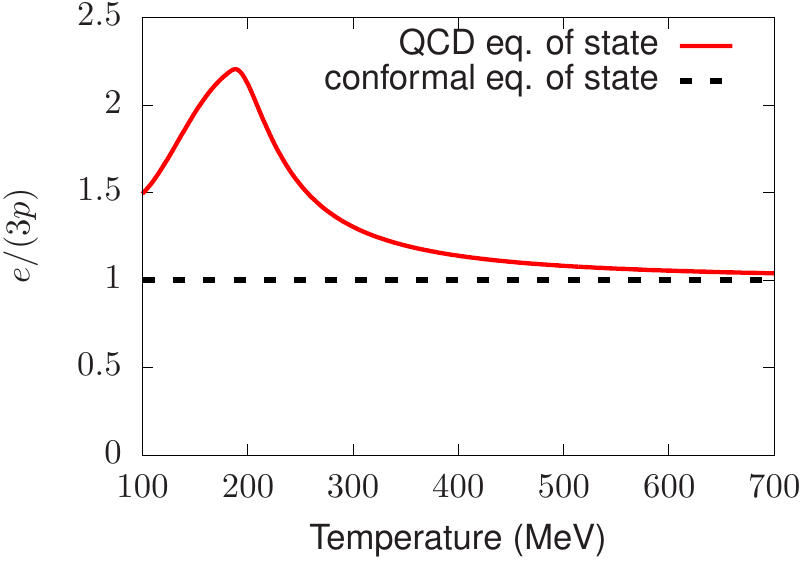}
        \caption{Deviation of the QCD equation of state~\cite{Huovinen:2009yb} from conformality, as 
        quantified by the ratio $e/(3 p)$.}
        \label{fig:conformality}
 \end{figure}

The effective kinetic theory approach used in this work assumes that the system is conformal, i.e. the pressure is $p(e)=\tfrac{1}{3}e$. 
Even though QCD is nearly  conformal at very high temperatures, deviations from conformality are expected for the range of temperatures of order $150{-}600$~MeV encountered in heavy ion collisions at the RHIC and LHC, as can be observed from Fig.~\ref{fig:conformality}, where the ratio 
$e/3p$ 
is shown for a QCD equation of state~\cite{Huovinen:2009yb}. Since hydrodynamic simulations of heavy ion collisions necessarily require a realistic equation of state,
this  leads to discontinuous matching of the energy-momentum tensor. Specifically, when the energy-momentum tensor of \kompost{} is decomposed and passed to the hydrodynamics, the energy-momentum tensor of the hydrodynamics becomes:
\begin{equation}
T^{\mu\nu}_{\textrm{hydro}}=e u^\mu u^\nu+ p_{\textrm{QCD}}(\epsilon) 
\Delta^{\mu\nu}+\pi^{\mu\nu}
\label{eq:TmunuIPG}
\end{equation} 
which is not equal to $T^{\mu\nu}$ of the effective kinetic theory because $p_{\textrm{QCD}}(e) \neq p_{\textrm{conformal}}(e)$. Unfortunately, there is no obvious way to improve on this procedure, as a better matching will ultimately require breaking the conformal symmetry in the pre-equilibrium phase which is of higher order in $\alpha_s$. 
On the other hand, it is straightforward to study and quantify the effects associated with this break of conformality, by replacing the QCD equation of state by a conformal one and reproducing \Fig{fig:average_glauber_qcd}(a-c). 
These results are presented in the bottom row of \Fig{fig:average_glauber_qcd}, where the different panels (d-f) again show the time evolution of $\left<\tau e^{3/4}\right>$, $\left< v_\perp \right>$ and $\varepsilon_p^\prime$. It is clear from these figures that the $\tauhydro$ dependence, which was already small for a QCD equation of state, is even smaller with the conformal equation of state.
In particular, all of the $\tauhydro$ dependence for $\left<\tau e^{3/4}\right>$ --- which amount to approximately $10\%$ between $\tauhydro=0.4$ and $1.2$~ --- is explained by the break of conformality between the initial conditions and the hydrodynamic evolution with  a QCD equation of state.
The flow observables $\left< v_\perp \right>$ and $\varepsilon_p^\prime$ also have a smaller dependence on $\tauhydro$ when a conformal equation of state is used, as can be observed by comparing \Fig{fig:average_glauber_qcd} (b) with (e), and (c) with (f). The effect is not as significant as for $\left<\tau e^{3/4}\right>$, however, in part because the $\tauhydro$ dependence was already small in the first place with the QCD equation of state.

\subsubsection{Transverse plane profiles of hydrodynamic fields}

\begin{figure*}
	\centering
	\subfig{a}{\includegraphics[height=0.25\linewidth]{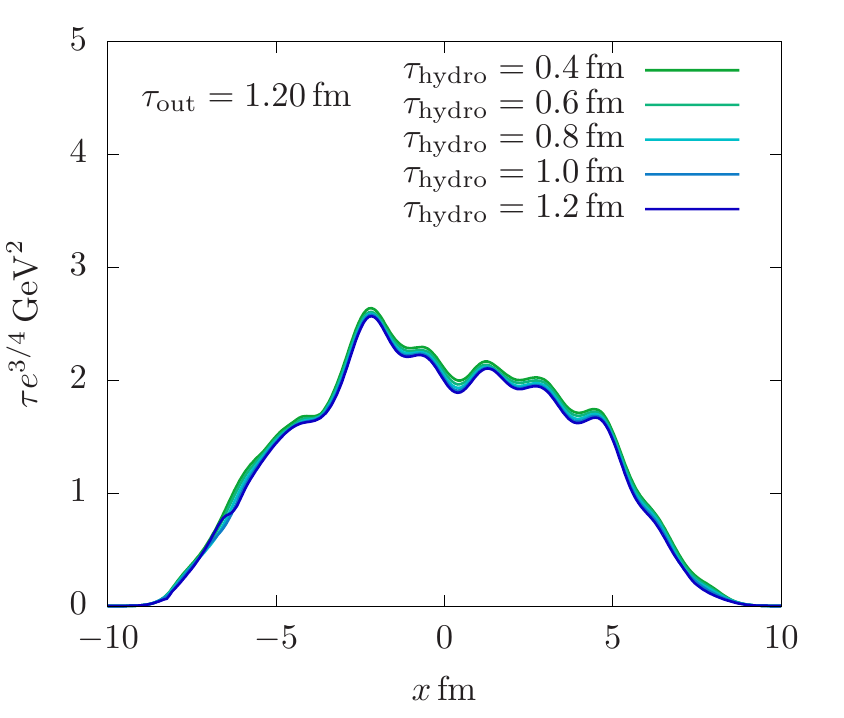}}\quad
	\subfig{b}{\includegraphics[height=0.25\linewidth,trim=25 0 0 0, clip]{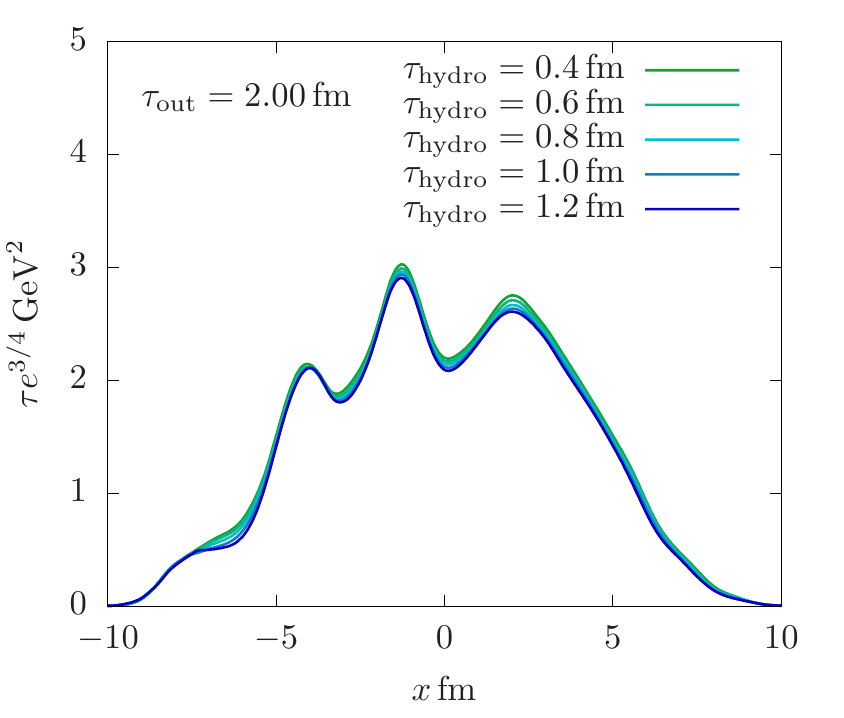}}\quad
	\subfig{c}{\includegraphics[height=0.25\linewidth,trim=25 0 0 0, clip]{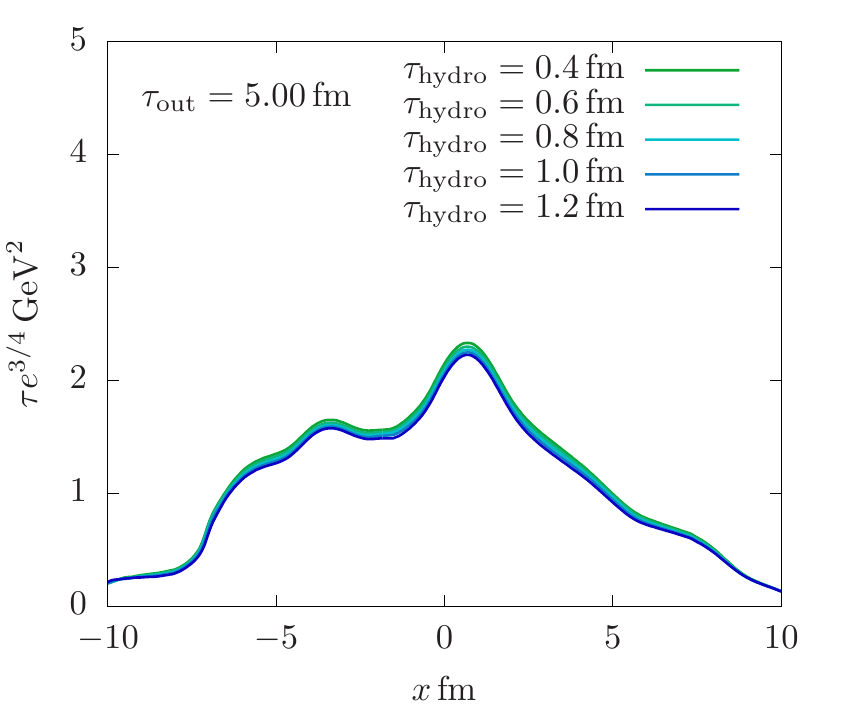}}
	\subfig{d}{\includegraphics[height=0.25\linewidth]{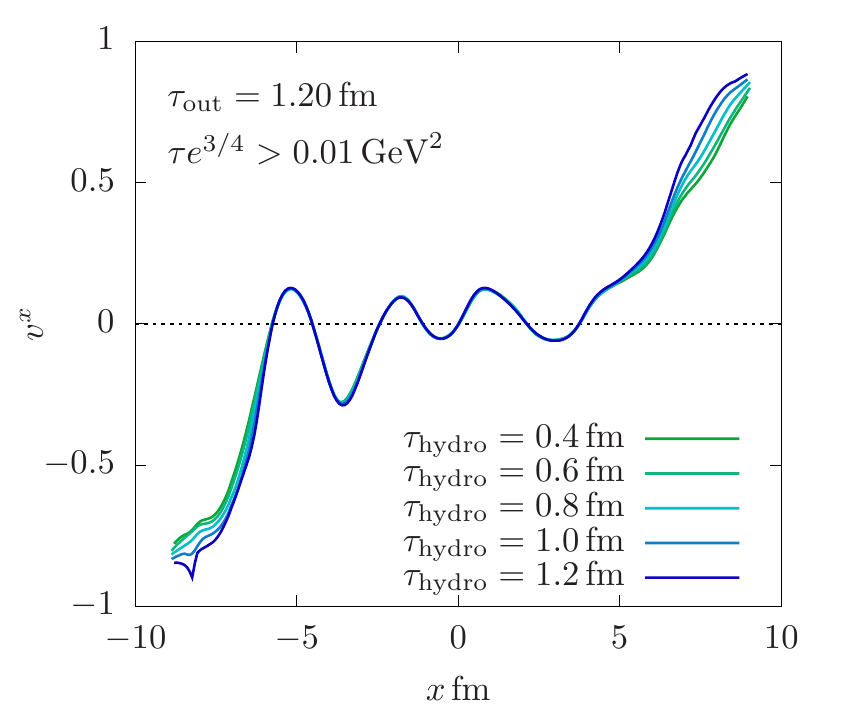}}\quad
	\subfig{e}{\includegraphics[height=0.25\linewidth,trim=28 0 0 0, clip]{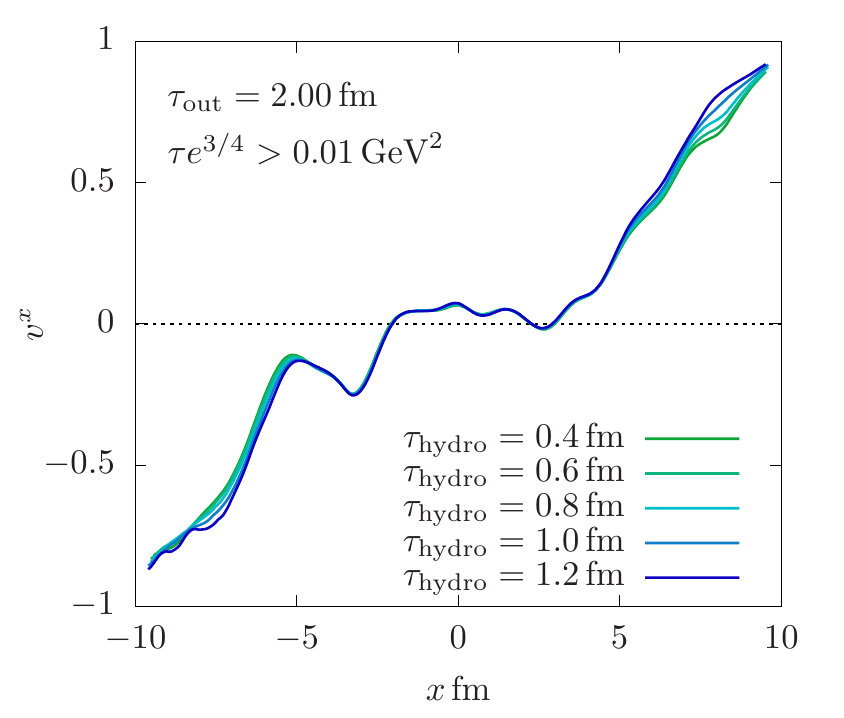}}\quad
	\subfig{f}{\includegraphics[height=0.25\linewidth,trim=28 0 0 0, clip]{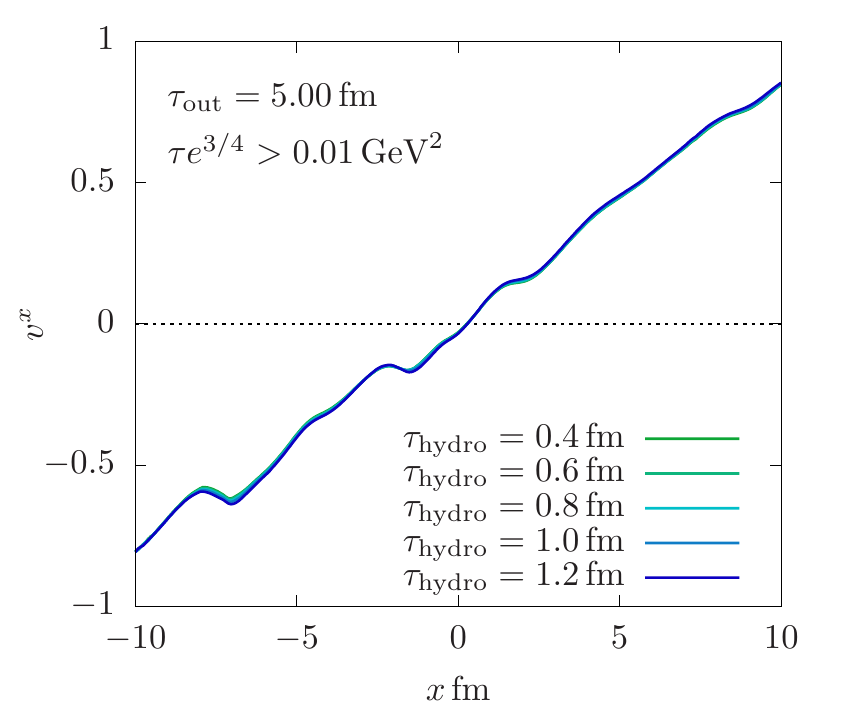}}%
	\caption{Single event profiles along $x$-axis ($y=0$) of $\tau e^{3/4}$ (top row) and velocity $v^x$ (bottom row) for different hydrodynamics transition times $\tauhydro$. Different columns correspond to three different times in hydrodynamic evolution: $\tau=1.2, 2.0$ and $5.0$~fm. The same EKT initialization time $\tauekt=0.1$~fm was used. The equation of state is a realistic QCD one. The transverse velocity is not shown for very low energy densities ($\tau e^{3/4}<0.01\,\text{GeV}^2$) where numerical errors can generate spurious values of velocity.	\label{fig:glauber_profiles}}
\end{figure*}

Based on our analysis in the previous section, the average energy, transverse velocity and momentum anisotropy were found to have a weak dependence on the value of $\tauhydro$. The reason for this insensitivity to $\tauhydro$ is that after a short evolution, the dynamics of energy-momentum tensor in the effective kinetic theory approaches that of hydrodynamics, and the two descriptions are approximately equivalent. However, the integrated quantities shown in the previous subsection have a significantly reduced sensitivity to any type of fluctuations in the hydrodynamic fields. In this section we show profiles of hydrodynamic fields in the transverse plane of the collisions, to emphasize that the smaller features of the fields do not show any significant sensitivity to the hydrodynamic initialization time $\tauhydro$ either. 
Since the effect of the transition from a conformal kinetic theory to a non-conformal hydrodynamics evolution was quantified in the previous section, it is not revisited again here and all results that follow were obtained with a QCD equation of state.

\begin{figure*}
	\centering
	\subfig{a}{\includegraphics[width=0.4\linewidth]{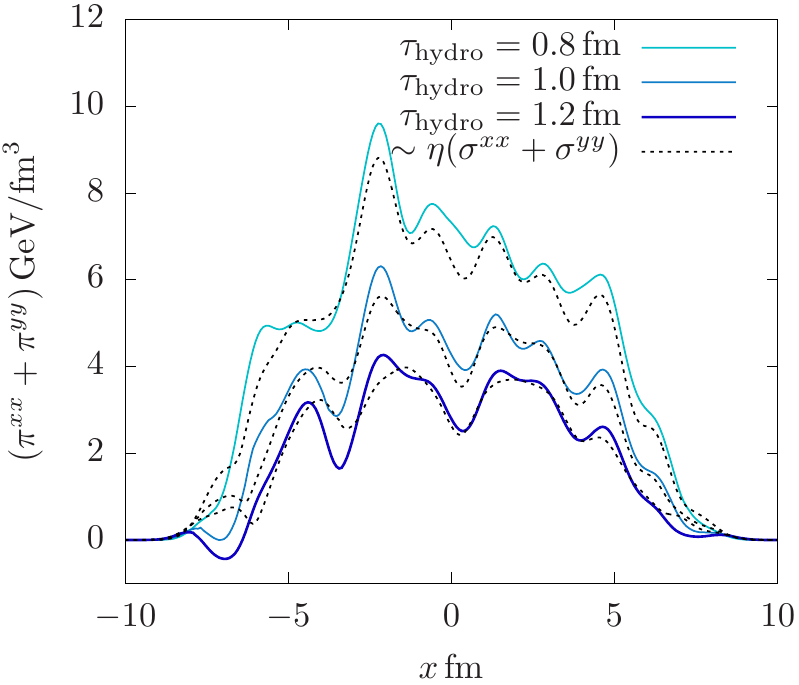}}\quad
	\subfig{b}{\includegraphics[width=0.4\linewidth]{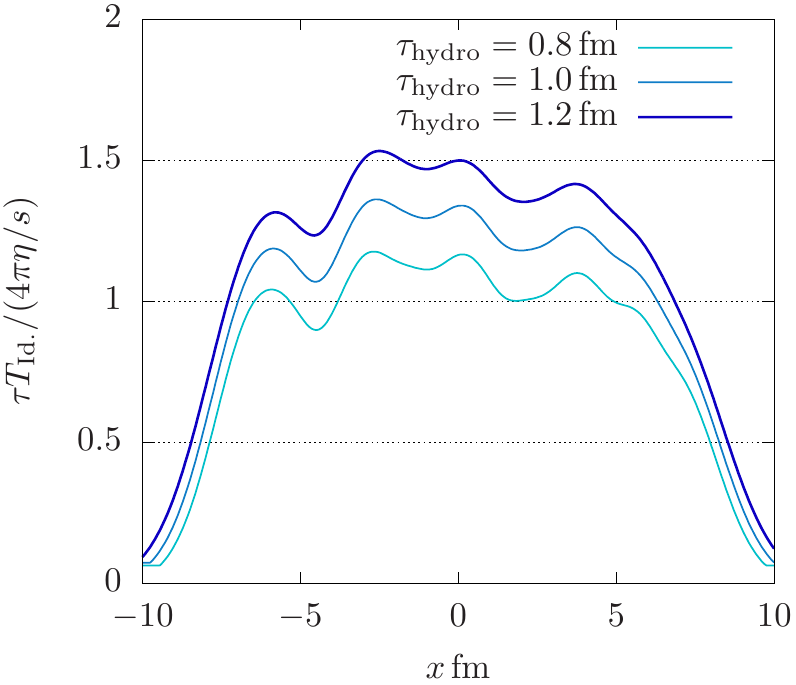}}
	\caption{(a) Comparison of the out-of-equilibrium shear stress tensor (c.f.\,\Eq{eq:TmunuIPG}) with the Navier-Stokes estimate at different hydrodynamics initialization times $\tauhydro=0.8,1.0,1.2$~fm. (b) Scaled evolution time variable $\xSc$ at different hydro starting times. 
   Values of $\tau \TId/(4\pi\eta/s)>1$ indicate that the system is close enough to local thermal equilibrium for hydrodynamics to become applicable (see \Sec{sec:hydtime}).
   }
	\label{fig:glauber_pimunu}
\end{figure*}

In \Fig{fig:glauber_profiles} we show profile plots of $\tau \epsilon^{3/4}$ (top row) and the flow velocity $v^x$ (bottom row) along the $x$-axis 
at midrapidity. Different panels (a-f) show the profiles at different hydrodynamic evolution times $\tau=1.2, 2.0$ and $5.0$~fm. The different curves on each panel correspond to a hydrodynamic \emph{initialization} times $\tauhydro$ from $0.4$ to $1.2$~fm. One observes that even for differential observables, the sensitivity to the initialization time of the hydrodynamic evolution is very small. Except for a few percent change in the overall normalization of $\tau \epsilon^{3/4}$ between the different curves, which can be attributed to the mismatch between the conformal equation of state in \kompost{} to the QCD equation of state in the hydrodynamic evolution, the profiles look essentially identical indicating a robust matching of the pre-equilibrium dynamics to viscous hydrodynamics. 

One can further probe the approach of kinetic theory towards a hydrodynamic evolution by comparing the out-of-equilibrium shear-stress tensor $\pi^{\mu\nu}$ 
from the kinetic theory evolution with an estimate from the Navier-Stokes value $\pi^{\mu\nu}=-\eta\sigma^{\mu\nu}$, where $\sigma^{\mu\nu}$ is calculated from 
the velocity profile, see \Eq{eq:sigmamunu}. In \Fig{fig:glauber_pimunu}(a) we plot the value of $\pi^{xx}+\pi^{yy}$ for $\tauhydro=0.8,1.0,1.2\,\text{fm}$ and Navier-Stokes 
estimate in dashed lines. We see that in most of the collision area the kinetic theory result approached hydrodynamic constitutive equations. One exception is 
the sharp edges of the fireball where the small gradient assumption breaks down. However, it is not clear that either hydrodynamics or linearized kinetic theory \`a la \kompost{} provide an accurate description of the space-time evolution of the edges. The regions where a good matching between kinetic theory and hydrodynamics is expected can be quantified with the typical momentum relaxation time $\tau_R(\tau)$ defined at \Eq{eq:tauR} in Section~\ref{subsec:background}, using $\tau/\tau_R(\tau) \gtrsim 4 \pi$, i.e. $\tau \TId/(4\pi\eta/s)\gtrsim  1$. This ratio can be calculated locally
  and indicates how the approach to hydrodynamics varies in the transverse plane. The result is shown in \Fig{fig:glauber_pimunu}(b). As estimated in \Sec{sec:hydtime}, we find that for starting times $\tauhydro>0.8\,\text{fm}$, most of the medium is at scaled 
time $\tau \TId/(4\pi\eta/s)>1$ where hydrodynamics becomes applicable. Since the local energy density at the edges is significantly smaller, the edges of the fireball remain at $\tau \TId/(4\pi\eta/s)<1$ for a longer time, quantifying the statement that the approach to hydrodynamic behavior does not occur isochronously.

\subsubsection{Hadronic observables}
\label{sec:glauber_hadronic}

\begin{figure*}
    \centering
\subfig{a}{       \includegraphics[width=0.31\linewidth]{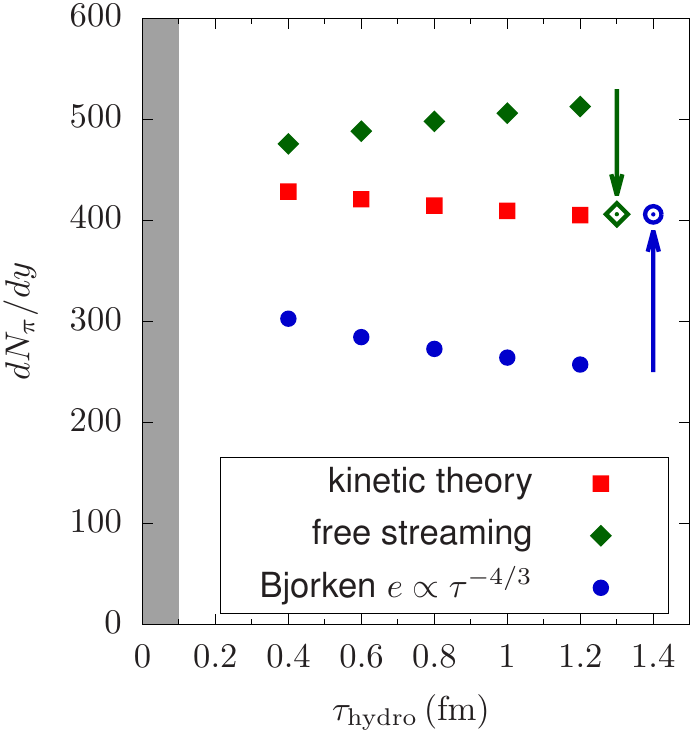}}%
\subfig{b}{\includegraphics[width=0.31\linewidth]{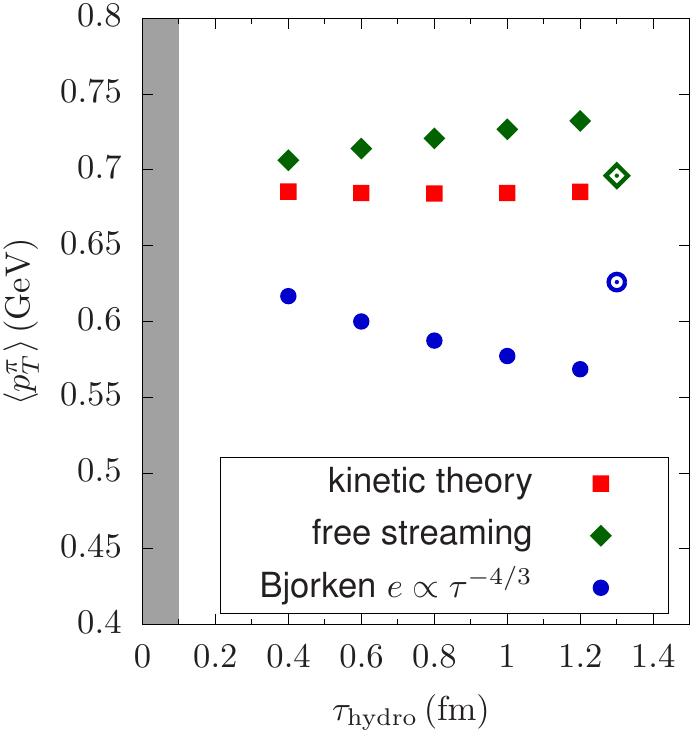}}%
\subfig{c}{\includegraphics[width=0.31\linewidth]{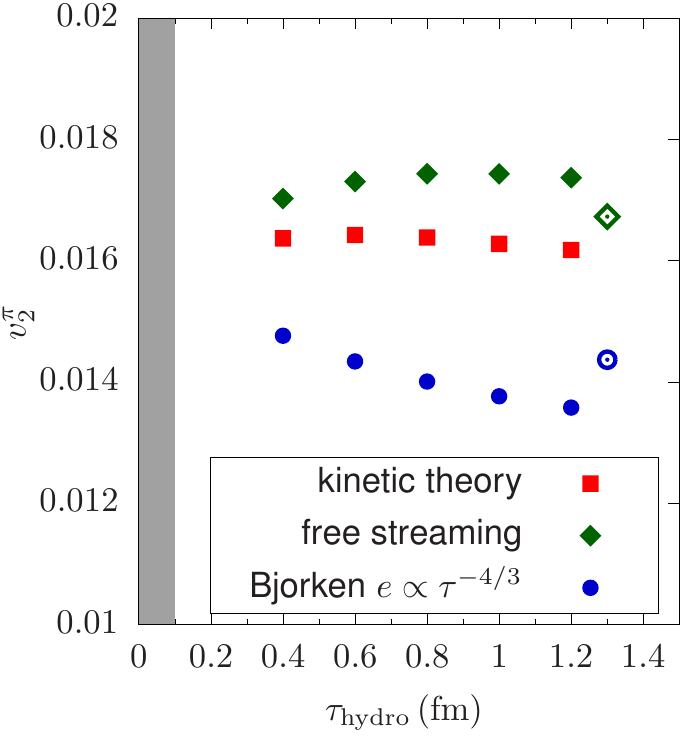}}
    \caption{
       The thermal freeze-out pion
   (a) multiplicity $dN_\pi/dy$ (b) radial flow $\langle p_T^\pi\rangle$  and
   (c) elliptic flow $v_2^\pi$ as a function of hydrodynamic initialization
   time $\tauhydro=0.4{-}1.2\,\text{fm}$, i.e. different duration of
   pre-equilibrium evolution. (Note the suppressed zero in (b) and (c).) The initial MC-Glauber conditions are specified at $\tauekt=0.1\,\text{fm}$ as indicated by the grey band. Different
   pre-equilibrium scenarios are: linearized kinetic theory evolution
   \kompost{}, interaction-less free streaming and simple Bjorken energy
   rescaling with time $\propto \tau^{-4/3}$ with no dynamics. The open symbols
   show the $\langle p_T^\pi \rangle$ and $v_2^\pi$ for the  free streaming and
   Bjorken pre-equilibrium evolution from 0.1\,fm to 1.2\,fm with energy
   density scaled to reproduce the same pion multiplicity as \kompost{} with
   $\tauhydro=1.2\,\text{fm}$.
\label{fig:v2WithMom}\label{fig:multWithMom}\label{fig:meanptWithMom}}
		\label{fig:glauber_hadronic_obs}
\end{figure*}

Based on the successful matching of the early time pre-equilibrium stage to the subsequent hydrodynamic regime discussed in the previous sections, we now investigate the impact of a consistent description of the early time dynamics on the final state hadronic observables computed after the freeze-out of the hydrodynamic evolution. We focus on the multiplicity $dN_{\pi}/dy$, the average transverse momentum $\langle p_T^{\pi} \rangle$ and the $v_2^{\pi}$ of thermal pions\footnote{We  reiterate, as noted in \Sec{sec:implementation}, that hadronic decays are not included.}, which can be thought of as analogues of the integrated hydrodynamic fields shown in \Fig{fig:average_glauber_qcd}. We note that, although only pion observables are shown in this section, we verified that similar results are found for heavier hadrons such as kaons and protons.

Starting with the same Glauber initial conditions at $\tauekt=0.1\,\text{fm}$, the effective kinetic theory is used to evolve the energy-momentum tensor up to five different times $\tauhydro$ from 0.4~fm to 1.2~fm, as in the previous sections. Subsequently, the hydrodynamic evolution is performed up to the isothermal freeze-out where hadronic observables are calculated. In \Fig{fig:glauber_hadronic_obs}, our results for the pion multiplicity $dN_{\pi}/dy$, the mean transverse momentum $\langle p_T^{\pi} \rangle$ and the $v_2^{\pi}$ are plotted as a function of $\tauhydro$. In addition to the results obtained from the effective kinetic theory pre-equilibrium evolution, the dependence of hadronic observables on $\tauhydro$ is also shown for two other types of pre-equilibrium evolution: free-streaming\footnote{Our results for free-streaming evolution are obtained by replacing the kinetic evolution of the background energy $\mathcal{E}(x)$ and the corresponding response functions $G^{\mu\nu}_{\alpha\beta}$ with their free-streaming counterparts (see  \app{sec:freestreaming}).} and simple Bjorken $\tau^{-4/3}$ scaling. 
The use of free-streaming to describe the pre-equilibrium dynamics has been studied previously in~\cite{Broniowski:2008qk,Liu:2015nwa}. Similarly, the procedure of scaling the energy density of a set initial condition with $\tau^{-4/3}$, as would be expected for a system undergoing ideal Bjorken hydrodynamic expansion, is also used regularly in heavy ion physics to rescale the energy density of the initial conditions when changing the initialization time of hydrodynamics.

In \Fig{fig:multWithMom}(a) we show pion multiplicity $dN_{\pi}/dy$ as a function of hydrodynamic initialization time $\tauhydro$ for the three different pre-equilibrium evolution scenarios. We find that for all pre-equilibrium scenarios, the multiplicity has an approximately linear dependence on $\tauhydro$. For the effective kinetic theory, the multiplicity is only approximately $\sim 5\%$ smaller if the hydrodynamics is initialized at $\tauhydro=1.2$~fm rather than $\tauhydro=0.4$~fm, while in the case of the Bjorken $\tau^{4/3}$, this figure is significantly larger, $\sim 15\%$. Conversely, for a free-streaming pre-equilibrium dynamics the longitudinal pressure is underestimated during the pre-equilibrium phase, such that the energy decreases much less rapidly than in hydrodynamics, and the multiplicity is $\sim 8\%$ \emph{larger} with $\tauhydro=1.2$~fm than with $\tauhydro=0.4\,\text{fm}$. We conclude that overall, the pion multiplicity has the smallest dependence on $\tauhydro$ when \kompost{} is used to describe the early stage of the medium evolution, although all curves are relatively flat.

In \Fig{fig:glauber_hadronic_obs}(b) we look at the radial flow dependence on $\tauhydro$, again for kinetic theory, free streaming and $\tau^{4/3}$ scaling. We find that for 
kinetic theory equilibration, the radial flow build-up is consistent with hydrodynamic evolution and the mean pion $\left<p_T^\pi\right>$ is independent of 
switching time. In contrast, for the free streaming evolution, the overall energy scale grows slightly too rapidly, leading to a $\sim 3\%$ change in $\langle p_T 
\rangle$ if the hydrodynamics is initialized at $\tauhydro=1.2$~fm rather than $\tauhydro=0.4$~fm. Conversely, for Bjorken $\tau^{4/3}$ scaling, no pre-flow is built up during the pre-equilibrium stage resulting in a decrease by $\sim 8\%$ of $\left<p_T\right>$ for $\tauhydro=1.2$~fm rather than $\tauhydro=0.4$~fm. Similar observations can be made for the second order flow harmonic $v_2$, presented in \Fig{fig:glauber_hadronic_obs}(c), albeit the overall magnitude of the variations is somewhat smaller in this case.

Besides the different $\tauhydro$-dependence of $\left<p_T^\pi\right>$ and $v_2$, it is also clear from \Fig{fig:glauber_hadronic_obs} that the $\left<p_T^\pi\right>$ and $v_2$ obtained after a kinetic pre-equilibrium evolution is different from that obtained through the free-streaming and $\tau^{4/3}$ scaling. Of course the same is also true for the multiplicity in \Fig{fig:glauber_hadronic_obs}(a): given the same initial conditions, $dN_\pi/dy$ is larger for a free-streaming evolution than a kinetic one, and smaller for $\tau^{4/3}$ scaling.
In practice, the normalization of the initial conditions of hydrodynamics simulations of heavy-ion collision is always adjusted so as to reproduce hadronic multiplicities. It is thus relevant to ask how $\left<p_T^\pi\right>$ and $v_2$ change for free-streaming and $\tau^{4/3}$ scaling if their respective initial condition normalizations are adjusted so that they produced the same pion multiplicity as the kinetic theory evolution. This result is shown with the open symbol on   \Fig{fig:glauber_hadronic_obs}, for a fixed $\tauhydro=1.2$~fm.
We see that even if the pion multiplicity is fixed by hand, different pre-equilibrium dynamics still lead to a difference in $\left<p_T\right>$ and $\left<v_2^\pi\right>$, although this difference becomes very small for free-streaming.
In the case of Bjorken $\tau^{4/3}$ scaling, the discrepancy remains relatively large, which we attribute in part of the lack of pre-equilibrium flow velocity resulting from this simplified scaling.

\subsection{Energy and momentum perturbations with IP-Glasma initial 
conditions\label{sec:ipglasma}}

\begin{figure}
	\centering
	\includegraphics[width=\linewidth]{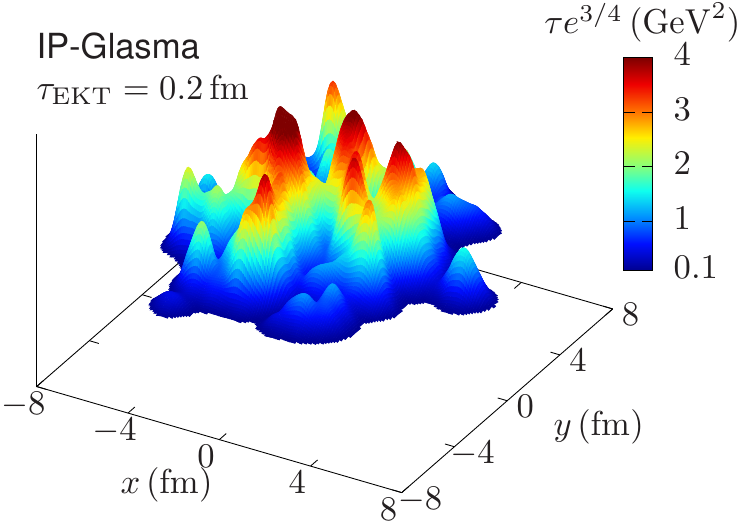}
	\caption{Transverse ``entropy'' density \ $\tau e^{3/4}\sim s\tau$ for a single  $\sqrt{s_{NN}}=2.76\,\text{TeV}$ central Pb-Pb  IP-Glasma 
	event at kinetic theory initialization  time $\tau_\text{EKT}=0.2\,\text{fm}$. }
	\label{fig:ipglasma3dtauekt}
\end{figure}

Besides being applicable to general initial condition ansatzs,
our framework of kinetic pre-equilibrium evolution can also be applied to microscopically 
motivated initial states. In this section we use the IP-Glasma model, where the 
evolution at early times of heavy ion collisions is described in terms of classical Yang-Mills (CYM)
dynamics~\cite{Schenke:2012wb,Schenke:2012fw}.   We note that in contrast to our previous 
discussion of the MC-Glauber model, the microscopic IP-Glasma model also 
includes initial momentum fluctuations $\delta T^{\tau i}$, which are propagated by our kinetic theory evolution. Therefore we need the complete set of response functions discussed in \Sec{sec:generalresponse}.

Since classical-statistical field theory and effective kinetic theory 
have an overlapping range of validity, the combination of the two 
 allows for a consistent weak coupling description of 
early time dynamics~\cite{Mueller:2002gd,Jeon:2004dh, York:2014wja, Kurkela:2015qoa}. In principle, such combined approach  describes particle production  from the partonic structure of nuclei at high 
energies to the onset of 
hydrodynamics in the quark-gluon plasma.

\begin{figure}
	\centering
	\subfig{a}{\includegraphics[width=0.95\linewidth]{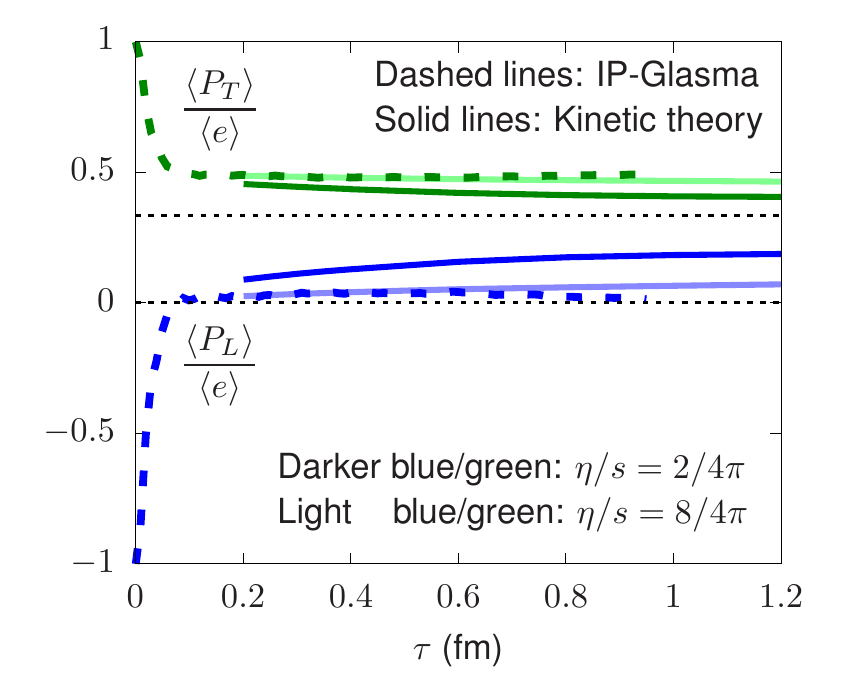}}
	\subfig{b}{\includegraphics[width=0.95\linewidth]{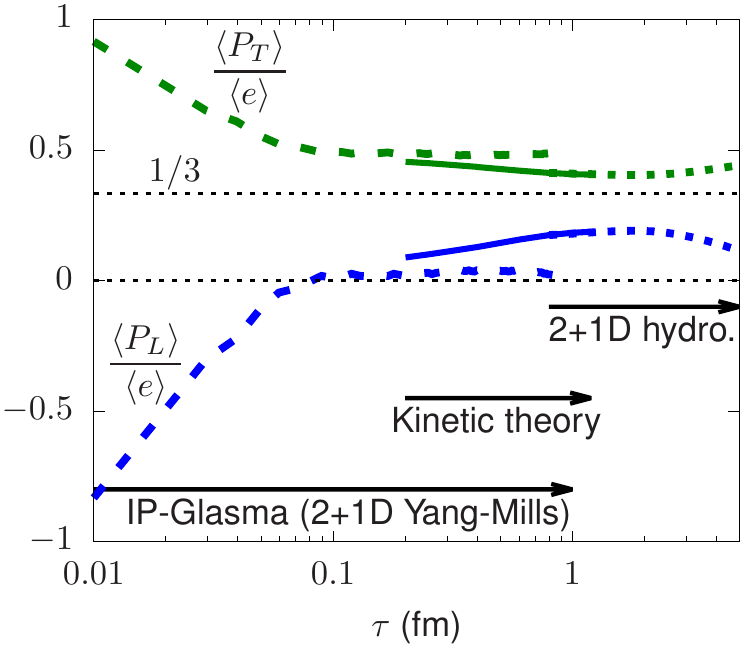}}%
   \caption{Time evolution of the longitudinal  and transverse 
   stress tensor components,
   $\llangle P_L \rrangle$ and $\llangle P_T \rrangle$,  
   averaged across the transverse plane, 
   relative to the average
   background energy density of a single IP-Glasma event shown in
   \Fig{fig:ipglasma3dtauekt}.    (a) 
   $\llangle P_L \rrangle$ and $\llangle P_T \rrangle$  
   in the 2+1D Yang Mills and kinetic theory codes for
   a range of $\eta/s$ at early times.  (b) 
   $\llangle P_L \rrangle$ and $\llangle P_T \rrangle$  
   in the 2+1D Yang Mills, kinetic theory,  and viscous hydrodynamics codes with $\eta/s{=}2/4\pi$ for a wide range of times.
   After $\sim 2\,{\rm fm}$ the 3D hydrodynamic expansion starts,  and the (integrated) estimate 
   $\llangle P_L \rrangle/\llangle e \rrangle \simeq \tfrac{1}{3}$ does
   not serve as a useful measure of local thermal equilibrium for a flowing fluid. }
	\label{fig:pressure_matching_ipglasma}
\end{figure}

In what follows, we use the original version of the IP-Glasma model, which is effectively 2+1D 
with boost-invariant fields in the longitudinal direction. 
Even though a qualitative matching between  classical-statistical field theory and effective kinetic theory has been 
demonstrated in previous works~\cite{Baier:2000sb,Kurkela:2011ub,Berges:2013fga,Kurkela:2015qoa}, no concrete implementation has been 
achieved to date for realistic full 2+1D simulation of heavy ion collisions.
Because of the reduction to 2+1D (boost-invariant) fields, the dynamics in IP-Glasma becomes 
effectively free-streaming after $\tau \sim 1/Q_s \sim 
\mathcal{O}(0.1~\textrm{fm})$ of classical Yang-Mills evolution~\cite{Schenke:2015aqa}. This is different from the first stage of ``bottom-up" equilibration, which is recovered in a full 3+1D classical-statistical simulation~\cite{Berges:2013fga,Berges:2013eia}.
However in practice,
we find that at weak coupling (larger values of $\eta/s$) the matching between classical Yang-Mills and kinetic theory is still rather 
smooth, as long as the switching to the kinetic description is performed at 
sufficiently early times $(1/Q_s \ll \tauekt \ll (4\pi \eta/s)/\TId)$ as demonstrated below.

In order to illustrate the smooth matching at early times, we analyze the evolution of transversely averaged longitudinal $\langle P_L\rangle$  and transverse $\langle P_T \rangle$ 
pressures defined by \Eq{eq:defPLPT}, relative to the mean background energy density $\langle e\rangle $ in a particular event shown in \Fig{fig:ipglasma3dtauekt}.
The IP-Glasma event used in our analysis is a central event with $N_p=404$ participant nucleons and an eccentricity $\epsilon_2=0.126$ at time $\tauekt=0.2$~fm.
The total energy in the event is
$
\int d^2{\x} \tauekt  T^{\tau\tau}
= 4.4 \times 10^4\, \frac{{\rm GeV}}{{\rm fm}} ,
$
corresponding to a charged hadron multiplicity of $dN_\text{ch}/d{\eta}=1419$. Fluctuations smaller than $\tauekt$ have been smeared with a Gaussian.

In \Fig{fig:pressure_matching_ipglasma} we present the time dependence of the pressure to energy 
ratio
$\left<P_{L,T}\right>/\langle e\rangle$ during  various stages of the evolution. 
In panel a) we see that at  
very early times ($\tau \ll 1/Q_s$) in the IP-Glasma evolution, the 
longitudinal 
pressure is negative due to the presence of strong longitudinal color fields~\cite{Lappi:2006fp}.
However, on a time scale $\tau\sim1/Q_s \sim 0.1\,\text{fm}$ the fields decay, and the longitudinal 
pressure  
approaches zero, except for a slight overshoot due to the 
residual pressure in the classical fields. At this time the 2+1D classical Yang-Mills dynamics becomes essentially free streaming, and one should 
then evolve the
system with QCD kinetics
in order to describe the subsequent approach towards equilibrium. 
Since at very early times, the expansion-dominated kinetic theory is also effectively free-streaming,  the classical Yang-Mills and kinetic theory evolutions
can be smoothly matched,  provided $\tauekt \gtrsim 1/Q_s$
is small in units of the relaxation time,  
$\tauekt \TId/(4\pi\eta/s)\ll 1$.  Indeed, 
in \Fig{fig:pressure_matching_ipglasma}(a) we see that  IP-Glasma to kinetic-theory transition becomes increasingly smooth 
as $4\pi\eta/s$ at early times is increased from $2$ to $8$. 
The energy density is high at such early times,
and this would  naturally lead to larger values of $\eta/s$ in a theory where the coupling runs.
Of course the physics of the running coupling is  absent in our leading order
analysis where $4\pi\eta/s \simeq 2$.

In  \Fig{fig:pressure_matching_ipglasma}(b) we  see that during 
kinetic phase the pressure anisotropy decreases, and 
after the system is sufficiently close to local equilibrium  $\tau 
\TId/(4\pi\eta/s)>1$, the subsequent evolution 
can be smoothly matched to second order viscous hydrodynamics,  as
discussed in \Sec{subsec:background}. 
\Fig{fig:pressure_matching_ipglasma} represents the integrated stress and
energies across a realistic IP-Glasma event,
which approximately follows a 1D  Bjorken expansion
for the  first $\tau\lesssim
2.5\,\text{fm/c}$. 
After this time the 3D hydrodynamic expansion starts, and the (integrated)
estimate $\llangle P_L \rrangle/\llangle e \rrangle \simeq \tfrac{1}{3}$ does
not provide a useful measure of local thermal equilibrium in the flowing fluid.

Having investigated the different stages of the evolution,  we will evaluate 
the transition between kinetic theory and hydrodynamics in 
greater detail  by monitoring the
stress tensor, 
and by checking that  hadronic
observables are independent of the crossover time $\tauhydro$. The analysis parallels the
MC-Glauber initial conditions described in \Sec{sec:glauber}, and the discussion will highlight the differences.

\subsubsection{Average hydrodynamic fields}
\begin{figure*}
	\centering
	(a-c) with latttice QCD equation of state\\
	\subfig{a}{\includegraphics[width=0.3\linewidth]{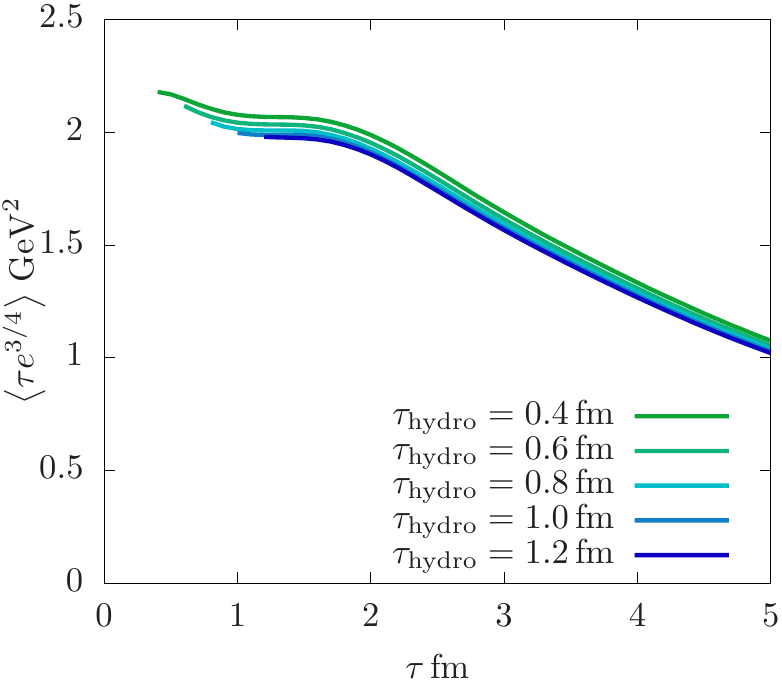}}\quad
	\subfig{b}{\includegraphics[width=0.3\linewidth]{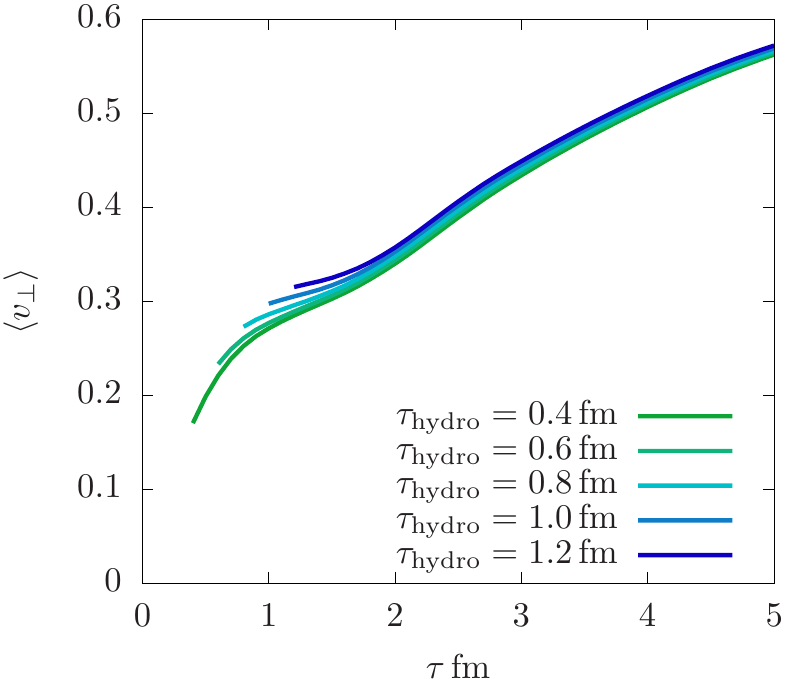}}\quad
	\subfig{c}{\includegraphics[width=0.3\linewidth]{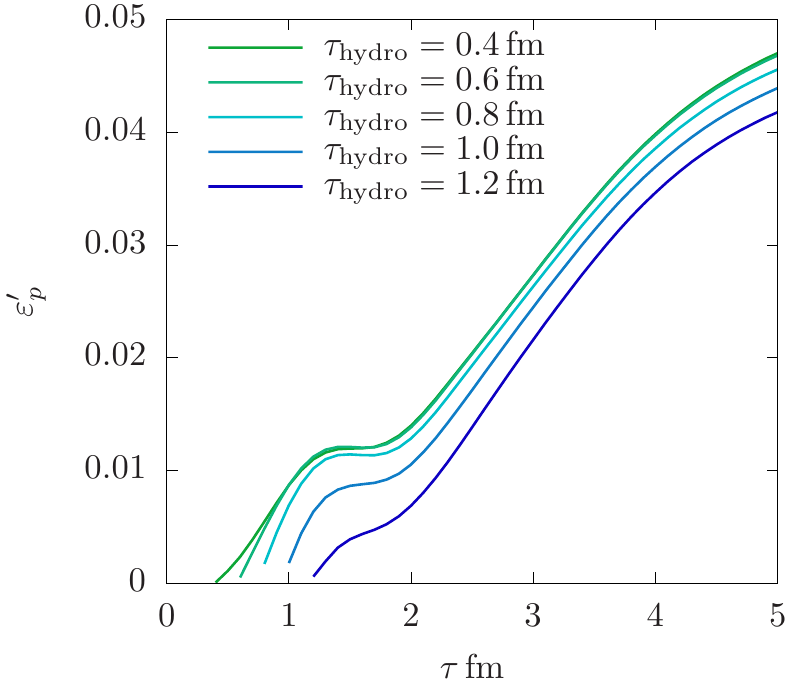}}
	(d-f) with conformal equation of state\\
	\subfig{d}{\includegraphics[width=0.3\linewidth]{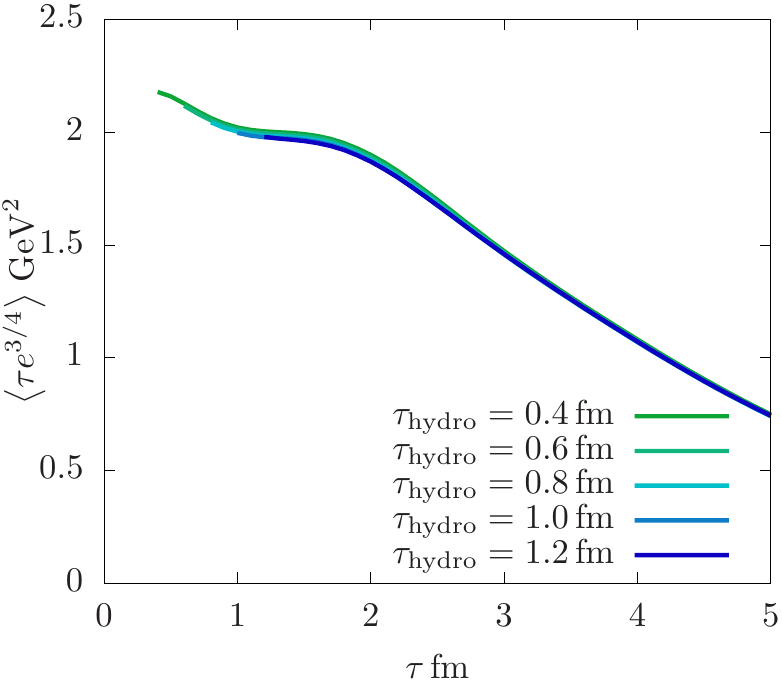}}\quad
	\subfig{e}{\includegraphics[width=0.3\linewidth]{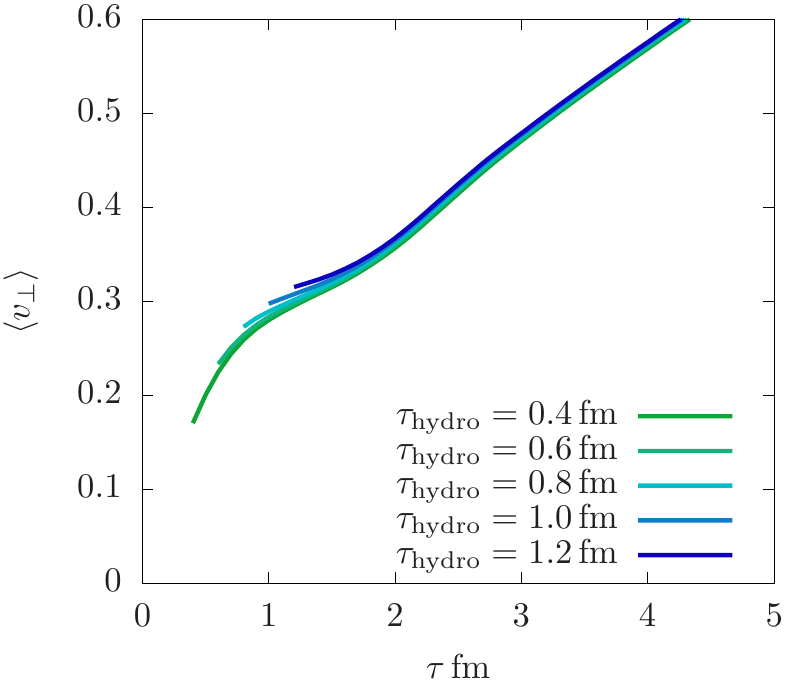}}\quad
	\subfig{f}{\includegraphics[width=0.3\linewidth]{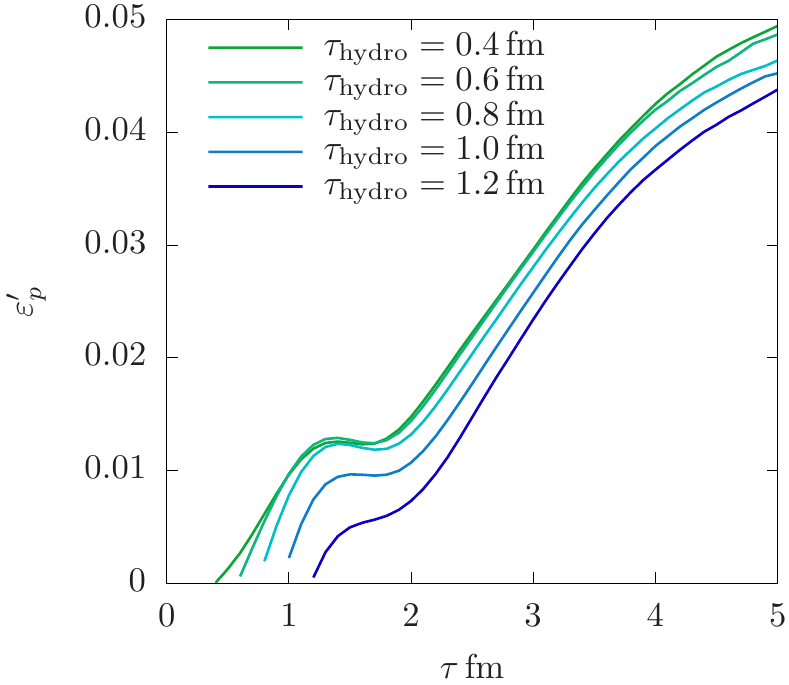}}
	\caption{(top row) Transverse average of $\tau e^{3/4}$ and of the transverse 
		velocity $v_\perp$, as defined in \Eq{eq:energy_average}, and momentum 
		eccentricity as defined by \Eq{eq:momentum_aniso}, as a function of  
		time $\tau$. The three average fields are plotted for different 
		hydrodynamic initialization time $\tauhydro$, i.e. different duration of kinetic pre-equilibrium evolution. The initial 
		condition of the effective kinetic theory at time 
		$\tauekt=0.2$~fm is a central IP-Glasma events normalized to 
		correspond to a $\sqrt{s_{NN}}=2.76$~TeV Pb-Pb collisions (see \Fig{fig:ipglasma3dtauekt}). (bottom row) the same as (a-c), except a conformal equation of state is used in the hydrodynamics evolution instead of a lattice QCD one. }
	\label{fig:average_ipglasma_conformal}	\label{fig:average_ipglasma_qcd}
\end{figure*}

As in \Sec{sec:Glauber_average}, we consider the transversely averaged hydrodynamic fields of energy $\langle \tau e^{3/4}\rangle$, velocity $\langle v_\perp \rangle$ and momentum eccentricity $\epsilon'_p$ (\Eq{eq:momentum_aniso}) after the linearized kinetic pre-equilibrium evolution \kompost{} until $\tauhydro=0.4,0.6,0.8,1.0,1.2\,\text{fm}$ starting from IP-glasma initial conditions at $\tauekt=0.2\,\text{fm}$.
The transversely averaged hydrodynamic fields are shown in Fig~\ref{fig:average_ipglasma_conformal}(a-c).
The transverse average of $\tau e^{3/4}$ and the transverse velocity $v_\perp$ are both showing an overall small dependence on $\tauhydro$ consistent with what was observed in \Sec{sec:Glauber_average} for MC-Glauber initial conditions.  The momentum eccentricity $\epsilon_p^\prime$ (\Eq{eq:momentum_aniso}), on the other hand, is showing a larger dependence on $\tauhydro$ than in the Glauber case. We verified that similar dependence is obtained for $\epsilon_p'$ in peripheral IP-Glasma collisions with appreciable background ellipticity (not shown). 
The initial momentum perturbations for IP-Glasma initial conditions (which are absent in MC-Glauber initialization) are also propagated by the kinetic theory evolution, but they only make a minor contribution to averaged energy density $\langle\tau e^{3/4}\rangle$ and radial velocity $\langle v_\perp \rangle$ at later times.

We note that both IP-Glasma and the kinetic theory are conformal, which means that the breaking of conformality discussed in Section~\ref{sec:Glauber_average} also occurs and increases the dependence on $\tau_{\textrm{hydro}}$ for hydrodynamic evolution with realistic equation of state.
In the bottom row of \Fig{fig:average_ipglasma_qcd} we verified that using a conformal  equation of state for the hydrodynamic phase slightly reduces  the $\tauhydro$  dependence of both the transverse average of $\tau e^{3/4}$ and of the transverse 
velocity $v_\perp$, while leaving  $\epsilon_p^\prime$ relatively unchanged, as discussed in the previous section.

\begin{figure*}
	\centering
\subfig{a}{\includegraphics[width=0.3\linewidth]{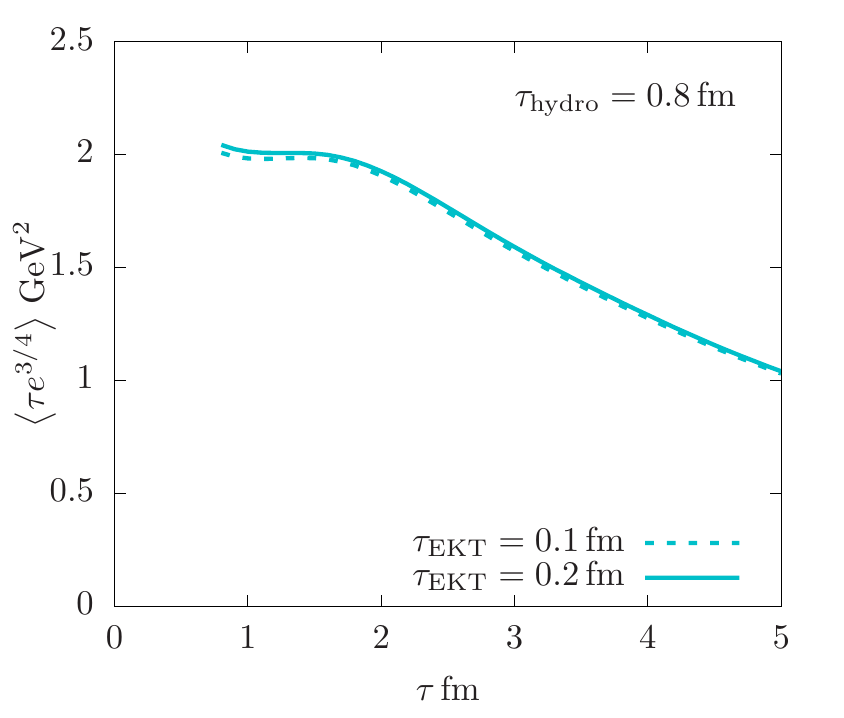}}%
\subfig{b}{\includegraphics[width=0.3\linewidth]{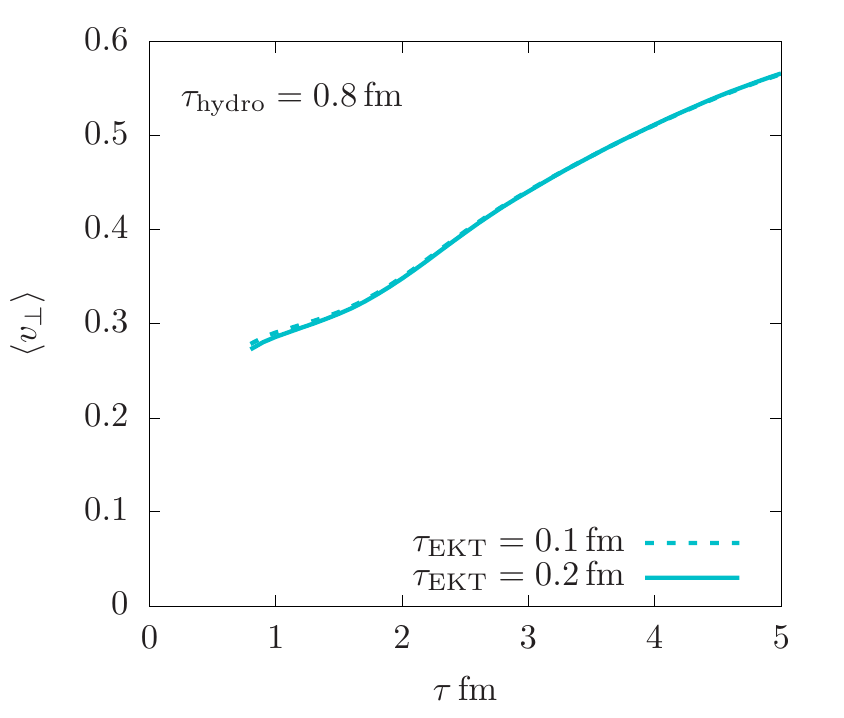}}%
\subfig{c}{\includegraphics[width=0.3\linewidth]{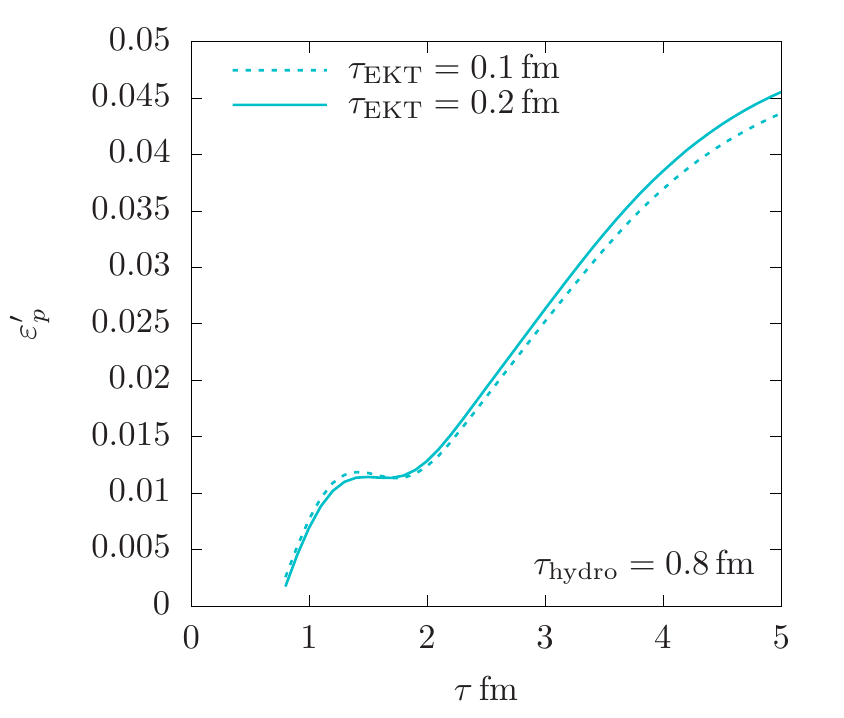}}
	\caption{Transverse average of $\tau \epsilon^{4/3}$ and of the transverse 
		velocity $v_\perp$, as defined in \Eq{eq:energy_average}, and momentum 
		eccentricity as defined by \Eq{eq:momentum_aniso}, as a function of  
		time $\tau$. The three average fields are plotted for a single hydrodynamic 
		initialization time $\tauhydro=0.8$~fm, and two kinetic theory 
		initialization time $\tauekt=0.1$~fm and 
		$\tauekt=0.2$~fm, for IP-Glasma initial conditions. A 
		realistic QCD equation of state is used in the hydrodynamics evolution.}
	\label{fig:average_ipglasma_qcd_tauEKTdep}
\end{figure*}

We also vary the transition time between IP-Glasma and the kinetic theory $\tauekt$, namely we do simulations with  $\tauekt=0.1$\,fm and $0.2$\,fm, while keeping the crossover time to $\tauhydro=0.8$\,fm fixed. As shown in Fig~\ref{fig:average_ipglasma_qcd_tauEKTdep}, all three averaged fields --- the transverse energy, velocity and momentum eccentricities --- show very little dependence on $\tauekt$. This is expected as at the crossover time the 2+1D Yang Mills evolution and kinetic theory are both close to free streaming. For the rest of this section we will use a fixed transition time of IP-Glasma to kinetic theory of  $\tauekt=0.2$\,fm.

\subsubsection{Transverse plane profiles of hydrodynamic fields}

\begin{figure*}
	\centering
	\subfig{a}{\includegraphics[height=0.25\linewidth]{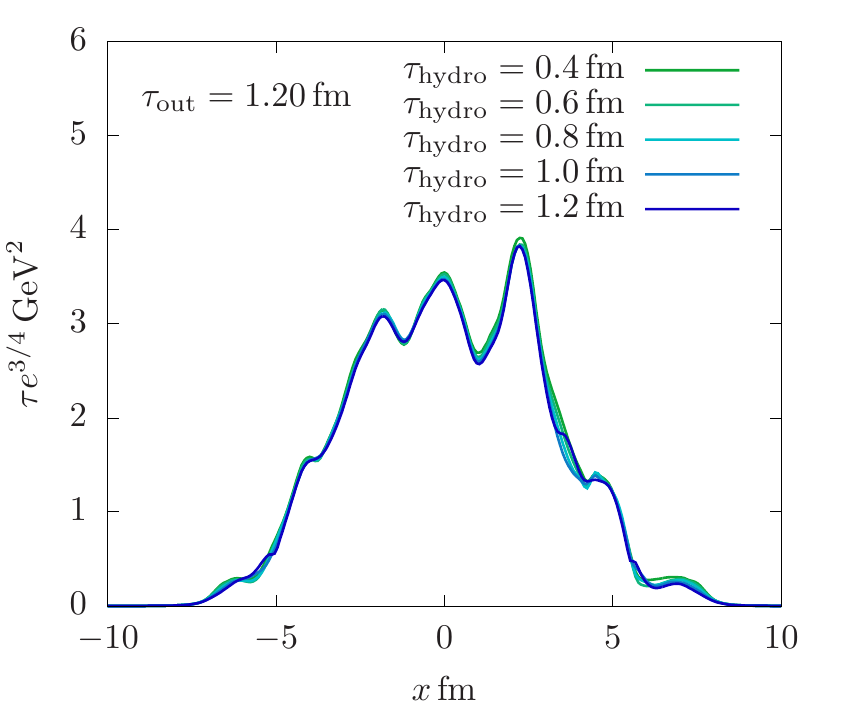}}%
	\subfig{b}{\includegraphics[height=0.25\linewidth,trim=25 0 0 0, clip]{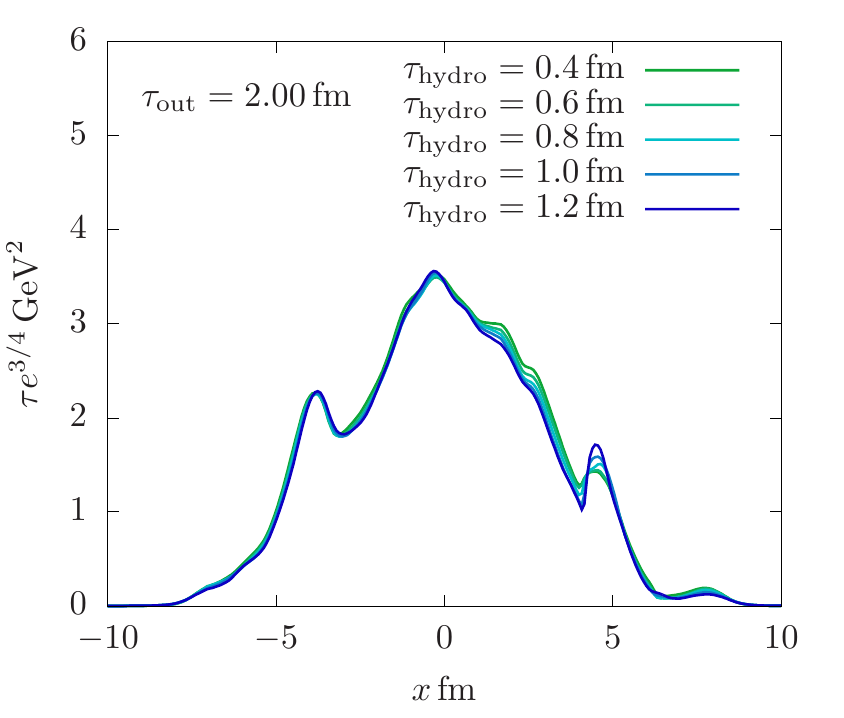}}%
	\subfig{c}{\includegraphics[height=0.25\linewidth,trim=25 0 0 0, clip]{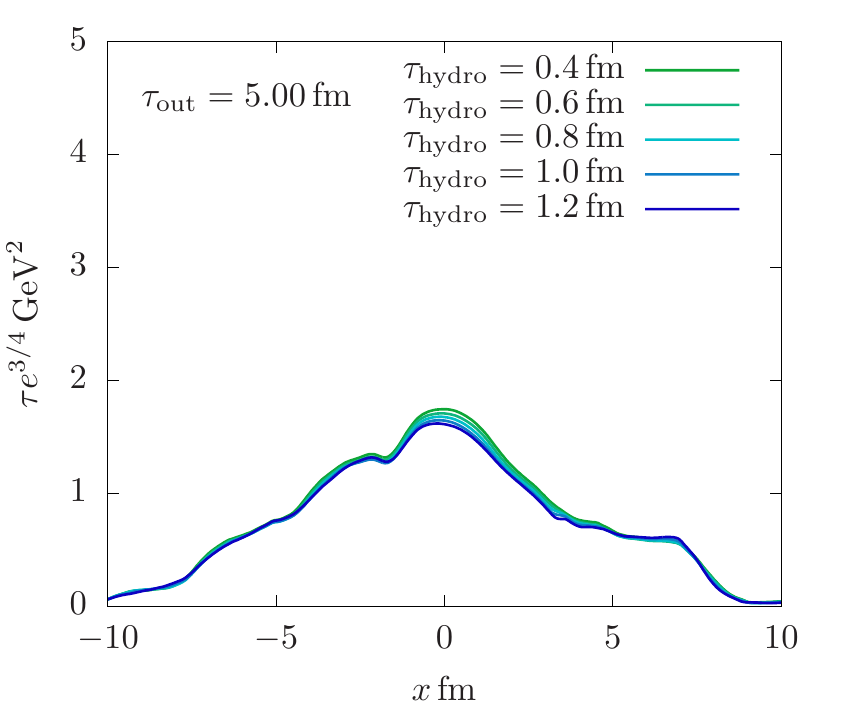}}
	\subfig{d}{\includegraphics[height=0.25\linewidth]{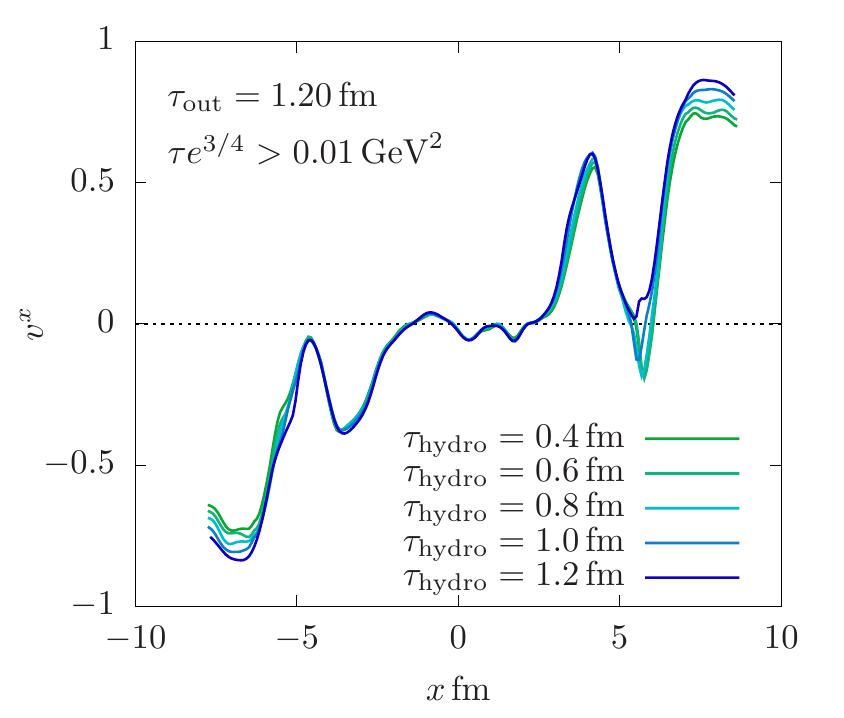}}%
	\subfig{e}{\includegraphics[height=0.25\linewidth,trim=28 0 0 0, clip]{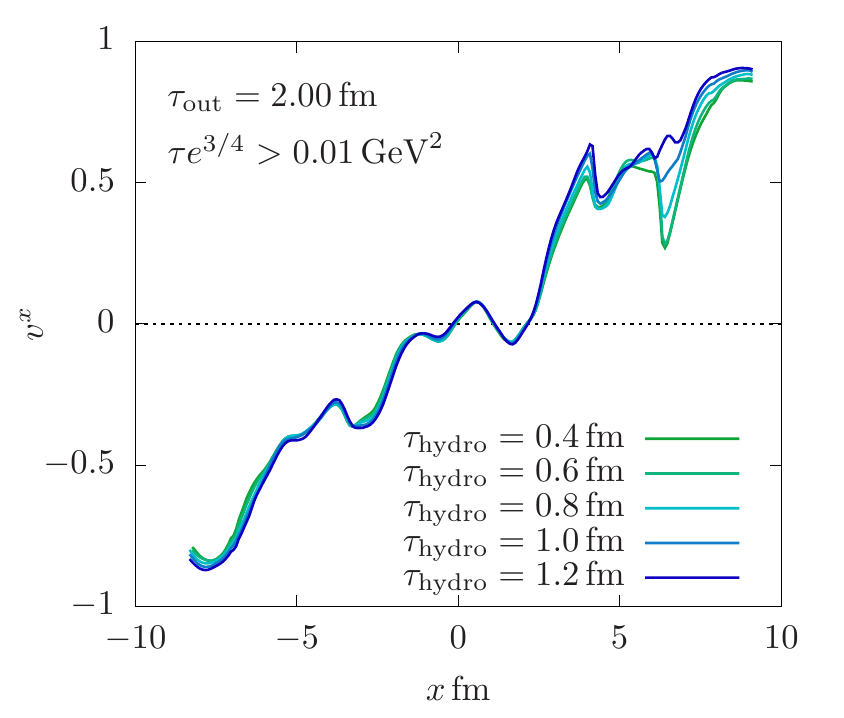}}%
	\subfig{f}{\includegraphics[height=0.25\linewidth,trim=28 0 0 0, clip]{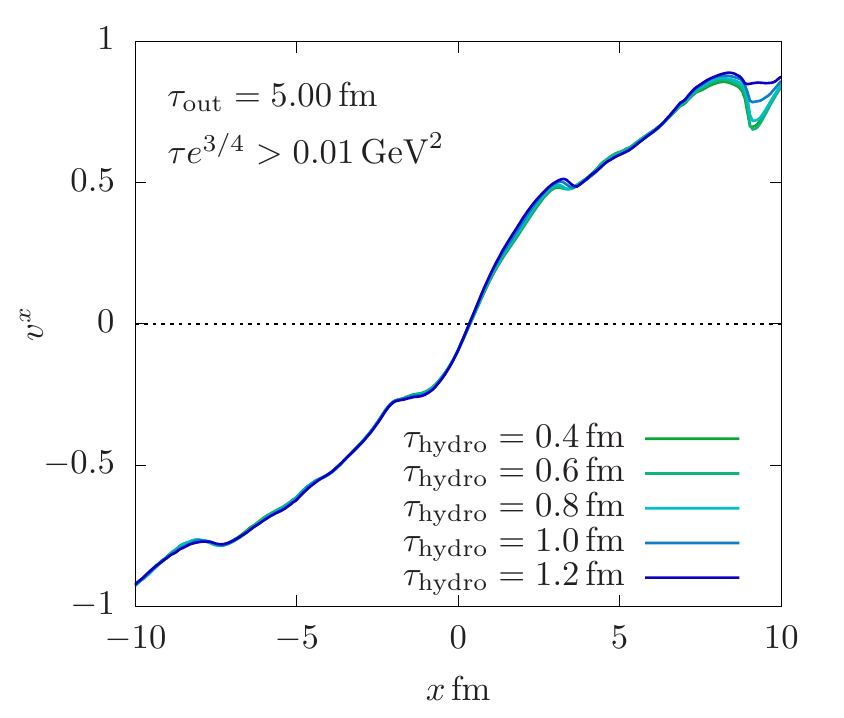}}
	\caption{Transverse profile of $\tau e^{3/4}$ (upper panels) and velocity $v^x$ (lower panels) for different hydrodynamics transition times $\tauhydro$ at $y=0\,\text{fm}$. The different columns correspond to three different hydrodynamic evolution time, $\tau=1.2, 2.0$ and $5.0$~fm. The same EKT initialization time $\tauekt=0.2$~fm was used. The equation of state is a realistic QCD one. The transverse velocity is not shown for very low energy densities ($\tau e^{3/4}<0.01\,\text{GeV}^{2}$) where numerical errors can generate spurious values of velocity.}
	\label{fig:ipglasma_profiles}
\end{figure*}

Next, we scrutinize the matching between \kompost{} with IP-Glasma initial conditions and hydrodynamic evolution by looking at the transverse energy and velocity profiles along the $x$-axis ($y=0$) in \Fig{fig:ipglasma_profiles}.
The upper row shows the transverse profile of energy density $\tau e^{3/4}$ at different times in hydrodynamic evolution. Different lines represent a varying length of kinetic theory pre-equilibrium evolution with crossover times $\tauhydro=0.4{-}1.2\,\text{fm}$. A good overlap of different curves indicates a smooth matching between kinetic theory and hydrodynamics. The small spread in energy can be attributed to the conformal breaking discussed in previous section, c.f. \Fig{fig:average_ipglasma_qcd}(a) and (d).
 In the bottom row of \Fig{fig:ipglasma_profiles}, we show the transverse velocity $v^x$ along $x$-axis. In the central region of the plasma we observe a smooth matching between kinetic theory and a full 2+1D relativistic hydrodynamics. For small energy densities at the edge of the medium ($|x|\gtrsim 5\,\text{fm}$), the velocities
are not as smoothly matched, but according to the discussion in \Sec{sec:hydtime}, these regions do not satisfy the criterion of hydrodynamization anyway. Although at later times the pre-equilibrium flow is dominated by the response to initial energy gradients, the IP-Glasma initial conditions have a non-zero initial velocity at $\tauekt=0.2\,\text{fm}$, which contribute to the fine details of profiles shown in \Fig{fig:ipglasma_profiles}.

\begin{figure*}
	\centering
\subfig{a}{\includegraphics[width=0.40\linewidth]{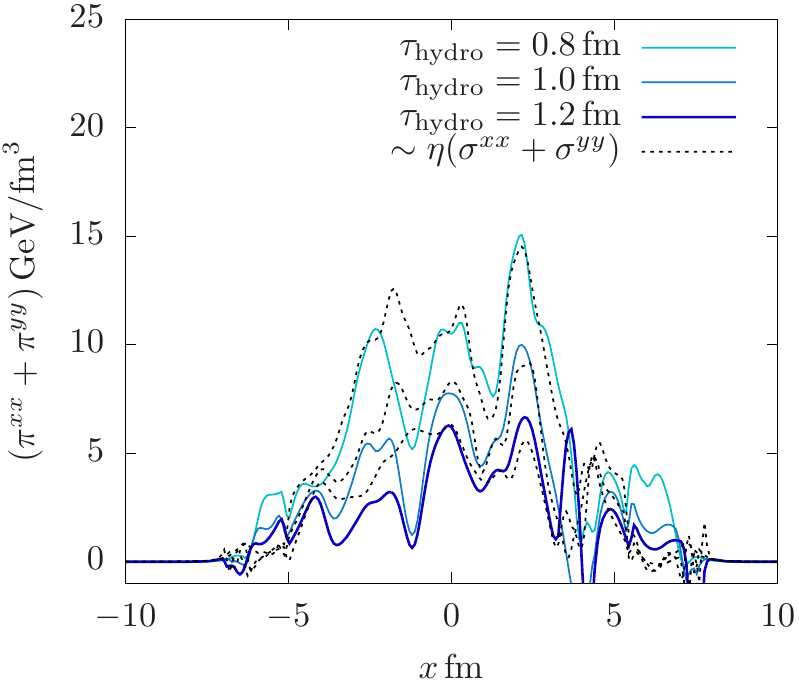}}\quad\quad
\subfig{b}{\includegraphics[width=0.40\linewidth]{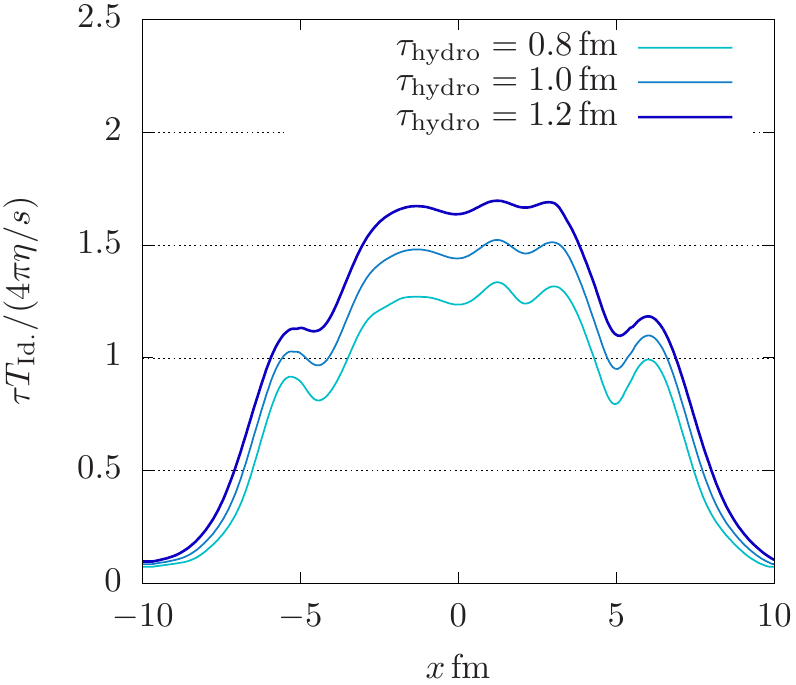}}
	\caption{(a)	Comparison of $\pi^{xx}+\pi^{yy}$ as defined in 
	\Eq{eq:TmunuIPG} with its Navier-Stokes counterpart at hydro
	initialization times $\tauhydro=0.8, 1.0, 1.2\,\text{fm}$ with IP-Glasma initial 
	conditions. (b) Scaled time variable for different locations in the transverse plane and at different crossover times $\tauhydro$. 
   Values of $\tau \TId/(4\pi\eta/s)>1$ indicate that the system is close enough to local thermal equilibrium for hydrodynamics to become applicable (see \Sec{sec:hydtime}).
   }
	\label{fig:ipglasma_pimunu}
\end{figure*}

To test the convergence of viscous components of energy momentum tensor  to hydrodynamic expectations, in \Fig{fig:ipglasma_pimunu}(a) we show the shear stress tensor $\pi^{\mu\nu}$ profile for three values of the hydrodynamic initialization time $\tauhydro= 0.8\,\text{fm}, 1.0\,\text{fm}$ and $1.2\,\text{fm}$, and compare to the estimated Navier-Stokes value $\sim \eta \sigma ^{\mu\nu}$ (obtained from the velocity profile). The agreement with the Navier-Stokes value is not as good as observed for the smoother Glauber initial conditions (\Fig{fig:glauber_pimunu}), although still reasonable. In panel (b) we show the local scaled evolution time $\xSc$ for different crossover times, which shows that the central part of the collision is within the hydrodynamic regime $\tau \TId/(4\pi\eta/s)>1$ at time $\tauhydro>0.8\,\text{fm}$ as stipulated in \Sec{sec:hydtime}.

\subsubsection{Hadronic observables}

\begin{figure*}
	\centering
\subfig{a}{\includegraphics[width=0.31\textwidth]{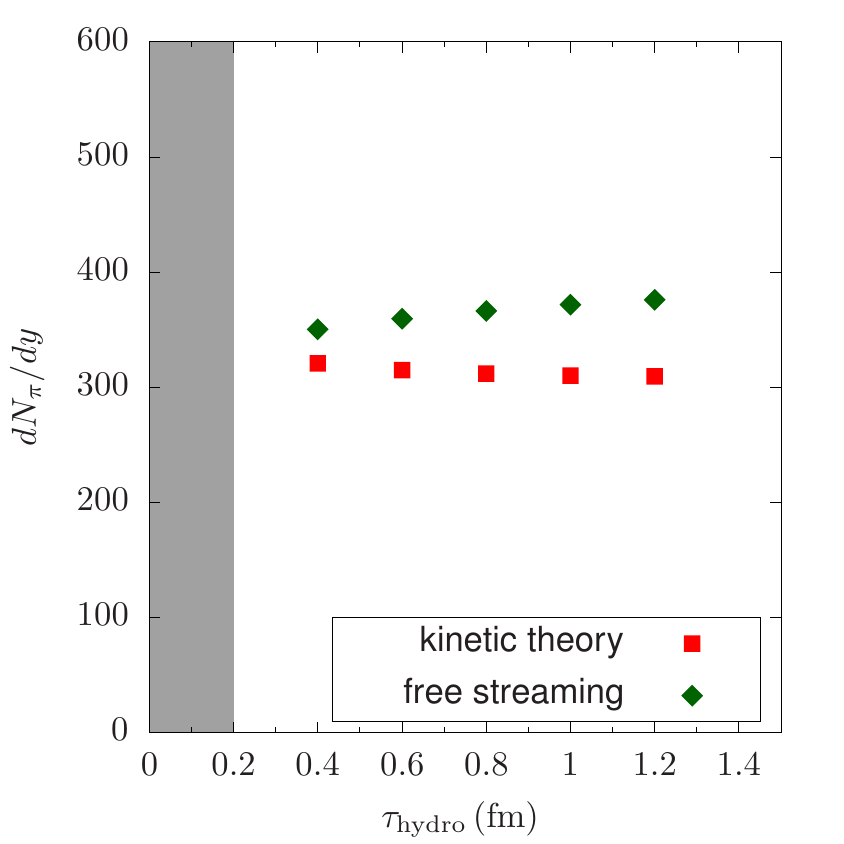}}%
\subfig{b}{\includegraphics[width=0.31\textwidth]{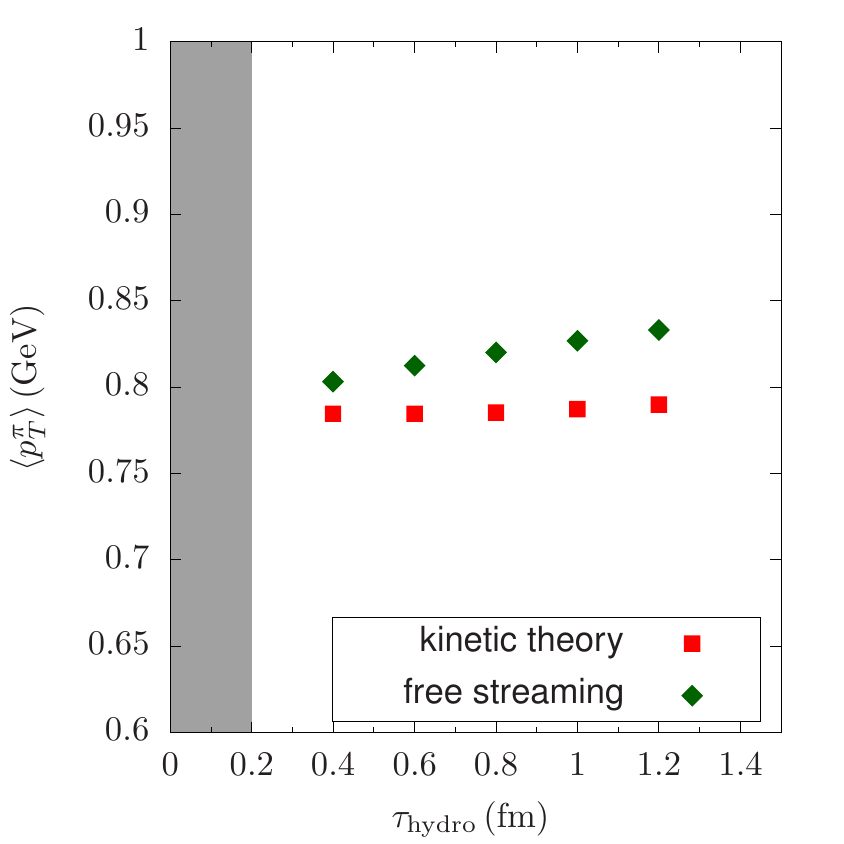}}%
\subfig{c}{\includegraphics[width=0.31\textwidth]{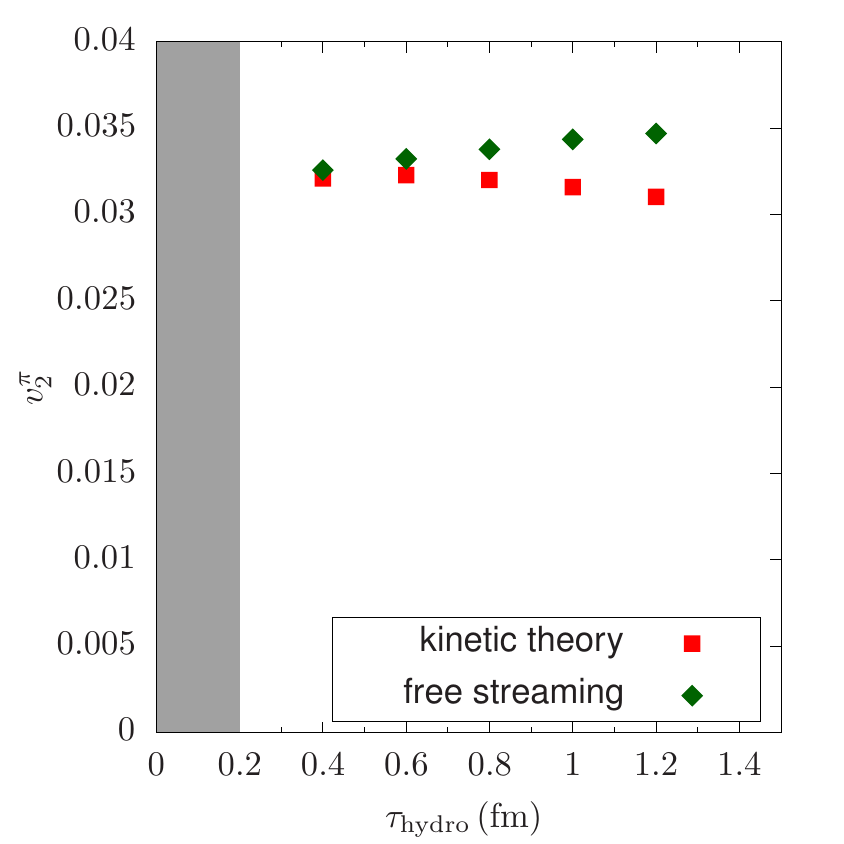}}
	\caption{
The thermal freeze-out pion
   (a) multiplicity $dN_\pi/dy$ (b) radial flow $\langle p_T^\pi\rangle$ and (c) elliptic flow $v_2^\pi$ as a function of hydrodynamic initialization time $\tauhydro=0.4{-}1.2\,\text{fm}$, i.e. different duration of pre-equilibrium evolution. The initial IP-Glasma conditions are specified at $\tauekt=0.2\,\text{fm}$. The different pre-equilibrium scenarios are linearized kinetic theory (i.e. \kompost{}) and free streaming.
 \label{fig:ipglasma_mult}\label{fig:ipglasma_meanpt}\label{fig:ipglasma_v2}}
	\label{fig:ipglasma_hadronic_obs}
\end{figure*}

After checking the smooth matching between individual stages of the evolution, we now test the effect of the hydrodynamic initialization time $\tauhydro$ on the final state observables. To recap, the IP-Glasma initial conditions are evolved until $\tauekt=0.2\,\text{fm}$ at which point the background energy density together with energy and momentum perturbations are passed to \kompost{}. After linearized kinetic theory pre-evolution the hydrodynamic fields like energy $e$, velocity $u^\mu$ and shear-stress tensor $\pi^{\mu\nu}$ are passed to viscous hydrodynamic simulation at time $\tauhydro=0.4{-}1.2\,\text{fm}$. Then thermal hadronic observables are computed at constant temperature $T_\text{FO}=145\,\text{MeV}$ freeze-out surface via the standard Cooper-Frye procedure. In \Fig{fig:ipglasma_hadronic_obs} we show the thermal pion multiplicity $dN_\pi/y$, the mean transverse momentum $\langle p_T^\pi\rangle$ and $v_2^\pi$, as a function of the hydrodynamic initialization time $\tauhydro$. For comparison, we replace the kinetic theory evolution with free-streaming background evolution and response functions. As observed in the Glauber case (see \Sec{sec:glauber_hadronic} and \Fig{fig:glauber_hadronic_obs}), hadronic observables show very little dependence on $\tauhydro$ when a kinetic theory pre-equilibrium evolution is used. The pion multiplicity changes by less than 4\%; for free-streaming pre-equilibrium the change is twice as large in the opposite direction. The mean radial $\langle p_T\rangle $ is essentially independent of $\tauhydro$, but is slightly increasing for free streaming pre-equilibrium. Finally, the elliptic flow slightly changes for both types of pre-equilibrium evolutions, but it is still smaller for kinetic theory description. All in all, the dependence on $\tauhydro$ is small for the \kompost{} pre-equilibrium evolution and the dependence increase when the kinetic theory is replaced by free-streaming description.
We highlight that the $\tauhydro$ dependence of the pion $v_2$ is still small, despite the somewhat larger dependence of $\epsilon_p^\prime$ seen in \Fig{fig:average_ipglasma_qcd}.

\section{Effective descriptions of early time dynamics}
So far we have demonstrated the practical performance of our framework to describe early time dynamics of high-energy collisions. We will now investigate in more detail theoretical relations and practical comparison to other approaches previously discussed in the literature~\cite{Broniowski:2008qk,Liu:2015nwa, Vredevoogd:2008id,vanderSchee:2013pia,Romatschke:2015gxa}.

\subsection{Long wavelength limit of kinetic theory response\label{sec:lowk}}

\begin{figure*}
	\centering
\subfig{a}{\includegraphics[width=0.4\linewidth]{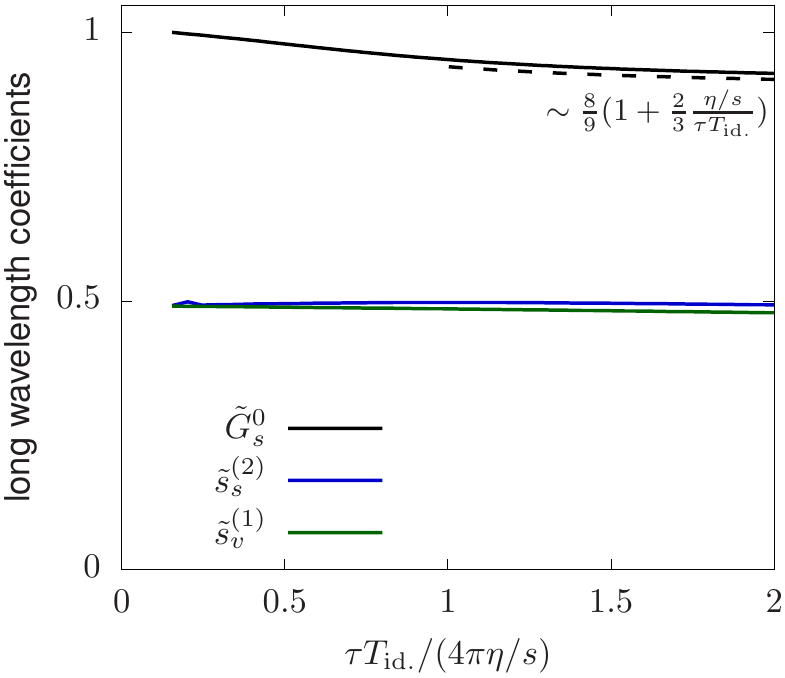}}	
\subfig{b}{\includegraphics[width=0.4\linewidth]{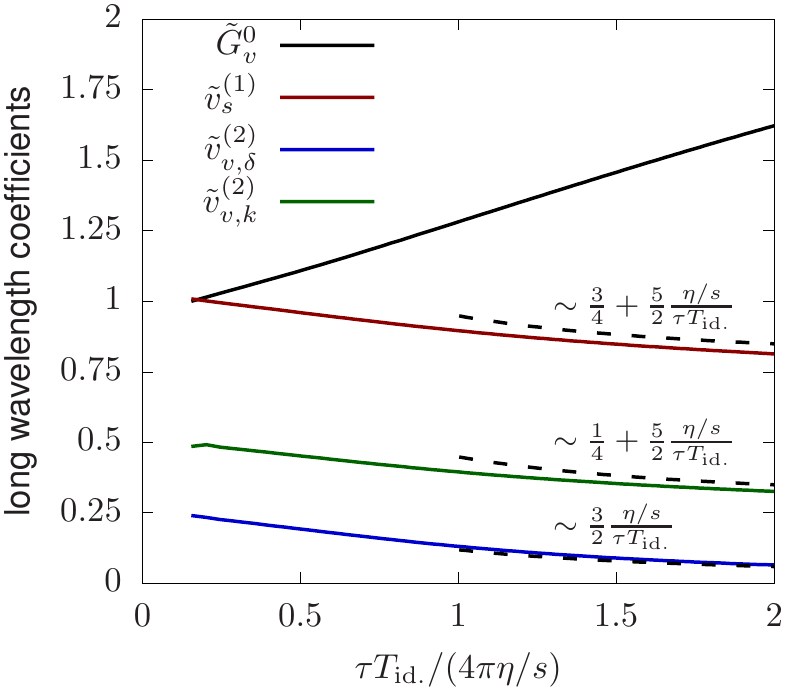}}
	\caption{\label{fig:coefde} Linear and quadratic long wavelength response coefficients to
		initial energy perturbations (a) and initial momentum perturbations (b) from non-equilibrium kinetic theory evolution. Dashed lines show comparison to viscous hydrodynamic asymptotic derived in \app{sec:hydrolimit}.}
\end{figure*}

In viscous hydrodynamics small scale fluctuations dampen rapidly, and many final state observables are only sensitive to long wavelength perturbations~\cite{Gardim:2011xv,Teaney:2012ke,Floerchinger:2013rya,Mazeliauskas:2015efa,Mazeliauskas:2015vea,Noronha-Hostler:2015coa,Gardim:2017ruc}. Consequently, one could expect that also the pre-equilibration discussed in previous sections could be well captured by the long wavelength response. We will now discuss how to formalize and test this idea, based on a low $\ktt$ expansion of our non-equilibrium linear response formalism. Interestingly, it will also be useful to establish the relation of our formalism to previous ideas related to the concept of a ``universal pre-flow" \cite{Vredevoogd:2008id}. Details of the derivations presented in this section are worked out in \App{app:lowk}.

In order to study the low $\ktt$ limit of kinetic theory pre-equilibrium evolution to initial conditions of heavy ion collisions, we first filter out the small wavelength perturbations by performing a Gaussian smearing of the initial energy-momentum tensor $T^{\mu\nu}(\tau_0,\x_0)$. Specifically, we define a coarse grained energy-momentum tensor
\begin{equation}
	\label{eq:smoothTmunu}
	\bar{T}^{\mu\nu}(\tau_0,\x_0)= \int 
	d^2\xt'_{0}~S_{\sigma}(\xt_0-\xt'_0)~T^{\mu\nu}(\tau_0,\xt_0')\;,
\end{equation}
with the same Gaussian smearing kernel $S_{\sigma}(\xt_{0}-\xt'_{0})$ used to define the local background energy $\bar{T}^{\mu\nu}_{\x}(\tau_0)$  (c.f. \Eq{eq:avgBG}).
Considering a space-time point $(\tau,\x)$ on the future hydrodynamic surface, we can then perform our usual decomposition of the energy-momentum tensor into local background $\bar{T}^{\mu\nu}_{\x}(\tau)$ and perturbations $\delta T^{\mu\nu}_{\x}(\tau_0,\x_0)$. However, instead of considering fluctuations on all scales, the initial energy-momentum tensor perturbation $\delta T^{\mu\nu}_{\x}(\tau_0,\x_0)$ can now be decomposed further into short and long-wavelength components as 
\begin{align}
	\delta T^{\mu\nu}_{\xt}(\tau_0,\xt_0)=&\underbrace{T^{\mu\nu}(\tau_0,\xt_0)-\bar{T}^{\mu\nu}(\tau_0,\x_0)}_{\text{short wavelength}} \nonumber \\
	&\qquad+\underbrace{\bar{T}^{\mu\nu}(\tau_0,\x_0)-\bar{T}^{\mu\nu}_{\x}(\tau_0)}_{\text{long wavelength}}\;,
\end{align}
where, recalling that the definition of the local background  $\bar{T}^{\mu\nu}_{\x}(\tau_0)$ involves the same coarse graining procedure, the long wave-length components of $\delta T^{\mu\nu}_{\xt}(\tau_0,\xt_0)$ are then entirely given in terms of smooth fields $\bar{T}^{\mu\nu}$. If the initial profile of the energy-momentum tensor 
$T^{\mu\nu}(\tau,\x)$ is sufficiently smooth on length scales smaller than $\sigma\sim |\tau-\tau_{0}|$ the short wave-length component is very small and can safely be ignored\footnote{Note that even if the initial energy-momentum tensor features fluctuations smaller on length scales smaller than $\sigma$, short wave length components on scales
less than $\sigma_{{\rm visc}} \sim (\tau-\tau_0)/\sqrt{\xSc}$, are subject to strong viscous damping and could still be neglected.}. Neglecting short wavelength components in the following, the energy-momentum tensor at later times is then given by
\begin{align}
	\label{eq:EvolutionLowKmain}
	\frac{\delta 
		T^{\mu\nu}(\tau,\xt)}{\bar{T}^{\tau\tau}_{\xt}(\tau)}&=\frac{1}{\bar{T}^{\tau\tau}_\xt(\tau_0)}\times\nonumber\\
	\int
	d^2\xt_0&G^{\mu\nu}_{\alpha\beta}\Big(\tau,\tau_0,\xt-\xt_0\Big)
	~(\bar{T}^{\alpha\beta}(\tau_0,\xt_0)-\bar{T}_\xt^{\alpha\beta}(\tau_0))\;.
\end{align}
By expressing the convolution in Fourier space, and expanding the response functions in powers of the wave number $\ktt$, as
\begin{align}
	\tilde{G}^{s}_{s}(\tau,\tau_0,\ktt)&=\tilde{G}_{s}^{0}(\tau,\tau_0)~\Big(\tilde{s}_s^{(0)}-\frac{1}{2}
	\ktt^2 (\tau-\tau_0)^2~\tilde{s}^{(2)}_{s}+...\Big)\;, \nonumber \\
	\tilde{G}^{v}_{s}(\tau,\tau_0,\ktt)&=\tilde{G}_{s}^{0}(\tau,\tau_0)~\Big(\ktt
	(\tau-\tau_0)~\tilde{s}^{(1)}_{v}+...\Big)\;,	\label{eq:lowkcoef}
\end{align}
and similarly for the other components, it is then straightforward to show that the long wave length response in \Eq{eq:EvolutionLowKmain} is proportional to the gradients of the 
\emph{smoothened} energy-momentum tensor $\bar{T}^{\mu\nu}(\tau,\x)$. Saving the details of the derivation for \app{app:lowk}, the energy-momentum response to initial energy gradients is then given by
\begin{subequations}
\label{eq:preflowedense}
\begin{align}
	\frac{\delta T^{\tau\tau}(\tau,\xt)}{\bar{T}_\xt^{\tau\tau}(\tau)}&\approx 
	\frac{\tilde{G}_s^{0}(\tau,\tau_0) }{\bar{T}_\xt^{\tau\tau}(\tau_0)} 
	\left[\frac{1}{2}\tilde{s}_s^{(2)}(\tau-\tau_0)^2 
	\partial_k\partial^k\right]\bar{T}^{\tau\tau}(\tau_0,\x)\\
	\frac{\delta T^{\tau i}(\tau,\xt)}{\bar{T}_\xt^{\tau\tau}(\tau)}&\approx 
	\frac{\tilde{G}_s^{0}(\tau,\tau_0) }{\bar{T}_\xt^{\tau\tau}(\tau_0)} 
	\left[- \tilde{s}_v^{(1)}(\tau-\tau_0) \partial^i 
	\right]\bar{T}^{\tau\tau}(\tau_0,\x) 
	\label{eq:pratt}
\end{align}
\end{subequations}
which is a generalization of the previous result for the transverse pre-flow derived in~\cite{Vredevoogd:2008id, Keegan:2016cpi}. \Eq{eq:preflowedense} says
that in addition to the ``pre-flow'' 
which develops during 
the equilibration process due to gradients~\cite{Vredevoogd:2008id}, 
local maxima  of the initial energy density, which have negative second derivatives,  are depleted during the evolution.
We can also obtain the long wavelength response to initial momentum perturbations
\begin{align}
	\frac{\delta T^{\tau\tau}(\tau,\xt)}{\bar{T}^{\tau\tau}_{\xt}(\tau)} &\approx \frac{\tilde{G}_{v}^{0}(\tau,\tau_0)}{\bar{T}^{\tau\tau}_\xt(\tau_0)} \Big[ - \tilde{v}^{(1)}_{s} (\tau-\tau_0)  \partial_{i} \Big]~\bar{T}^{\tau i}(\tau_0,\x) \;, \\
	\frac{\delta T^{\tau i}(\tau,\xt)}{\bar{T}_\xt^{\tau\tau}(\tau)}&\approx  \frac{\tilde{G}_{v}^{0}(\tau,\tau_0)}{\bar{T}^{\tau\tau}_\xt(\tau_0)} \Big[ 1+  \frac{(\tau-\tau_0)^2}{2}  \tilde{v}^{(2)}_{\delta} \partial_k\partial^k\Big] \bar{T}^{\tau i}(\tau_0,\xt) \nonumber \\
&+\frac{\tilde{G}_{v}^{0}(\tau,\tau_0)}{\bar{T}^{\tau\tau}_\xt(\tau_0)} \Big[ \frac{(\tau-\tau_0)^2}{2} \tilde{v}^{(2)}_{k} \partial^{i}\partial_{j} \Big]~\bar{T}^{\tau j}(\tau_0,\xt) \;. \nonumber \\
	\label{eq:prattmomentum}
\end{align}

We extract the $1$st and $2$nd order coefficients in $\ktt(\tau-\tau_0)$ 
by performing a polynomial 
fit to the first few $\ktt$ values of kinetic theory response functions, e.g.\ 
\Fig{fig:rescaledresponse19de}.
Our results for the zero momentum response $G^{0}_{s/v}$ as well as the long wave-length coefficients $s^{(n)}$,$v^{(n)}$ from effective kinetic theory simulations, are compactly summarized in Fig.~\ref{fig:coefde}, where we plot the various response coefficients as a function of  the scaling variables $\xSc$. We find that the time dependence of the coefficients is rather weak, such that in practice the long wave length response can be approximated rather well by the hydrodynamic asymptotics, see \app{sec:hydrolimit}.

Before we compare the long wavelength results to the full treatment in \kompost, it is important to point out that the above separation into long and short wavelength modes introduces an artificial regulator dependence not present the full treatment. Specifically for the long wave-length filter $\tilde{S}_{\sigma}(\k)\propto e^{-\frac{\sigma^2}{2} \k^2}\simeq 1-\frac{\sigma^2}{2}\k^2$ in \Eq{eq:smoothTmunu}, the first difference formally appears at quadratic order $\mathcal{O}(\k^2)$. However, the $\mathcal{O}(\k^2)$ can always be absorbed into an additive renormalization of the long wavelength coefficients, i.e.\ by subtracting $\sigma^2/(\tau-\tau_0)^2$ from the quadratic  $\tilde{s}^{(2)}_s$ coefficient in energy response, as explained in \app{app:regsigma}.
 Based on this renormalization scheme, the regulator dependence enters only at cubic order $O(\ktt^3)$ and the residual dependence can be used to quantify the systematic uncertainties of the long wavelength approximation.

\subsection{Comparison of effective kinetic theory, long wavelength response \& free streaming}
\begin{figure*}
	\centering
\subfig{a}{\includegraphics[width=0.3\linewidth]{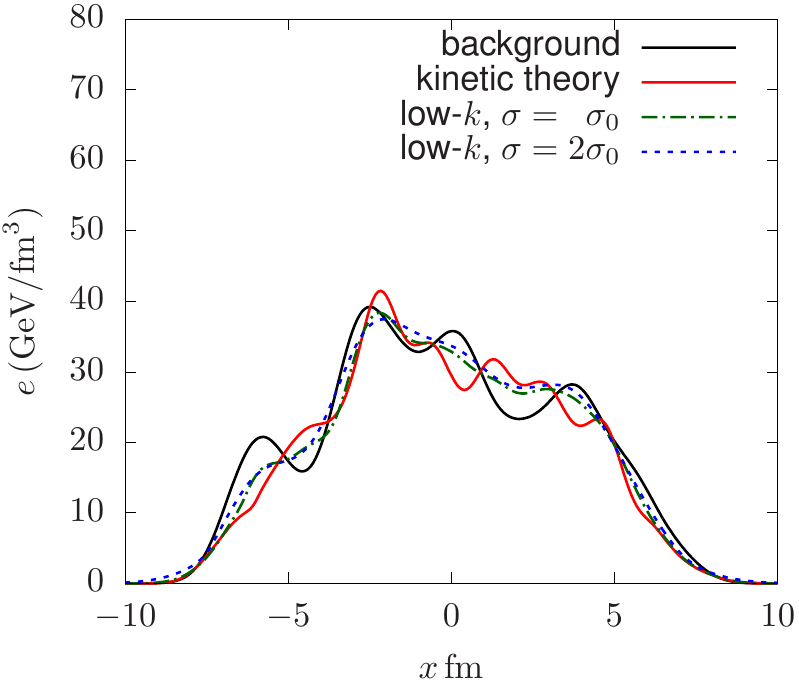}}%
\subfig{b}{\includegraphics[width=0.3\linewidth]{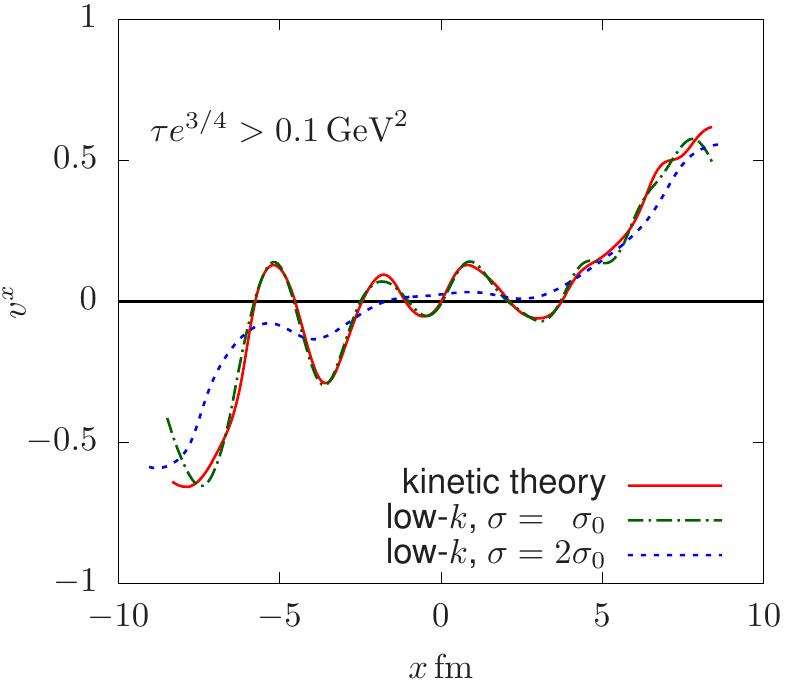}}%
\subfig{c}{\includegraphics[width=0.3\linewidth]{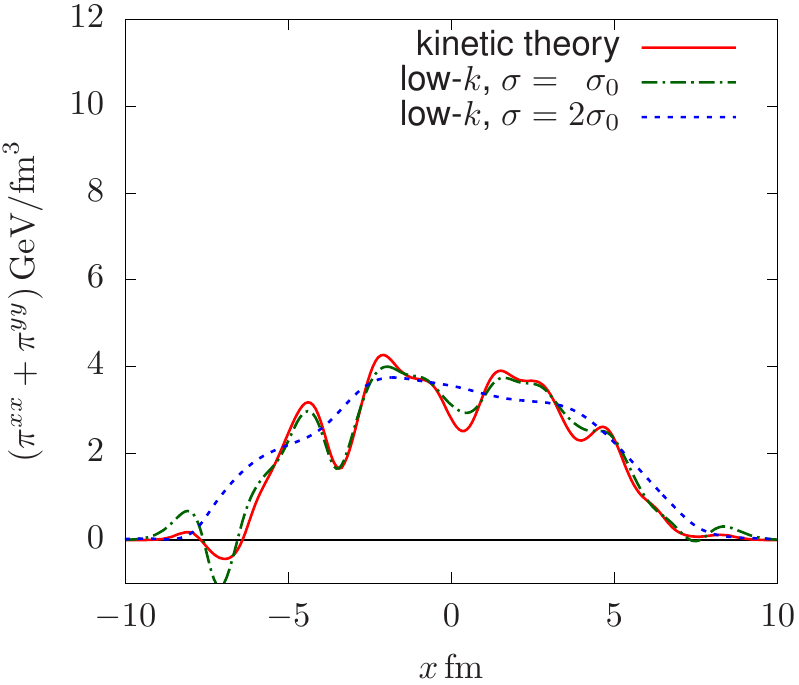}}
\subfig{d}{\includegraphics[width=0.3\linewidth]{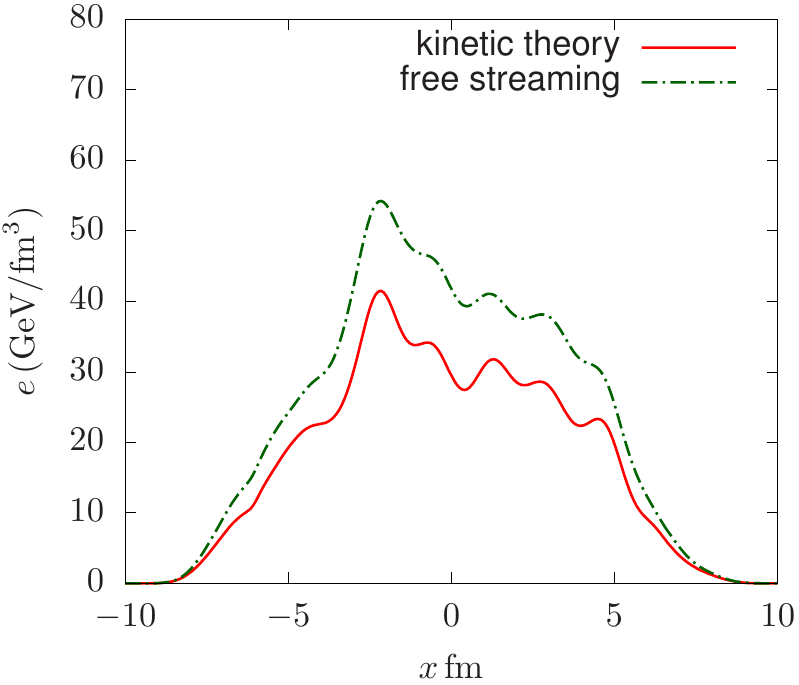}}%
\subfig{e}{\includegraphics[width=0.3\linewidth]{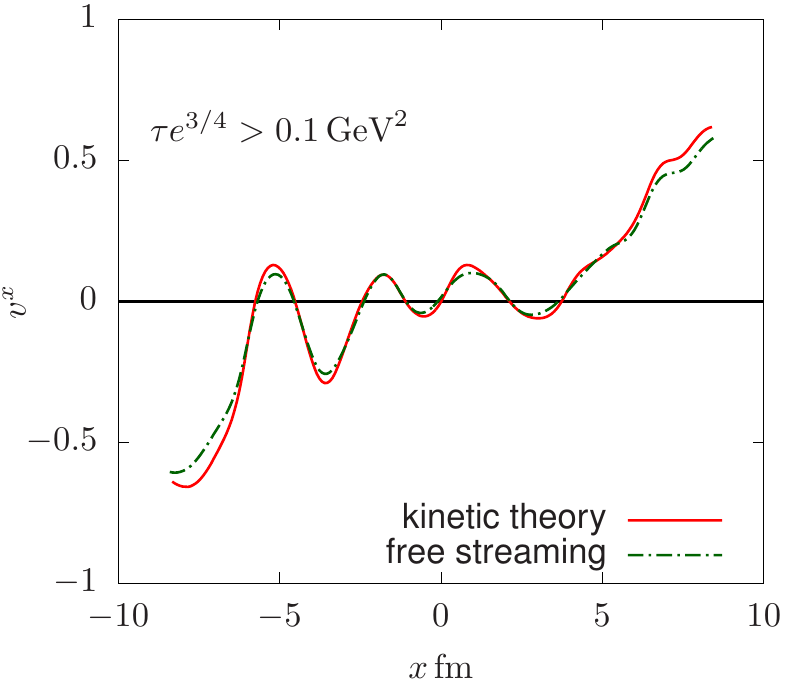}}%
\subfig{f}{\includegraphics[width=0.3\linewidth]{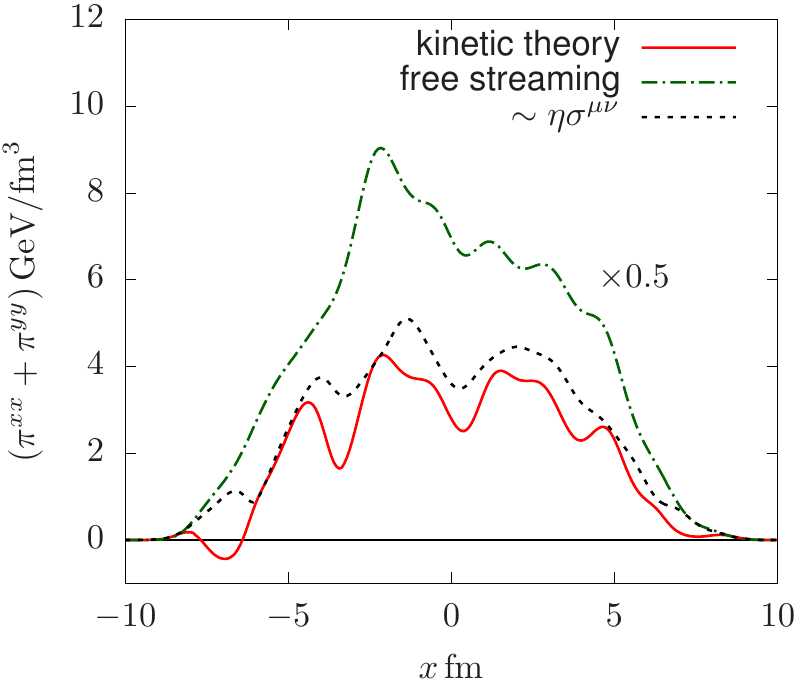}}%
	\caption{Comparison of hydrodynamic fields obtained in the long-wavelength (top) and free streaming limits (bottom), with the results of full kinetic theory pre-equilibrium (\kompost{}). Different columns  show
profiles of energy density $e$ (left),  flow velocity $v^x$  (center) and shear stress $(\pi^{xx}+\pi^{yy})$ (right).  }
	\label{fig:lowk}
\end{figure*}

We now turn to the comparison between the full kinetic theory and the low-$\ktt$ limit described in previous section. Our results are summarized  in the top row of \Fig{fig:lowk}, where we compare profiles of the energy-momentum tensor at the end of the pre-equilibrium evolution at $\tauhydro=1.2\,\text{fm}$ initialized with the same MC-Glauber initial conditions at   $\tauekt=0.1\,\text{fm}$. We assess the robustness of the long wavelength results by using two different smearing widths $\sigma_1=\sigma_0$ and  $\sigma_2=2\sigma_0$, where $\sigma_0=(\tau_\text{hydro}-\tauekt)/2$ is the same Gaussian width used to define the local averaged background in the full kinetic theory response. 

As it is visible from  \Fig{fig:lowk}(a) the simplified low-$k$ evolution accurate to quadratic order in small $\ktt$ reproduces the energy density profile rather well  and is largely 
insensitive to the smearing width. However, most of the energy is evolved as a background according to universal scaling curves, \Eq{eq:universalE}, which is the the same 
for both the full kinetic and low $\ktt$ response.

In \Fig{fig:lowk}(b) we compare the transverse velocity profile in kinetic theory and long wavelength limit. Formally, the low-$k$ limit given by \Eq{eq:pratt} is only accurate to linear order in 
$k$ (c.f.\ \cite{Vredevoogd:2008id}). However for the particular choice of the regulator $\sigma_1=\sigma_0$, the long wavelength limit approximately reproduces the cubic order 
flow response (see \App{app:regsigma} for details) and the resulting curve is very close to the full kinetic theory result. Despite this accidental agreement, it is also evident from the comparison with the curve for $\sigma_2=2\sigma_0$ that the long wavelength result exhibits a strong regulator dependence, which points to the fact that the actual leading order $(\sim k)$ velocity profile provides a less accurate approximation of the full kinetic theory result.

Similar features can be observed in \Fig{fig:lowk}(c), where we compare the kinetic theory evolution of the shear-stress tensor $\pi^{\mu\nu}$ with the corresponding long wavelength result obtained for simplicity through hydrodynamic constitutive equations. At leading viscous order $\pi^{\mu\nu}\propto \eta \sigma^{\mu\nu}$ is mainly dominated by velocity gradients, therefore the agreement is better for the low-$\ktt$ response with accidental cubic accuracy in velocities, but is also reasonably good for the different choice of coarse-graining.

It is constructive to compare in the same fashion the kinetic theory response with a simple model of the pre-equilibrium evolution prescription based on free streaming~\cite{Broniowski:2008qk,Liu:2015nwa}. For completeness the free-streaming response functions are summarized in \app{sec:freestreaming}.
 The comparison is shown in \Fig{fig:lowk}(d-f).  In free streaming evolution the longitudinal pressure is completely neglected and thus the energy density decreases more slowly as a function of time, overshooting the kinetic theory results (see \Fig{fig:lowk}(d)). For the 
special case of initial (scalar) energy perturbations, the free streaming 
response functions for transverse \emph{velocity} have approximately the same 
low-$\ktt$ expansion as the full kinetic theory evolution, see \Eq{eq:cubic}. 
Therefore the transverse velocity profile in \Fig{fig:lowk}(e) between kinetic 
theory and free streaming response is almost indistinguishable. However, the physical momentum 
is not correct, because of the incorrect background energy evolution,  
(c.f.\ \Fig{fig:lowk}(d)).

Finally in \Fig{fig:lowk}(f) we compare the shear-stress tensor $\pi^{\mu\nu}$ in free streaming (the dashed-dotted line) to the full kinetic theory result (the solid line). In 
free streaming  the longitudinal pressure is completely neglected, and thus 
the transverse  stress $\pi^{xx} {+} \pi^{yy}$ is 
too large when the system is passed to hydrodynamics. (We have reduced the free $\pi^{xx}{+}\pi^{yy}$ by a factor of two in \Fig{fig:lowk} for visibility.) If instead of 
the free streaming result for $\pi^{xx}{+}\pi^{yy}$, one just uses the free energy and 
velocity profiles, \Fig{fig:lowk}(d-e), together with  the hydrodynamic  constitutive 
relations,  the shear-stress tensor is closer to the kinetic theory 
result, as indicated by a Navier-Stokes estimate $\sim \eta
\sigma^{\mu\nu}$ in the figure.

To summarize, 
the low-$|\k|$ limit does a good job at describing the velocity, but it makes no predictions for the background energy density as a function of time. Therefore it must 
be supplemented  by a theory (such as QCD kinetics) 
which describes the longitudinal pressure at early times.
Free streaming 
 also correctly describes the velocity response,
but in this case the energy density and second order  hydrodynamic variables
must be continually readjusted in order to have a smooth transition 
to hydrodynamics.  On a practical note,  using 
{\kompost  } is computationally no more expensive
than the other alternatives, and it should  become the 
pre-hydro engine of choice.

\section{Summary \& outlook \label{sec:summary}}

In this paper we developed a linear response framework to describe the non-equilibrium evolution of the energy-momentum tensor during the early stages of high-energy heavy-ion collisions. Based on microscopic input from effective kinetic theory simulations, we presented a practical implementation of the ``bottom-up`` thermalization scenario with realistic fluctuating heavy-ion initial conditions and demonstrated a consistent matching between the pre-equilibrium and the hydrodynamic evolution on an event-by-event basis. 
Our linear kinetic pre-equilibrium propagator \kompost{}~\cite{kompost_github}
provides a practical implementation of a systematic procedure to propagate initial out-of-equilibrium perturbations to the hydrodynamic initialization time~\cite{Keegan:2016cpi}. Crucially, for short evolution times the equilibration dynamics can be described as sum of local background equilibration and linear response to fluctuations of initial conserved energy and momentum density, \Eq{one}. We developed a concrete realization of the general linear response formalism presented in \Sec{sec:macro} and \Sec{sec:generalresponse} using QCD kinetic theory with gluonic degrees of freedom and extrapolation to moderate values of the coupling constant $\lambda$~\cite{Arnold:2002zm,Keegan:2016cpi,Kurkela:2015qoa}. 
Within our framework, the entire kinetic equilibration of $T^{\mu\nu}$ is compactly summarized by a single evolution curve of the homogeneous background (c.f.\ \Fig{fig:scaledTmunu}) and a handful of response functions (c.f.\ Figs.~\ref{fig:plot_grgss} and \ref{fig:plot_grgvs} ). Interestingly, we find that for the relevant range of coupling constants, corresponding to realistic values of $\eta/s$, the kinetic equilibration only depends on the scaling time $\sim \tau T/(\eta/s)$ such that pre-tabulated kinetic response functions can be used for event-by-event simulations.

We found that for typical QGP parameters the leading order kinetic theory predicts a hydrodynamization time  around  $\tauhydro\sim 1\,\text{fm}$ (c.f.\ \Eq{eq:hydrotime} and Ref.~\cite{Keegan:2016cpi}). During this time the initial gluon number density per rapidity roughly doubles and, thus, significantly alters the
relation between final particle multiplicities and the entropy density per rapidity in the initial state (see \Fig{fig:entropya}).

We applied our formalism to two widely used initial state descriptions: a phenomenological MC-Glauber ansatz for transverse energy density deposition in \Sec{sec:glauber} and the first-principle dynamical IP-Glasma model (which contains both energy and momentum fluctuations) in \Sec{sec:ipglasma}.  In both cases we demonstrated a smooth matching between the kinetic theory and hydrodynamics, both in the hydrodynamics fields, e.g.\ \Fig{fig:glauber_profiles}, and in the final hadronic observables, see \Figs{fig:glauber_hadronic_obs} and \ref{fig:ipglasma_hadronic_obs}, largely eliminating the dependence on the crossover time $\tauhydro$. Note that the kinetic response functions reproduce the spatial components of the energy-momentum tensor $\delta T^{ij}$ without explicitly imposing the constitutive relations, (c.f. \Figs{fig:glauber_pimunu}(a) and \ref{fig:ipglasma_pimunu}(a)).

Finally we studied the kinetic response in the ``low wavelength'' (hydrodynamic) and ``fast expansion'' (free-streaming) limits to establish
 connections with the existing literature~\cite{Broniowski:2008qk,Liu:2015nwa, Vredevoogd:2008id,vanderSchee:2013pia,Romatschke:2015gxa}. We find that the first few terms in small $|\k|$-expansion are sufficient to capture most of kinetic response (\Fig{fig:lowk}). Incidentally, the leading order velocity response in free-streaming limit agrees with kinetic theory, but energy density and shear-stress tensor evolution are described incorrectly.
 
Our publicly available kinetic propagator package \kompost{}~\cite{kompost_github} offers a simple, yet non-trivial description of hydrodynamization in heavy ion collisions. The reduced sensitivity on the hydrodynamic initialization time $\tauhydro$ brings us closer to ab initio initial conditions of heavy ion collisions. Moreover, there are multiple possibilities to build on this work to achieve an even better description of the early stage of heavy ion collisions. The inclusion of quarks as degrees of freedom would lead to a more realistic description of the pre-equilibration stage of collisions, which should provide a better matching with the QCD equation of state. Studying non-trivial background evolution in kinetic theory, e.g.\ radial gradients, could improve the applicability of presented framework at the edges of the QGP fireball and in small systems.  

More importantly, the formalism derived in this work to propagate linear perturbations on top of a smoother background can be used with response functions computed in limits other than weakly-coupled effective QCD kinetic theory. By using this framework to compare systematically the macroscopic description of equilibration from a weakly-coupled regime and a strongly-coupled one, one can hope to better constrain the real dynamics of the medium produced in heavy ion collisions. %

\begin{acknowledgments}

The authors would like to thank Bj\"orn Schenke for insightful discussions and for his help adapting the hydrodynamics code MUSIC for this work, and
Liam Keegan for his contributions at the beginning of this project. Useful discussions with J\"urgen Berges, Stefan Fl\"orchinger, Yacine Mehtar-Tani, Klaus Reygers, Raju Venugopalan are gratefully acknowledged.
Results in this paper were obtained using the high-performance computing system 
at the Institute for Advanced Computational Science at Stony Brook University. 
This work was supported in part by the U.S. Department of Energy, Office of 
Science, Office of Nuclear Physics  under Award Numbers 
DE\nobreakdash-FG02\nobreakdash-88ER40388 (A.M., J.-F.P., D.T.), DE-FG02-05ER41367 (J.-F.P.) and 
DE-FG02-97ER41014 (S.S.). This work was supported in part by the German Research Foundation (DFG) 
Collaborative Research Centre (SFB) 1225 (ISOQUANT) (A.M.). Finally, A.M.
would like to thank CERN Theoretical Physics
Department for the hospitality during the short-term visit.
\end{acknowledgments}

\bibliography{master}

\appendix

\section{Background evolution in  kinetic theory\label{app:parameters}}

\subsection{Extraction of transport coefficients\label{sec:transp}}

In this section we summarise the procedure of extracting the transport coefficients in kinetic theory for different values of the coupling constant $\lambda$. At asymptotically weak coupling, the transport coefficients can be obtained from perturbative calculations~\cite{Arnold:2003zc,York:2008rr}, however in this work we consider moderate values of $\lambda$. So instead of using perturbative formulas, we extract the transport coefficients directly from the evolution of homogeneous and longitudinally expanding plasma. Effectively we treat $\lambda$  as a dial parameter in our leading order kinetic theory description to tune to different values of transport coefficients (which, perhaps, correspond to different value of the coupling constant in a higher order treatment).

\begin{figure}
\centering
\includegraphics[width=0.9\linewidth]{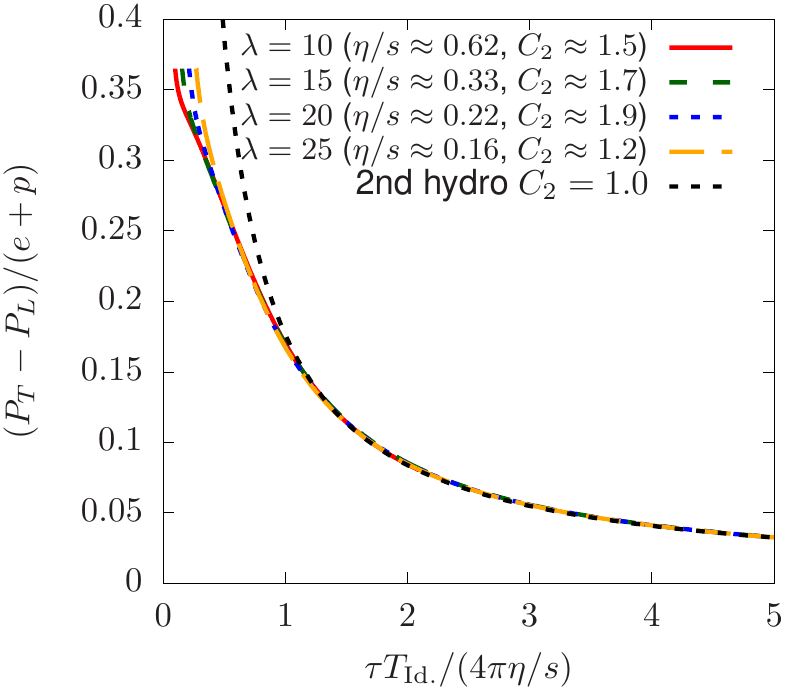}
\caption{Evolution of pressure anisotropy, \Eq{eq:pressanis}, in kinetic theory with initial distribution function \Eq{eq:init_cond}, and fitted transport coefficients $\eta/s$ and $C_2$ (see text).}
\label{fig:paperetas}
\end{figure}

First, the transport coefficients are obtained from the relaxation of pressure anisotropy. For Bjorken expansion and in second order  conformal hydrodynamics one can show that  it takes the following form (see Ref.~\cite{Baier:2007ix} and \app{sec:hydroresp}) 
\begin{equation}
\frac{P_T-P_L}{e+p} = 
2\left(\frac{\eta/s}{\tau T(e)}\right)+\frac{4}{3}C_2\left(\frac{\eta/s}{\tau T(e)}\right)^2\label{eq:pressanis}
\end{equation}
where $T(e)$ is the equilibrium temperature obtained from the Landau matching condition, \Eq{eq:Landau}, and $C_2$ is a constant combination of first and second order transport coefficients, see \Eq{eq:C2}. The shear over entropy ratio $\eta/s$ and $C_2$ are then fitted to the late time tail of pressure anisotropy.  The fit results for different values of $\lambda$ is shown in \Fig{fig:paperetas}. Note that $C_2$ is only modestly dependent on the coupling constant and rather close to the weak coupling estimate $C_2\approx 1.0$ ($\tau_\pi \sim 5.1\eta/sT$, and $\lambda_1 \sim 0.8 \eta \tau_\pi$\cite{York:2008rr}), so that in practice the same calculated value $C_2\approx 1.0$ can be used for the entire range of the coupling constants. 

As discussed in \Sec{subsec:background}, the hydrodynamic evolution collapses to the same scaling curve if time is counted in relaxation time units, see \Eq{eq:tauR}. In order to define a simple single parameter temperature scale at all times  we introduced temperature $\TId(\tau;\Lambda_T)$ in  \Eq{eq:TId}. Relating $\TId$ to
the equilibrium temperature $T(e)$  at first viscous order
\begin{equation}
\Lambda_T^{2/3}\equiv\TId \tau^{1/3}=\tau^{1/3}T(e)\left(1+\frac{2}{3}\frac{\eta/s}{\tau T(e)}\right)\label{eq:LambdaTfit}
\end{equation} 
gives a simple and efficient way of determining the energy scale $\Lambda_T$.
Note that 
the pressure equilibration, \Eq{eq:pressanis}, can be now written explicitly  
\begin{equation}
\frac{P_T-P_L}{e+p} = 
2\left(\frac{\eta/s}{\tau \TId(\tau; \Lambda_T)}\right)+\frac{4}{3}(1+C_2)\left(\frac{\eta/s}{\tau \TId(\tau; \Lambda_T)}\right)^2
\end{equation}
as a function of time and three fit parameters: $\eta/s$, $C_2$ and $\Lambda_T$.
Remarkably, the pressure anisotropy evolution in \Fig{fig:paperetas} collapses to the same curve even  at earlier times than when the second order asymptotics becomes applicable.

Finally for completeness we list the  asymptotic expansions for temperature, energy and entropy at second order in terms of scaling variable $\xSc$:
\begin{align}
\frac{T(\tau)}{\TId} &= 1-\frac{2}{3}\xScInv-\frac{2}{9}C_2\left(\xScInv\right)^2,\\
\frac{e(\tau)}{\nu_g \frac{\pi^2}{30}\TId^4}&=1-\frac{8}{3}\xScInv+\frac{8}{9}(3-C_2)\left(\xScInv\right)^2\label{eq:ehyd},\\
\frac{s(\tau)}{\nu_g \frac{2\pi^2}{45}\TId^3} &= 1-2\xScInv+\frac{2}{3}(2-C_2)\left(\xScInv\right)^2.
\end{align}

\subsection{Parametrization of background evolution}\label{sec:paramtr}

Below we provide explicit parametrizations of the evolution of the background components of the energy momentum tensor. Based on our discussion in \Sec{subsec:background}, the kinetic theory evolution can be parametrized in terms of a single scaling function $\mathcal{E}(x)$
for the energy density 
\begin{align}
e(\tau)&=\nu_g \frac{\pi^2}{30}\TId^4\,\mathcal{E}\left[x=\frac{\tau \TId}{\eta/s}\right].
\end{align}
We parametrize the scaling function $\mathcal{E}(x)$ by the following ansatz
\begin{align}
\mathcal{E}(x) &=\frac{1}{1+\frac{8}{3}\frac{1}{x}+\frac{8}{9}(5+C_2)\frac{1}{x^2}}\text{step}(x)+\nonumber\\
&+\sqrt{\tanh(F_0^2 x (1+b x +c x^2)^2)}\Big(1-\text{step}(x)\Big),\label{eq:fit}
\end{align}
where $\text{step}(x)$ is a smooth function interpolating between $\text{step}(x\gg 1)=1$ and $\text{step}(x\ll1)=0$. Explicitly we take
\begin{eqnarray}
\text{step}(x)\equiv \frac{1}{2} \Big( 1+ \tanh \Big[\tfrac{x^2-x^2_\text{Switch}}{x x_\text{Range}} \Big] \Big)\;.
\end{eqnarray}
At late times we recover the hydrodynamic limit \Eq{eq:ehyd}
\begin{align}
x\gg1\quad \mathcal{E}(x) &\approx 1-\frac{8}{3}\frac{1}{x}+\frac{8}{9}(3-C_2)\frac{1}{x^2}.
\end{align}
which is quite well described by $C_2=1.02$ \cite{York:2008rr}. 
At early times, the energy evolution is matched to a free-streaming behavior
\begin{equation}
x\ll1\quad \mathcal{E}(x)\approx F_0\sqrt{x}.
\end{equation}

The remaining fit parameters are
\begin{align}
\sqrt{4\pi} F_0=1.0767,\quad 4\pi b = -0.284,\quad (4\pi)^2c =0.134
\end{align}
while the step function is parametrized by
\begin{align}
	x_\text{Switch}/(4\pi)&=0.65,\quad &x_\text{Range}/(4 \pi) &= 0.25
\end{align}

The transverse and longitudinal pressure is determined by functions
\begin{align}
P_L(\tau)&=\nu_g \frac{\pi^2}{90}\TId^4\,\mathcal{P}(x),\\
P_T(\tau)&=\nu_g \frac{\pi^2}{90}\TId^4\,\Big(\frac{3}{2}\mathcal{E}(x)-\frac{1}{2}\mathcal{P}(x)\Big)\;.
\end{align}
Here the auxiliary function $\mathcal P(x)$ is determined by the equation  of motion for Bjorken expansion, i.e. $\partial_\tau (\tau e)=-P_L$, 
\begin{eqnarray}
\mathcal{P}(x)=\mathcal{E}(x) - 2 x \frac{d}{dx} \mathcal{E}(x)\;.
\end{eqnarray}

\begin{figure*}
\centering
\subfig{a}{\includegraphics[width=0.4\linewidth]{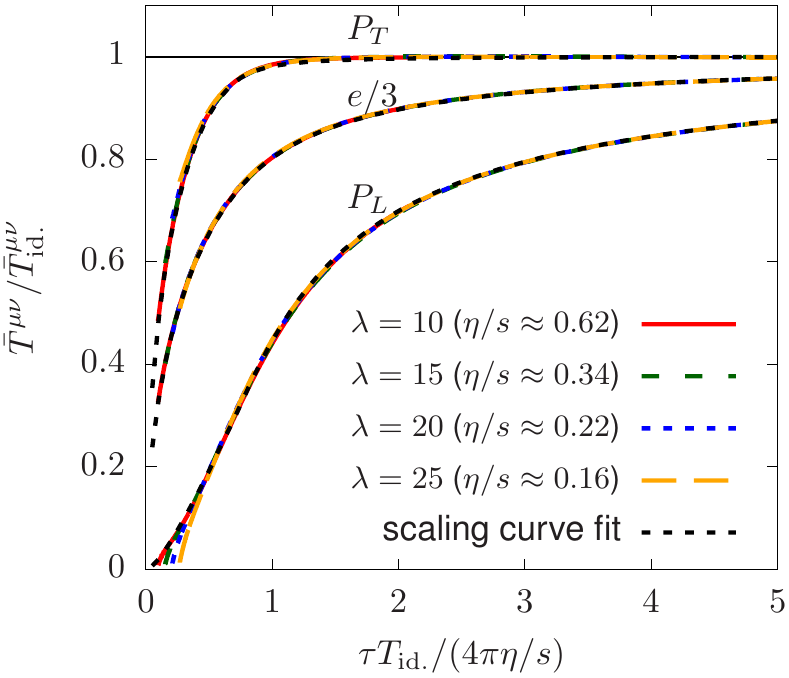}}
\subfig{b}{\includegraphics[width=0.4\linewidth]{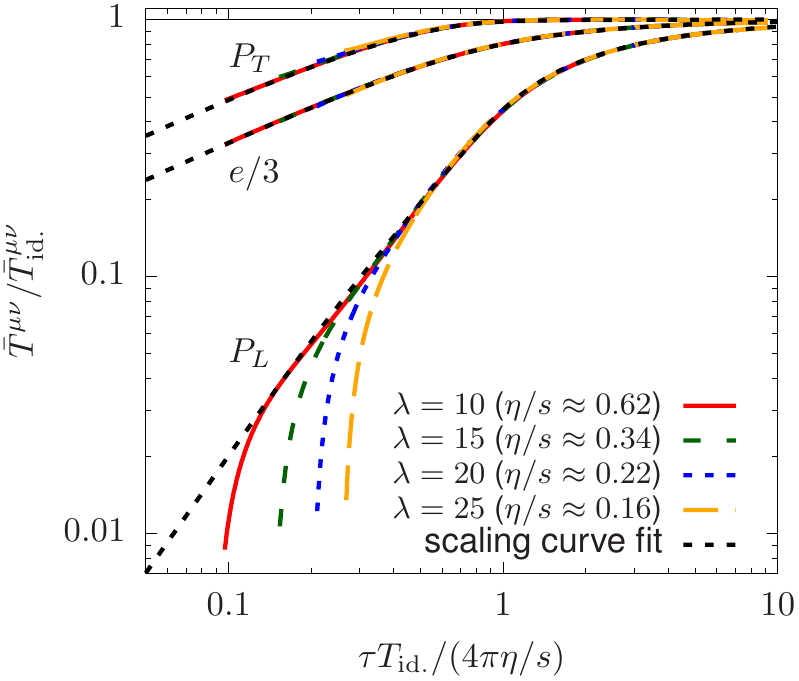}}
\caption{\label{fig:FitFunctions} The universal scaling function fit, \Eq{eq:fit}, to the evolution of background energy density on (a) linear and (b) log scale.}
\end{figure*}

 A comparison of the parametrization with the simulation data is compiled in Fig.~\ref{fig:FitFunctions}. We note that since the scaling of kinetic evolution curves is not perfect, especially for the longitudinal pressure at very early 
times, there is a systematic error in this parametrization at the order of a few percent.

\section{Decomposition of response functions\label{app:Tmunu_decompose}}

\subsection{General response functions in coordinate space}\label{sec:Hankel}

Similar to the discussion in \Sec{sec:generalresponse}, the coordinate space Green functions defined by \Eq{eq:dTmunuconvolv} can be decomposed into a complete set of tensor structures. Specifically, for (scalar) energy perturbations, the decomposition takes the form 
\begin{align}
G_{s}^{\tau\tau}(\tau,\tau_0,\rt)&= G_{s}^{s}(\tau,\tau_0,\rtt)\;,\nonumber\\
G_{s}^{\tau i}(\tau,\tau_0,\rt)&=\frac{\rt^{i}}{\rtt} G_{s}^{v}(\tau,\tau_0,\rtt)\;, \nonumber\\
G_{s}^{ij} (\tau,\tau_0,\rt)&=G_{s}^{t,\delta}(\tau,\tau_0,\rtt)~\delta^{ij}   +G_{s}^{t,r}(\tau,\tau_0,\rtt)~\frac{\rt^{i} \rt^{j}}{\rtt^2}\;,\label{eq:Gr_energy_decom}
\end{align}
and the coordinate space Green functions are related to their momentum space counterparts listed in \Eq{eq:G_energy_decomp} according to the following Fourier-Hankel transform
\begin{align}
G_{s}^{s}(\tau,\tau_0,\rtt) &= \frac{1}{2\pi} \int d\ktt \ktt J_{0}(\ktt \rtt)~ 
\tilde{G}_{s}^{s}(\tau,\tau_0,\ktt)\;,\nonumber \\
G_{s}^{v}(\tau,\tau_0,\rtt) &=\frac{1}{2\pi}  \int d\ktt \ktt J_{1}(\ktt 
\rtt)~\tilde{G}_{s}^{v}(\tau,\tau_0,\ktt)\;,\nonumber \\
G_{s}^{t,\delta}(\tau,\tau_0,\rtt) &=\frac{1}{2\pi}  \int d\ktt \ktt \bigg(   
J_{0}(\ktt\rtt) \tilde{G}_{s}^{t,\delta}(\tau,\tau_0,\ktt)\nonumber \\
&+ \frac{J_{1}(\ktt\rtt)}{\ktt \rtt} \tilde{G}_{s}^{t,k}(\tau,\tau_0,\ktt) 
\bigg)\;, \nonumber\\
G_{s}^{t,r}(\tau,\tau_0,\rtt) &=\frac{-1}{2\pi}  \int d\ktt \ktt 
J_{2}(\ktt\rtt) \tilde{G}_{s}^{t,k}(\tau,\tau_0,\ktt) \;, \label{eq:energy_hankel}
\end{align}
Similarly, for (vector) momentum perturbations, the decomposition take the form
\begin{align}
G_{v}^{\tau\tau,k}(\tau,\tau_0,\rt)&= \frac{\rt^{k}}{\rtt} G_{v}^{s}(\tau,\tau_0,\rtt)\;,\nonumber \\
G_{v}^{\tau i,k}(\tau,\tau_0,\rt)&= \delta^{ik} 
G_{v}^{v,\delta}(\tau,\tau_0,\rtt) + \frac{\rt^{i}\rt^{k}}{\rtt^2} 
G_{v}^{v,r}(\tau,\tau_0,\rtt)\;, \nonumber\\
G_{v}^{ij,k}(\tau,\tau_0,\rt)&= \delta^{ij} \frac{\rt^{k}}{\rtt} 
G_{v}^{t,\delta}(\tau,\tau_0,\rtt) + \nonumber\\
\frac{1}{2} \Big( \frac{\rt^{i}}{\rtt} 
\delta^{jk}& + \frac{\rt^{j}}{\rtt} \delta^{ik} \Big) 
G_{v}^{m,r}(\tau,\tau_0,\rtt) 
+\frac{\rt^{i}\rt^{j}\rt^{k}}{\rtt^3} G_{v}^{t,r}(\tau,\tau_0,\rtt)\;,\label{eq:Gr_momentum_decom}
\end{align}
with the different components obtained from \Eq{eq:G_momentum_decomp} according to following transformations
\begin{widetext}
\begin{align}
&G_{v}^{s}(\tau,\tau_0,\rtt) = \frac{1}{2\pi} \int d\ktt \ktt  J_{1}(\ktt 
\rtt)~\tilde G_{v}^{s}(\tau,\tau_0,\kt)\;,\nonumber \\
&G_{v}^{v,\delta}(\tau,\tau_0,\rtt) = \frac{1}{2\pi}  \int d\ktt \ktt \left(   
J_{0}(\ktt\rtt) \tilde{G}_{v}^{v,\delta}(\tau,\tau_0,\ktt) + 
\frac{J_{1}(\ktt\rtt)}{\ktt \rtt} \tilde{G}_{v}^{v,k}(\tau,\tau_0,\ktt) 
\right)\;, \nonumber \\
&G_{v}^{v,r}(\tau,\tau_0,\rtt) = \frac{-1}{2\pi}  \int d\ktt \ktt 
J_{2}(\ktt\rtt) \tilde{G}_{v}^{v,k}(\tau,\tau_0,\ktt) \;,  \nonumber \\
&G_{v}^{t,\delta}(\tau,\tau_0,\rtt) = \frac{1}{2\pi} \int d\ktt \ktt \left( 
J_{1}(\ktt\rtt)  \tilde{G}_{v}^{t,\delta}(\tau,\tau_0,\ktt)  + 
\frac{J_{2}(\ktt\rtt)}{\ktt\rtt} \tilde{G}_{v}^{t,k}(\tau,\tau_0,\ktt)   
\right)\;, \nonumber \\
&G_{v}^{t,m}(\tau,\tau_0,\rtt) = \frac{1}{2\pi} \int d\ktt \ktt \left( 
J_{1}(\ktt\rtt)  \tilde{G}_{v}^{t,m}(\tau,\tau_0,\ktt)  + 2 
\frac{J_{2}(\ktt\rtt)}{\ktt\rtt} \tilde{G}_{v}^{t,k}(\tau,\tau_0,\ktt)   
\right)\;, \nonumber \\
&G_{v}^{t,r}(\tau,\tau_0,\rtt) = \frac{-1}{2\pi}  \int d\ktt \ktt 
J_{3}(\ktt\rtt) \tilde{G}_{v}^{t,k}(\tau,\tau_0,\ktt)\;. \label{eq:momentum_hankel}
\end{align}
\end{widetext}

\subsection{Response functions from the energy-momentum tensor evolution}

Different components of the out-of-equilibrium Green functions are determined from the kinetic evolution of  the perturbations 
defined in \Eqs{eq:dfe} and \eq{eq:dfg},  and corresponding to initial  energy and momentum perturbations $\delta 
T^{\mu\nu}_{\kt,(s)}(\tau_0)$ and $\delta T^{\mu\nu,i}_{\kt,(v)}(\tau_0)$.

If initial amplitude of perturbations relative to the background is chosen such that 
\begin{align}
	 \frac{\delta T^{\tau\tau}_{\kt,(s)}(\tau_0) }{ 
		\TBg^{\tau\tau}(\tau_0)}=1\;,  \quad 
 \frac{\frac{1}{2}\delta_{ij}\delta 
	T^{\tau 
			i,j}_{\kt,(v)}(\tau_0) }{ \TBg^{\tau\tau}(\tau_0)}=1\;.  
\end{align}
then the response functions can be simply written as the following projections of the perturbation energy momentum tensor $\delta T^{\mu\nu}_{\k,(s/v)}$ at later time
\begin{align}
\tilde{G}_{s}^{s}(\tau,\tau_0,\kt)&=  
\frac{\delta T^{\tau\tau}_{\kt,(s)}(\tau)}{ \TBg^{\tau\tau}(\tau)}  \;,\nonumber \\ 
\tilde{G}_{s}^{v}(\tau,\tau_0,\kt)&=\left[\frac{+i \kt_{i}}{\ktt}\right]~\frac{\delta 
T^{\tau i}_{\kt,(s)}(\tau)}{ \TBg^{\tau\tau}(\tau)}\;, \nonumber\\ 
\tilde{G}_{s}^{t,\delta}(\tau,\tau_0,\kt)&= \left[ 
\delta_{ij} - \frac{\kt_{i}\kt_{j}}{\ktt^2} \right]  \frac{\delta 
T^{ij}_{\kt,(s)}(\tau)}{ \TBg^{\tau\tau}(\tau)}   \;, \nonumber \\
\tilde{G}_{s}^{t,k}(\tau,\tau_0,\kt)&= \left[  2 
\frac{\kt_{i}\kt_{j}}{\ktt^2} - \delta_{ij} \right] \frac{\delta 
T^{ij}_{\kt,(s)}(\tau)}{ \TBg^{\tau\tau}(\tau)} \;,
\end{align}
while for momentum perturbations the decomposition takes the form
\begin{align}
&\tilde{G}_{v}^{s}(\tau,\tau_0,\kt)=\left[\frac{+i 
\kt_{i}}{\ktt}\right]~\frac{\delta T^{\tau\tau, i}_{\kt,(v)}(\tau)}{ 
\TBg^{\tau\tau}(\tau)}\;,\nonumber  \\
&\tilde{G}_{v}^{v,\delta}(\tau,\tau_0,\kt)=\left[ 
\delta_{ij} - \frac{\kt_{i}\kt_{j}}{\ktt^2} \right] \frac{\delta T^{\tau i, 
j}_{\kt,(v)}(\tau)}{ \TBg^{\tau\tau}(\tau)}  \;,\nonumber \\
&\tilde{G}_{v}^{v,k}(\tau,\tau_0,\kt)=  
\left[2\frac{\kt_{i}\kt_{j}}{\ktt^2} -  \delta_{ij} \right] \frac{\delta 
T^{\tau i;, j}_{\kt,(v)}(\tau)}{ \TBg^{\tau\tau}(\tau)}  \;,\nonumber \\
&\tilde{G}_{v}^{t,\delta}(\tau,\tau_0,\kt)= \left[ 
\delta_{ij} - \frac{\kt_{i}\kt_{j}}{\ktt^2} \right]\left(\frac{+i 
\kt_{k}}{\ktt}\right)\frac{\delta T^{ij, k}_{\kt,(v)}(\tau)}{ 
\TBg^{\tau\tau}(\tau)},\nonumber \\
&\tilde{G}_{v}^{t,m}(\tau,\tau_0,\kt)=
2\left(\frac{+i \kt_{i}}{\ktt}\right)\left[ 
  \delta_{jk} - 
\frac{\kt_{j}\kt_{k}}{\ktt^2} \right]\frac{\delta T^{ij, k}_{\kt,(v)}(\tau)}{ 
	\TBg^{\tau\tau}(\tau)} \;, \nonumber
\\
&\tilde{G}_{v}^{t,k}(\tau,\tau_0,\kt)=\frac{\delta 
T^{ij,		k}_{\kt,(v)}(\tau)}{ \TBg^{\tau\tau}(\tau)}\times \nonumber\\
&\left[\left[2\frac{\kt_{i}\kt_{j}}{\ktt^2} - \delta_{ij} \right]
\left(\frac{+i \kt_{k}}{\ktt}\right) -2\left(\frac{+i \kt_{i}}{\ktt}\right)\left[ 
\delta_{jk} - 
\frac{\kt_{j}\kt_{k}}{\ktt^2} \right] \right]\;,
\end{align}
Based on the rotational symmetry of the background distribution, the Green 
functions only depend on $\ktt$ and it is sufficient to calculate the response 
for $\kt=(\ktt,0)$ oriented e.g.\ along the $x$-direction. We follow precisely 
this strategy, and in practice calculate the response to initial energy $\delta 
T^{\tau\tau}_{\kt,(s)}(\tau_0)$ perturbations,  longitudinal $\delta T^{\tau 
x;x}_{\kt,(v)}(\tau_0)$ and  transverse $\delta T^{\tau y;y}_{\kt,(s)}(\tau_0)$ 
momentum perturbations. Also, the kinetic theory simulation were actually performed with slightly different normalization
\begin{align}
	\frac{\delta T^{\tau\tau}_{\kt,(s)}(\tau_0) }{ 
		\TBg^{\tau\tau}(\tau_0)}=i\;,  \quad 
	\frac{\frac{1}{2}\delta_{ij}\delta 
		T^{\tau 
			i,j}_{\kt,(v)}(\tau_0) }{ \TBg^{\tau\tau}(\tau_0)}=i\;.  
\end{align}

Specifically, for $\kt=(\ktt,0)$ the decomposition for 
energy perturbations then takes the form
\begin{align}
\tilde{G}_{s}^{s}(\tau,\tau_0,\kt)&=  
\frac{\text{Im}\,\delta T^{\tau\tau}_{\kt,(s)}(\tau)}{ \TBg^{\tau\tau}(\tau)}  \;,\nonumber \\ 
\tilde{G}_{s}^{v}(\tau,\tau_0,\kt)&=\frac{\text{Re}\, \delta 
T^{\tau x}_{\kt,(s)}(\tau)}{ \TBg^{\tau\tau}(\tau)}\;,\nonumber \\ 
\tilde{G}_{s}^{t,\delta}(\tau,\tau_0,\kt)&=   \frac{\text{Im}\,  \delta 
T^{yy}_{\kt,(s)}(\tau)}{ \TBg^{\tau\tau}(\tau)}   \;, \nonumber \\
\tilde{G}_{s}^{t,k}(\tau,\tau_0,\kt)&= \frac{\text{Im}\,\delta 
T^{xx}_{\kt,(s)}(\tau)}{ \TBg^{\tau\tau}(\tau)} -\frac{\text{Im}\,\delta 
T^{yy}_{\kt,(s)}(\tau)}{ \TBg^{\tau\tau}(\tau)}\;,
\end{align}
while for momentum perturbations it is
\begin{align}
& \tilde{G}_{v}^{s}(\tau,\tau_0,\kt)=\frac{\text{Re}\, \delta 
 T^{\tau\tau, x}_{\kt,(v)}(\tau)}{ \TBg^{\tau\tau}(\tau)}\;,\nonumber  \\
&\tilde{G}_{v}^{v,\delta}(\tau,\tau_0,\kt)=\frac{\text{Im}\,\delta 
T^{\tau y, y}_{\kt,(v)}(\tau)}{ \TBg^{\tau\tau}(\tau)}  \;,\nonumber \\
&\tilde{G}_{v}^{v,k}(\tau,\tau_0,\kt)= \frac{\text{Im}\,\delta T^{\tau x, x}_{\kt,(v)}(\tau)}{ \TBg^{\tau\tau}(\tau)} - 
\frac{\text{Im}\,\delta T^{\tau y, y}_{\kt,(v)}(\tau)}{ \TBg^{\tau\tau}(\tau)}
\;,\nonumber \\
&\tilde{G}_{v}^{t,\delta}(\tau,\tau_0,\kt)= 
\frac{\text{Re}\, \delta T^{yy, x}_{\kt,(v)}(\tau)}{ \TBg^{\tau\tau}(\tau)}\;, \\
&\tilde{G}_{v}^{t,m}(\tau,\tau_0,\kt)= ~ 2 \frac{\text{Re}\,\delta 
T^{xy, y}_{\kt,(v)}(\tau)}{ \TBg^{\tau\tau}(\tau)}\;,  \\
&\tilde{G}_{v}^{t,k}(\tau,\tau_0,\kt)=\nonumber\\
&\left[ \frac{\text{Re}\,\delta T^{xx, x}_{\kt,(v)}(\tau)}{ 
\TBg^{\tau\tau}(\tau)} -2\frac{\text{Re}\,\delta T^{xy, y}_{\kt,(v)}(\tau)}{ 
\TBg^{\tau\tau}(\tau)} - \frac{\text{Re}\,\delta T^{yy, x}_{\kt,(v)}(\tau)}{ 
\TBg^{\tau\tau}(\tau)} \right]
\end{align}
which we use to determine the individual components of kinetic response functions. 

\begin{figure*}
\centering
\subfig{a}{\includegraphics[width=0.4\linewidth]{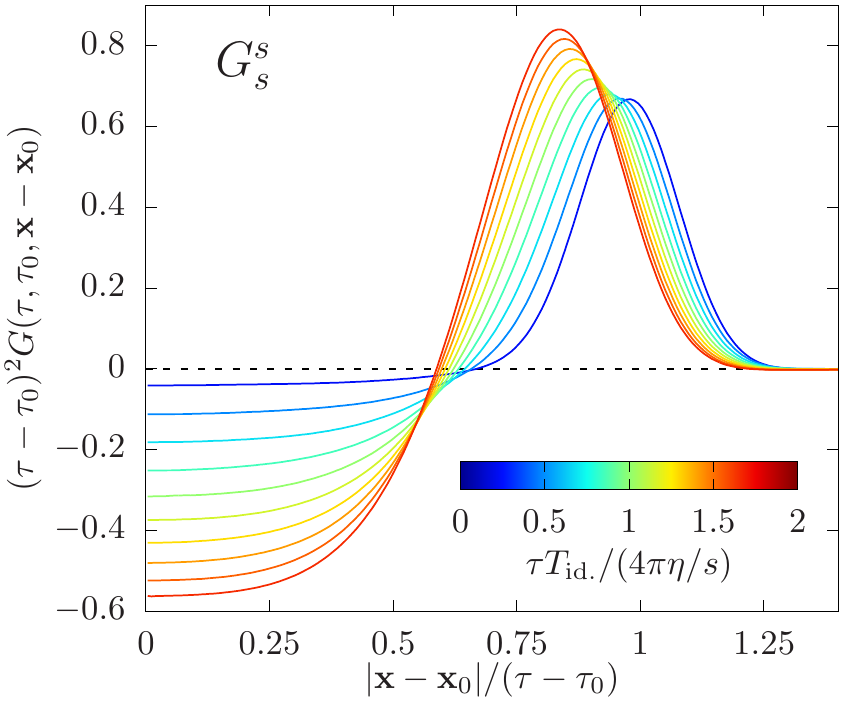}}
\subfig{b}{\includegraphics[width=0.4\linewidth]{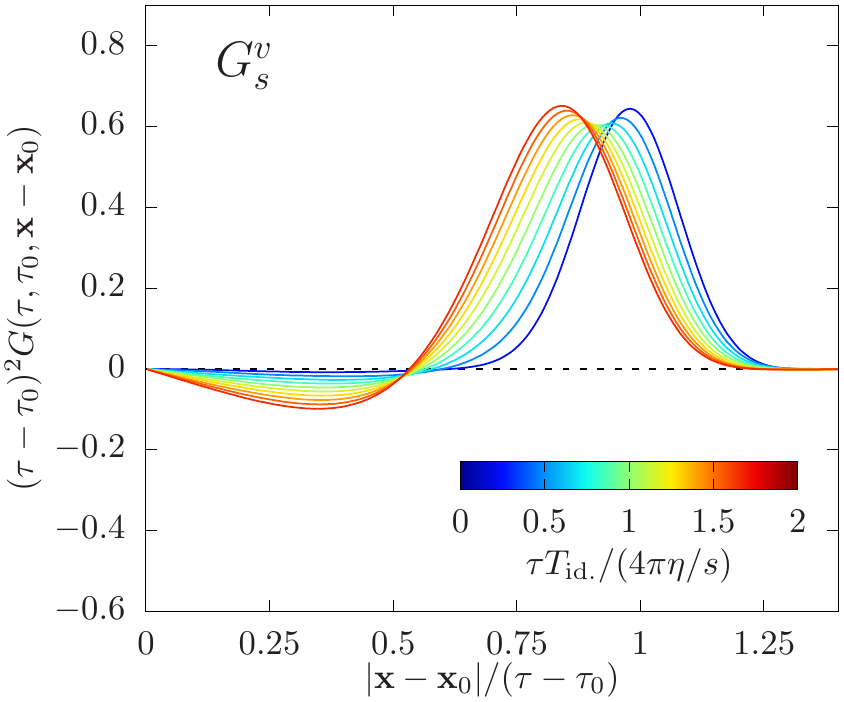}}\\
\subfig{c}{\includegraphics[width=0.4\linewidth]{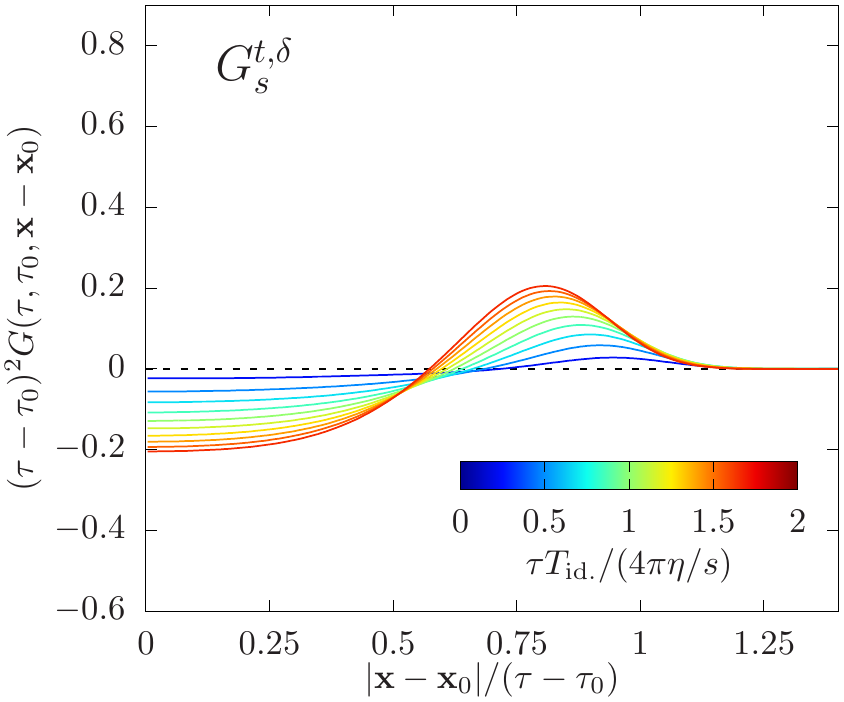}}
\subfig{d}{\includegraphics[width=0.4\linewidth]{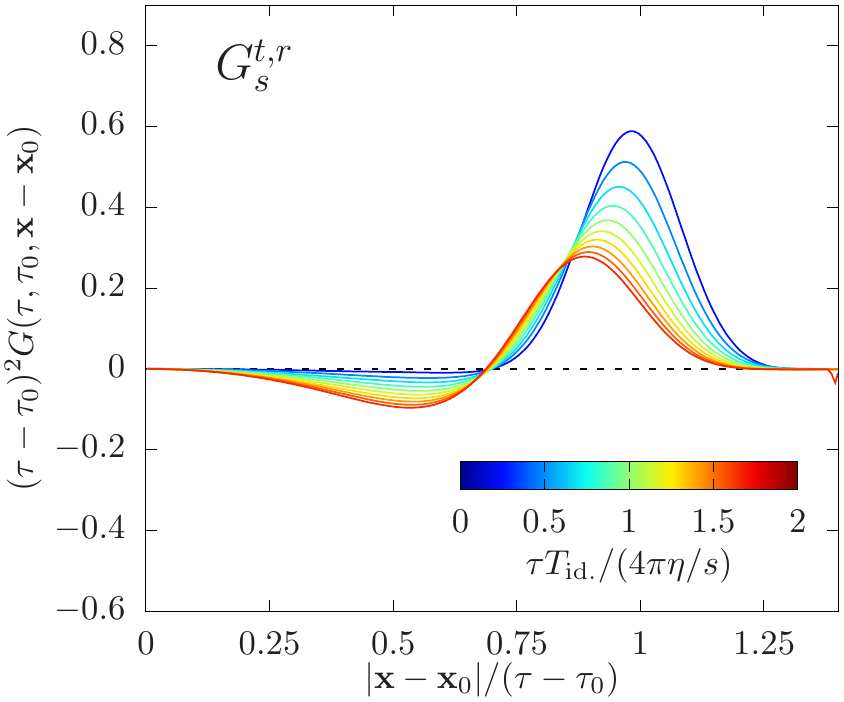}}
\caption{Independent tensor components of kinetic theory coordinate space response functions to initial energy (scalar) perturbations, \Eq{eq:Gr_energy_decom}. Different lines correspond to response functions at different \emph{scaled times} $\xSc$, \Eq{eq:scaledtime}. 
 }
\label{fig:plot_grgss}
\end{figure*}

\begin{figure*}
\centering
\subfig{a}{\includegraphics[width=0.3\linewidth]{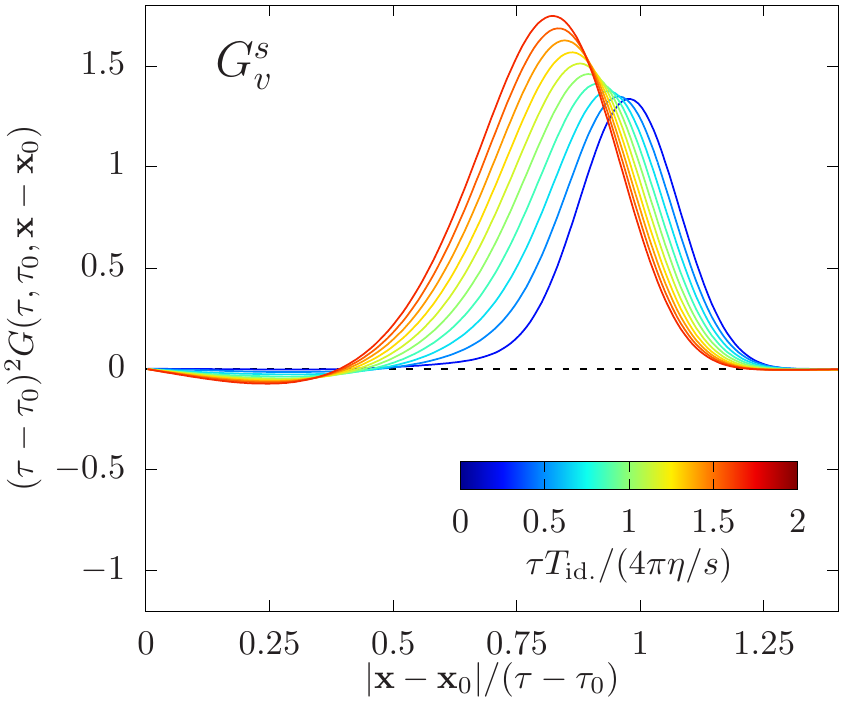}}
\subfig{b}{\includegraphics[width=0.3\linewidth]{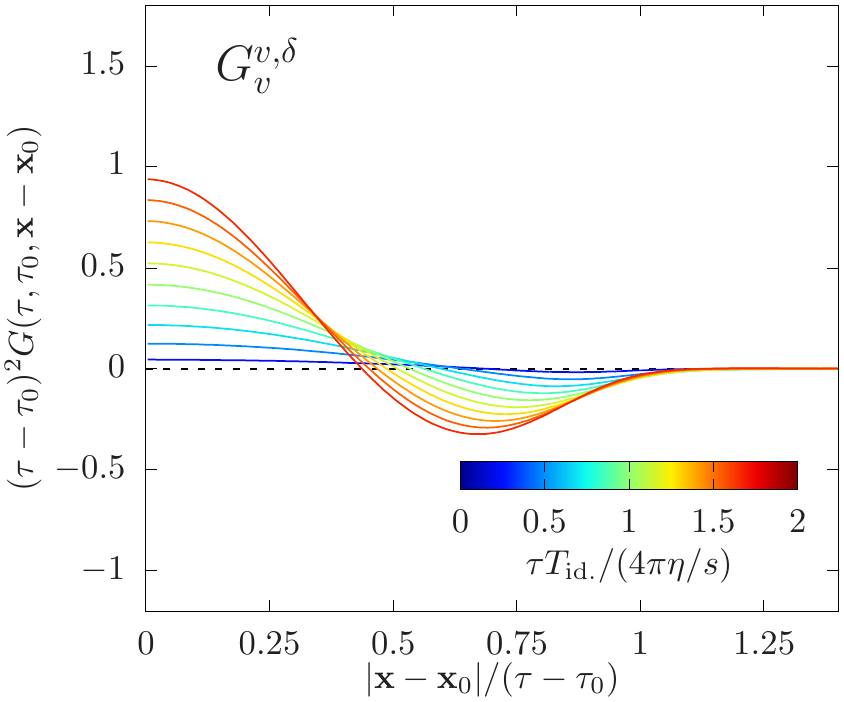}}
\subfig{c}{\includegraphics[width=0.3\linewidth]{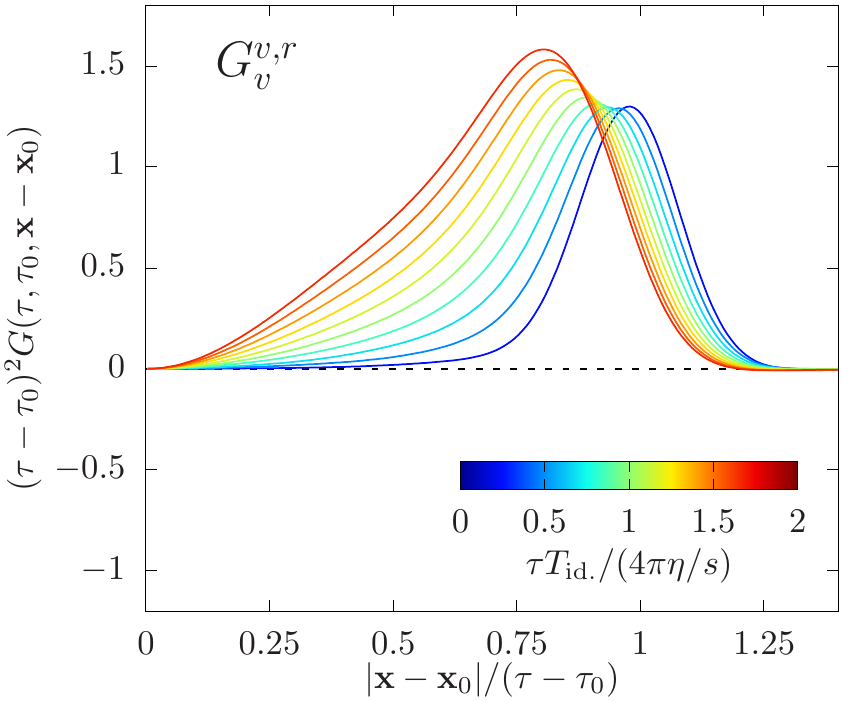}}
\subfig{d}{\includegraphics[width=0.3\linewidth]{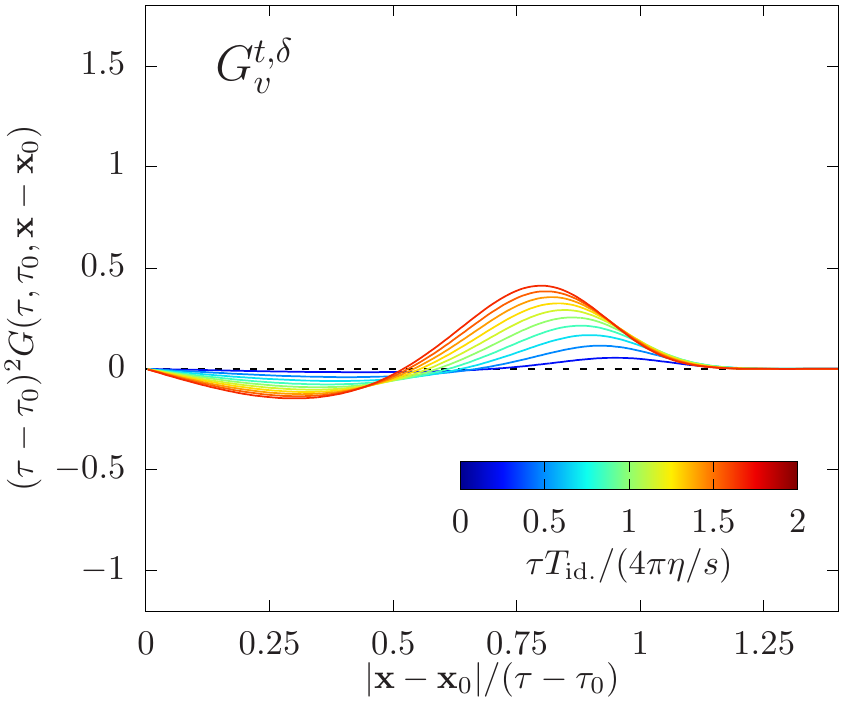}}
\subfig{e}{\includegraphics[width=0.3\linewidth]{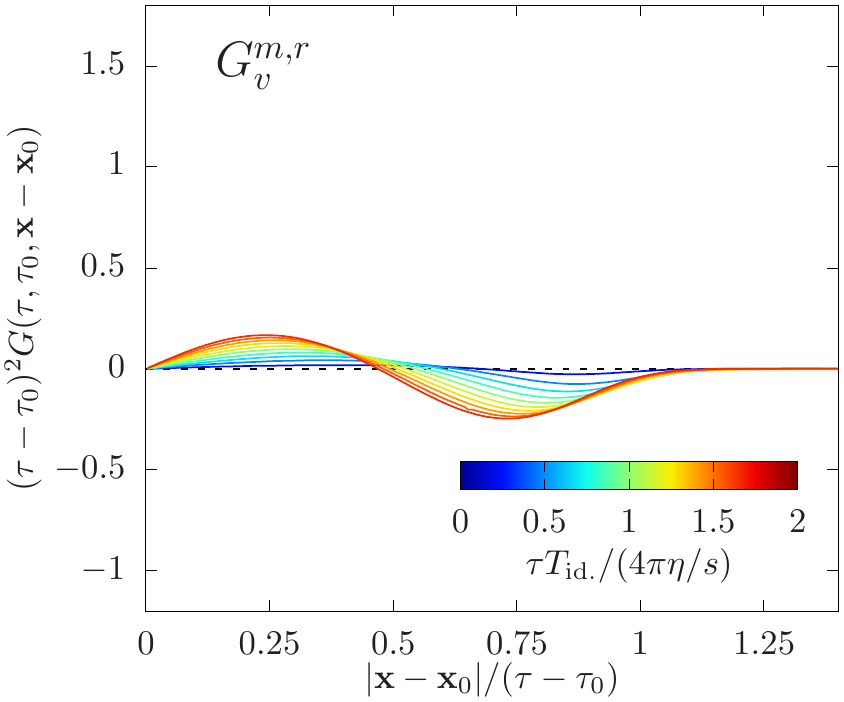}}
\subfig{f}{\includegraphics[width=0.3\linewidth]{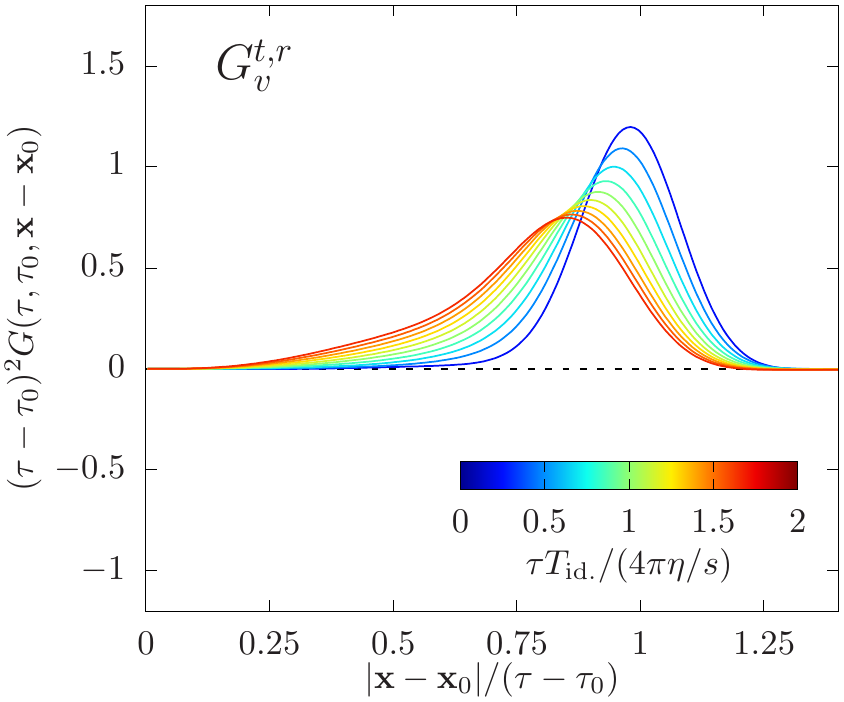}}
\caption{Independent tensor components of kinetic theory coordinate space response functions to initial momentum (vector) perturbations, \Eq{eq:Gr_momentum_decom}. Different lines correspond to response functions at different \emph{scaled times} $\xSc$, \Eq{eq:scaledtime}. }
\label{fig:plot_grgvs}
\end{figure*}

Finally, the coordinate space response functions are obtained by Fourier transform according to  \Eqs{eq:energy_hankel}  and \eq{eq:momentum_hankel} and the results are summarized in \Figs{fig:plot_grgss} and \ref{fig:plot_grgvs}. To recap, the  distribution function $\delta f_{\k,\p}$ corresponding to initial $\k$-wavenumber perturbation in energy or  momentum density are evolved according to a linerized Boltzmann  equation and the response functions are constructed from the energy-momentum tensor evolution.
 As discussed in Ref.~\cite{Keegan:2016cpi}\footnote{Note that Ref.~\cite{Keegan:2016cpi} considered an incomplete set of response function components for momentum perturbations as opposed to \Eq{eq:Gr_momentum_decom}.}, only a finite number of $\k$-space points are calculated in kinetic theory evolution, so for a smooth Fourier transform the momentum space response functions are extended with free-streaming approximation. Furthermore, the uncertainty of kinetic theory applicability for large values of wavenumber $\ktt$ can be included by regulating the large $\ktt$-tail by a Gaussian envelope $\exp(-\sigma^2 \ktt^2/2)$, where we take $\sigma\approx 0.1\,\text{fm}$, which corresponds to $\sigma/(\tau-\tau_0)=0.1$ smearing for the functions shown in \Figs{fig:plot_grgss} and \ref{fig:plot_grgvs}.

Because the momentum space response functions $\tilde{G}$, are, to a good approximation, universal functions of scaling variables $\ktt (\tau-\tau_0)$ and $\xSc$, the relations in \Eqs{eq:energy_hankel}  and \eq{eq:momentum_hankel} imply that the coordinate space response functions $(\tau-\tau_0)^2 G(\tau,\tau_0,|\r|)$ are also universal when expressed in terms of the scaling variables $\rtt / (\tau-\tau_0)$ and $\xSc$. Therefore the response functions in \Figs{fig:plot_grgss} and \ref{fig:plot_grgvs} computed for one value of the coupling constant, e.g.\ $\lambda=10$ ($\eta/s\approx 0.62$), and some initial background energy density can be reused in calculating the non-equilibrium kinetic response for a different value of effective $\eta/s$ or local energy density.

\section{Free-Streaming response functions\label{sec:freestreaming}}

Simple analytic results for the response functions can be obtained in the 
free-streaming limit, for which $P_L= 0$. This situation similar to early times of the kinetic evolution, when the expansion terms dominates over the collisions. The Bjorken expanding 
background energy density then simply scales with time 
\begin{equation}
	e(\tau)=\frac{e_0 \tau_0}{\tau}
\end{equation}
The response functions for linearized perturbations
around 
such boost invariant 
background can be straightforwardly found by solving the Boltzmann equation in 
the absence of 
interactions in Fourier space
\begin{align}
	\partial_\tau f_{\k_\perp,\p}+ i \frac{\p\cdot \k_\perp}{|\p|} 
	f_{\k_\perp,\p} - \frac{p^z}{\tau}\partial_{p^z} f_{\k_\perp,\p}=0,
\end{align} 
which is solved by
\begin{equation}
	 f_{\k_\perp} (\tau, \p_\perp, p_z) =  f_{\k_\perp}(\tau_0,\p_\perp, 
	 p_z\frac{\tau}{\tau_0})e^{-i \frac{\k_\perp\cdot 
	 \p_\perp }{|\p_\perp|}(\tau-\tau_0 \frac{\sqrt{\tau_0^2 
	 p_\perp^2+p_z^2\tau^2}}{\sqrt{\tau_0^2 p_\perp^2+p_z^2\tau_0^2}})}
\end{equation}

The independent structures of 
response functions for energy 
perturbations in Fourier space, \Eq{eq:G_energy_decomp}, can be expressed in 
terms of Bessel functions as
\begin{align}
	\tilde{G}_{s}^{s}(\tau,\tau_0,\kt)&= J_{0}\Big(\ktt 
	(\tau-\tau_0)\Big)\;,\nonumber \\ 
	\tilde{G}_{s}^{v}(\tau,\tau_0,\kt)&= J_{1}\Big(\ktt (\tau-\tau_0)\Big)\;, 
	\nonumber \\
	\tilde{G}_{s}^{t,\delta}(\tau,\tau_0,\kt)&= \frac{J_{1}\Big(\ktt 
		(\tau-\tau_0)\Big)}{\ktt (\tau-\tau_0)} \;, \nonumber \\
	\tilde{G}_{s}^{t,k}(\tau,\tau_0,\kt)&= -J_{2}\Big(\ktt (\tau-\tau_0)\Big)\;.
\end{align}
However, it is more insightful to consider the solutions in coordinate space, 
which are 
\begin{align}
	G^{s}_{s}(\tau,\tau_0,\rtt)&=	G^{v}_{s}(\tau,\tau_0,\rtt)=	
	G^{t,r}_{s}(\tau,\tau_0,\rtt)=\nonumber\\
	&=\frac{1}{2\pi 
		(\tau-\tau_0)}\delta(\rtt-(\tau-\tau_0))\nonumber\\
	G^{t,\delta}_{s}(\tau,\tau_0,\rtt)&=0
\end{align}
with characteristic concentric waves traveling at the speed of 	
light. Then
\begin{align}
	G^{\tau\tau}_{\tau\tau}(\xt,\xt_0,\tau,\tau_0)&=\frac{1}{2\pi 
		(\tau-\tau_0)}\delta(|\xt-\xt_0|-(\tau-\tau_0))\nonumber\\
	G^{\tau i}_{\tau\tau}(\xt,\xt_0,\tau,\tau_0)&= 
	\frac{(\xt-\xt_0)^{i}}{|\xt-\xt_0|}~G^{\tau\tau}_{\tau\tau}(\xt-\xt_0,\tau,\tau_0)\;,\nonumber
	\\
	G^{ij}_{\tau\tau}(\xt,\xt_0,\tau,\tau_0)&= 
	\frac{(\xt-\xt_0)^{i}(\xt-\xt_0)^{j}}{|\xt-\xt_0|^2}~G^{\tau\tau}_{\tau\tau}(\xt-\xt_0,\tau,\tau_0)\;,\label{eq:Jlate}
\end{align}
and analogous expressions for the response of the energy-momentum tensor to 
initial momentum perturbations are 
\begin{align}
	\tilde{G}_{v}^{s}(\tau,\tau_0,\kt)&= 2J_{1}\Big(\ktt 
	(\tau-\tau_0)\Big)\;,\nonumber \\ 
	\tilde{G}_{v}^{v,\delta }(\tau,\tau_0,\kt)&= 2\frac{J_{1}\Big(\ktt 
	(\tau-\tau_0)\Big)}{\ktt (\tau-\tau_0)}\;, 
	\nonumber \\
		\tilde{G}_{v}^{v,k }(\tau,\tau_0,\kt)&= -2 J_{2}\Big(\ktt 
		(\tau-\tau_0)\Big) \;, 
	\nonumber \\
	\tilde{G}_{v}^{t,\delta}(\tau,\tau_0,\kt)&= 2\frac{J_{2}\Big(\ktt 
		(\tau-\tau_0)\Big)}{\ktt (\tau-\tau_0)} \;, \nonumber \\
	\tilde{G}_{v}^{t,m}(\tau,\tau_0,\kt)&=  4 \frac{J_{2}\Big(\ktt 
		(\tau-\tau_0)\Big)}{\ktt (\tau-\tau_0)}) \;,\nonumber\\
		\tilde{G}_{v}^{t,k}(\tau,\tau_0,\kt)&= - 2 J_{3}\Big(\ktt 
		(\tau-\tau_0)\Big) \;.
\end{align}
In coordinate space
\begin{align}
	G^{s}_{v}(\tau,\tau_0,\rtt)&=	G^{v,r}_{v}(\tau,\tau_0,\rtt)=	
	G^{t,r}_{v}(\tau,\tau_0,\rtt)=\nonumber\\
	&=\frac{2}{2\pi 
		(\tau-\tau_0)}\delta(\rtt-(\tau-\tau_0))\nonumber\\
		G^{v,\delta}_{v}(\tau,\tau_0,\rtt)&=	
		G^{t,\delta}_{v}(\tau,\tau_0,\rtt)=	G^{t,m}_{v}(\tau,\tau_0,\rtt)=0
\end{align}

We also note for later comparison the long wavelength limit of the free streaming 
response functions for initial energy
\begin{align}
	\tilde{G}_{s}^{s}(\tau,\tau_0,\kt)&= 1 - \frac{1}{4} \ktt^2 
	(\tau-\tau_0)^2+ \frac{1}{64} \ktt^4 
	(\tau-\tau_0)^4\ldots , \nonumber\\ 
	\tilde{G}_{s}^{v}(\tau,\tau_0,\kt)&=\frac{1}{2}\ktt 
	(\tau-\tau_0) -\frac{1}{16} \ktt^3 
	(\tau-\tau_0)^3+\ldots\label{eq:freestrhighorder}
\end{align}
and momentum perturbations
\begin{align}
	\tilde{G}_{v}^{s}(\tau,\tau_0,\kt)&= \ktt 
	(\tau-\tau_0)  - \frac{1}{8} \ktt^3 	(\tau-\tau_0)^3 \ldots \nonumber, \\ 
	\tilde{G}_{v}^{v,\delta}(\tau,\tau_0,\kt)&=1 - \frac{1}{8} \ktt^2 	(\tau-\tau_0)^2\ldots \nonumber,\\
	\tilde{G}_{v}^{v,k}(\tau,\tau_0,\kt)&=-\frac{1}{4}\ktt^2 
	(\tau-\tau_0)^2+\ldots\label{eq:freestrhighordermom}
\end{align}

\section{Viscous Hydrodynamic response\label{sec:hydroresp}}
Below we detail the derivation of the hydrodynamic limit of the response functions for energy and momentum perturbations. We employ co-moving coordinates denotes as
\begin{align}
x^{\mu}=(\tau,x^{i},\eta)\;,
\end{align}
with $i=1,2$ labeling the transverse Lorentz indices. Our convention for the metric is mostly positive, i.e.
\begin{align}
g_{\mu\nu}&=\text{diag}\,(-1,+1,+1,+\tau^2)\;, \nonumber \\
g^{\mu\nu}&=\text{diag}\,(-1,+1,+1,+1/\tau^2)\;,
\end{align}
such that the only non-vanishing Christoffel symbols
\begin{eqnarray} 
\Gamma^{\alpha}_{\beta \gamma}=\frac{1}{2} g^{\alpha \mu }\Big( \frac{\partial 
g_{\mu\beta}}{\partial x^\gamma} + \frac{\partial g_{\mu\gamma}}{\partial 
x^\beta} - \frac{\partial g_{\beta\gamma}}{\partial x^\mu} \Big)\;,
\end{eqnarray}
are given by
\begin{align}
\Gamma^{\eta}_{\eta\tau}&=\frac{1}{\tau}\;, \qquad 
\Gamma^{\eta}_{\tau\eta}=\frac{1}{\tau}\;, \qquad 
\Gamma^{\tau}_{\eta\eta}=\tau\;.
\end{align}

\subsection{Second order viscous constitutive relations\label{app:constitutive}}

Hydrodynamic constitutive relations take the form
\begin{eqnarray}
T^{\mu\nu}=(e+p)~u^{\mu} u^{\nu} +p~g^{\mu\nu} + \pi^{\mu\nu}\;,
\end{eqnarray}
where at second order the shear stress tensor $\pi^{\mu\nu}$ is given 
by\footnote{We neglect the term associated with vorticity $\sim\lambda_2\left< 
	\sigma_{\phantom{\mu}\lambda}^{\mu} \Omega^{\nu\lambda} \right>$.}
\begin{align}
\pi^{\mu\nu} &= -\eta \sigma^{\mu\nu} +\eta  \tau_{\pi} \left[ \left< 
u^\lambda \nabla_\lambda\sigma^{\mu\nu} 
\right>  + \frac{1}{3} \sigma^{\mu\nu} \nabla_\lambda u^\lambda \right] 
\nonumber\\
&\phantom{= -\eta \sigma^{\mu\nu} +}+\lambda_1  \left< 
\sigma_{\phantom{\mu}\lambda}^{\mu} \sigma^{\nu 
	\lambda} \right> ,\\
\sigma^{\mu\nu}&= 2\left<\nabla^\mu u^\nu\right>\nonumber\\
&=\Big(\Delta^{\mu\alpha} 
\Delta^{\nu\beta} +\Delta^{\mu\beta} 
\Delta^{\nu\alpha} - \frac{2}{3} \Delta^{\mu\nu} \Delta^{\alpha\beta} \Big) 
\nabla_{\alpha} u_{\beta}\;,\label{eq:sigmamunu}
\end{align}
where $\nabla_{\alpha} u_{\beta}=\partial_{\alpha} u_{\beta} - 
\Gamma^{\mu}_{\alpha\beta} u_{\mu}$ denotes the covariant derivative and 
$\Delta^{\mu\nu}=g^{\mu\nu}+u^{\mu} u^{\nu}$ is the usual projector to the rest 
frame. Based on 
these definitions, the components of the shear-stress tensor can be evaluated 
according to the following relations (no summation over $\eta$ index).
\begin{eqnarray}
&&
\text{Background:} \nonumber \\
&&
\Delta^{\tau\tau}=0\;, \qquad
\Delta^{\tau i}=0\;, \qquad
\Delta^{ij}=\delta^{ij}\;, \qquad
\Delta^{\eta\eta}=1/\tau^2\;. 
\nonumber \\
&&
\nabla_{\alpha} u_{\beta}=+\Gamma^{\tau}_{\alpha\beta}\;, \quad
\Delta^{\alpha\beta}  \nabla_{\alpha} u_{\beta}= \nabla_\lambda u^\lambda = 
+\frac{1}{\tau}\;, 
\nonumber \\
&& 
\sigma^{\tau\tau}=0\;, \quad
\sigma^{\tau i}=0\;, \quad
\sigma^{ij}=-\frac{2}{3} \frac{1}{\tau}~\delta^{ij}\;, \quad
\sigma^{\eta}_{~\eta}=\frac{4}{3} \frac{1}{\tau}\;. \nonumber \\
&& 
\left< 
u^\lambda \nabla_\lambda\sigma^{\tau\tau} 
\right>  + \frac{1}{3} \sigma^{\tau \tau} \nabla_\lambda u^\lambda =0\;, 
\nonumber\\
&&
\left< 
u^\lambda \nabla_\lambda\sigma^{\tau i} 
\right>  + \frac{1}{3} \sigma^{\tau i} \nabla_\lambda u^\lambda =0\;, 
\nonumber\\
&&
\left< 
u^\lambda \nabla_\lambda\sigma^{ij} 
\right>  + \frac{1}{3} \sigma^{ij} \nabla_\lambda u^\lambda 
=\frac{4}{9}\delta^{ij}\frac{1}{\tau^2}\;, 
\nonumber\\
&&
\left< 
u^\lambda \nabla_\lambda\sigma^{\eta}_{\phantom{\eta}\eta} 
\right>  + \frac{1}{3} \sigma^{\eta}_{\phantom{\eta}\eta} \nabla_\lambda 
u^\lambda 
=-\frac{8}{9}\frac{1}{\tau^2}\;, 
\nonumber\\
&&
 \left< 
\sigma_{\phantom{\mu}\lambda}^{\tau} \sigma^{\tau 
	\lambda} \right> = 0\;,\quad  \left< 
\sigma_{\phantom{\mu}\lambda}^{\tau} \sigma^{i 
\lambda} \right> =0\;,\nonumber\\
&&
  \left< 
\sigma_{\phantom{\mu}\lambda}^{i} \sigma^{j 
\lambda} \right> =-\frac{4}{9}\frac{1}{\tau^2}\delta^{ij}\;,\quad   \left< 
\sigma_{\phantom{\mu}\lambda}^{\eta} \sigma^{\phantom{\eta} 
\lambda}_\eta \right> =\frac{8}{9}\frac{1}{\tau^2}
\end{eqnarray}

\begin{eqnarray}
&&
\text{Linearized transverse perturbations $(u^{\mu}\delta u_{\mu}=0)$:}   
\nonumber \\
&& 
\delta \Delta^{\tau\tau}=0\;, \quad
\delta \Delta^{\tau i}=\delta u^{i}\;, \quad
\delta \Delta^{ij}=0\;, \quad
\delta \Delta^{\eta\eta}=0\;. 
\nonumber \\
&&
\nabla_{\alpha} \delta u_{\beta}=\partial_{\alpha} \delta u_{\beta}\;, \qquad
\Delta^{\alpha\beta}  \nabla_{\alpha} \delta u_{\beta} + \delta 
\Delta^{\alpha\beta}  \nabla_{\alpha} u_{\beta}=\partial_{i}u^{i}\;, \nonumber 
\\
&& 
\delta \sigma^{\tau\tau}= 0 \;, \qquad
\delta \sigma^{\tau i}= - \frac{2}{3} \frac{1}{\tau} \delta u^{i}\;, \qquad
\nonumber \\
&&
\delta \sigma^{ij}= \Big( \partial^{i} \delta u^{j} +\partial^{j} \delta 
u^{i} -\frac{2}{3} \delta^{ij} \partial_{k} \delta u^{k}\Big) 
\;, \nonumber \\
&&
\delta \sigma^{\eta}_{~\eta}=-\frac{2}{3}~\partial_{k} \delta u^{k} 
 \;,\nonumber\\
 && 
 \delta \left[\left< 
 u^\lambda \nabla_\lambda\sigma^{\tau\tau} 
 \right>  + \frac{1}{3} \sigma^{\tau \tau} \nabla_\lambda u^\lambda\right] 
 =0\;, 
 \nonumber\\
 &&
 \delta \left[ \left< 
 u^\lambda \nabla_\lambda\sigma^{\tau i} 
 \right>  + \frac{1}{3} \sigma^{\tau i} \nabla_\lambda u^\lambda 
 \right]=\frac{4}{9}\frac{1}{\tau^2}v^i\;, 
 \nonumber\\
 &&
 \delta \left[ \left< 
 u^\lambda \nabla_\lambda\sigma^{ij} 
 \right>  + \frac{1}{3} \sigma^{ij} \nabla_\lambda u^\lambda \right]
 =(\frac{1}{3\tau}+\partial_\tau) 
 \delta\sigma^{ij}-\frac{4}{9\tau}\delta^{ij}\partial_k \delta u^k\;, 
 \nonumber\\
 &&
\delta \left[\left< 
 u^\lambda \nabla_\lambda\sigma^{\eta}_{\phantom{\eta}\eta} 
 \right>  + \frac{1}{3} \sigma^{\eta}_{\phantom{\eta}\eta} \nabla_\lambda 
 u^\lambda \right]
 =(-\frac{1}{3\tau}+\partial_\tau) 
 \delta\sigma^{\eta}_{\phantom{\eta}\eta}\;, 
 \nonumber\\
 &&
\delta \left< 
 \sigma_{\phantom{\mu}\lambda}^{\tau} \sigma^{\tau 
 	\lambda} \right> = 0\;,\quad \delta \left< 
 \sigma_{\phantom{\mu}\lambda}^{\tau} \sigma^{i 
 	\lambda} \right> =-\frac{4}{9}\frac{1}{\tau^2}v^i\;,\nonumber\\
 &&
\delta \left< 
 \sigma_{\phantom{\mu}\lambda}^{i} \sigma^{j 
 	\lambda} \right> =-\frac{4}{3}\frac{1}{\tau}\delta 
 	\sigma^{ij}+\frac{8}{9\tau}\delta^{ij}\partial_k \delta u^k\;,\nonumber\\
 &&   \left< 
 \sigma_{\phantom{\mu}\lambda}^{\eta} \sigma^{\phantom{\eta} 
 	\lambda}_\eta \right> =-\frac{8}{9}\frac{1}{\tau}\partial_k \delta u^k 
\end{eqnarray}
So the energy-momentum tensor takes the form
\begin{eqnarray}
&&
\text{Background:} \nonumber \\
&&
\bar T^{\tau\tau}=e\;, \qquad
\bar T^{ij}=\Big(p+\frac{2}{3} \frac{\eta}{\tau}+\frac{4}{9}\frac{\tau_\pi \eta 
-\lambda_1}{\tau^2}\Big) \delta^{ij}\;,  \nonumber \\
&&
\bar T^{\tau i}=0\;, \qquad
\bar T^{\eta}_{~\eta}=\Big(p-\frac{4}{3} 
\frac{\eta}{\tau}-\frac{8}{9}\frac{\tau_\pi 
\eta 
	-\lambda_1}{\tau^2}\Big)\;, \qquad
 \\
&&
\text{Linearized perturbations:}   \nonumber \\
&&
\delta T^{\tau\tau}=\delta e\;, \qquad
\delta T^{\tau i}=\Big(e+p+ \frac{2}{3} 
\frac{\eta}{\tau}+\frac{4}{9}\frac{\tau_\pi \eta 
	-\lambda_1}{\tau^2}\Big) \delta u^{i}\;, \qquad
\nonumber \\
&&
\delta T^{ij}=\frac{\partial \bar T^{ij}}{\partial e}\delta e +(-\eta+\eta 
\tau_\pi(\partial_\tau +\frac{1}{3\tau})-\frac{4}{3\tau}\lambda_1 )( 
\partial^{i} \delta 
u^{j} 
+\partial^{j} \delta u^{i})  \;, \nonumber \\
&&
\qquad \qquad-\frac{2}{3}\delta^{ij}  (-\eta +\eta\tau_\pi(\partial_\tau 
+\frac{2}{3\tau})-\frac{8}{3\tau}\lambda_1     ) \partial_k \delta u^k 
\nonumber\\
&&
\delta T^{\eta}_{~\eta}= \frac{\partial \bar T^{\eta}_{~\eta}}{\partial 
e}\delta e 
- \frac{2}{3}(-\eta+\eta 
\tau_\pi(\partial_\tau -\frac{1}{3\tau})+\frac{4}{3\tau}\lambda_1  
)~\partial_{k} u^{k}  \;, 
\end{eqnarray}

Specifically for a conformal system the speed of sound $c_s^2=1/3$,
shear viscosity to entropy density ratio ${\eta}/{s}$, and seconder order 
transport terms $\tau_\pi T/(\eta/s)$ and $\lambda_1/(\tau_\pi \eta)$ are all 
constant. Then
\begin{align}
\frac{\partial \eta}{\partial e}= \frac{\eta}{sT},\quad \frac{\partial 
\tau_\pi}{\partial e}=-\frac{c_s^2\tau_\pi}{sT},\quad 
\frac{\partial\lambda_1}{\partial e}=\frac{(1-c_s^2)\lambda_1}{sT}.
\end{align}
Also the time derivatives in the second order terms can be replaced by using 
the leading order (ideal) equations of motion for perturbations, namely,
\begin{equation}
\partial_\tau \delta u^i= \frac{c_s^2}{\tau}\delta 
u^i-\frac{c_s^2}{sT}\partial^i \delta e.\quad \text{ideal order}
\end{equation}

By closer inspection of the relations, one immediately observes that the perturbation of the flow velocity can be inferred as
\begin{eqnarray}
\delta u^{i}=\frac{\delta T^{\tau i}}{\bar T^{\tau\tau}+\frac{1}{2}\bar 
T^{k}_{~k}}
\end{eqnarray}
which can be used to eliminate the variable $\delta u^{i}$ from the equations 
of motion.

\begin{eqnarray}
&&
\text{Linearized perturbations:}   \nonumber \\
&&
\delta T^{ij}=\delta T^{\tau\tau} \Big[ (c_s^2+ \frac{2}{3}\frac{\eta/s}{\tau 
T}+ \frac{4(1-c_s^2)}{9}\frac{\tau_\pi \eta-\lambda_1}{sT\tau^2} 
-\frac{2}{3} c_s^2 k^2 \frac{\tau_\pi \eta}{sT})\delta^{ij}\nonumber\\
&&
\hspace{6cm}+2c_s^2 k^ik^j  
\frac{\tau_\pi\eta}{sT} \Big]+
\nonumber \\
&& \qquad \qquad+ \frac{ik^i\delta T^{\tau j}+ik^j\delta T^{\tau i}}{\bar 
T^{\tau\tau}+\frac{1}{2}\bar 
	T^{k}_{~k}}\Big[-\eta + \eta\tau_\pi 
	\frac{c_s^2+\frac{1}{3}}{\tau}-\frac{4}{3\tau}\lambda_1\Big] 
\nonumber \\
&& \qquad \qquad - \frac{2}{3} \frac{ik_k\delta T^{\tau k} \delta^{ij}}{\bar 
	T^{\tau\tau}+\frac{1}{2}\bar 
	T^{k}_{~k}}\Big[-\eta + \eta\tau_\pi 
\frac{c_s^2+\frac{2}{3}}{\tau}-\frac{8}{3\tau}\lambda_1\Big] \;, 
\nonumber \\
&&
\delta T^{\eta}_{~\eta }=\delta T^{\tau\tau} \Big[ c_s^2- 
\frac{4}{3}\frac{\eta/s}{\tau 
	T}- \frac{8(1-c_s^2)}{9}\frac{\tau_\pi \eta-\lambda_1}{sT\tau^2} 
-\frac{2}{3} c_s^2 k^2 \frac{\tau_\pi \eta}{sT}\Big]
\nonumber \\
&& \qquad \qquad - \frac{2}{3} \frac{ik_k\delta T^{\tau k} }{\bar 
	T^{\tau\tau}+\frac{1}{2}\bar 
	T^{k}_{~k}}\Big[-\eta + \eta\tau_\pi 
\frac{c_s^2-\frac{1}{3}}{\tau}+\frac{4}{3\tau}\lambda_1\Big] \;,
\end{eqnarray}

By comparing the hydrodynamic constitutive relation with the 
explicit expressions in terms of the non-equilibrium Greens functions
\begin{widetext}
\begin{eqnarray}
&&
\text{Energy perturbations:}   \nonumber \\
&& \delta T^{\tau\tau}(\tau,\kt)=\quad\left( \frac{\TBg(\tau)}{\TBg(\tau_0)} 
\right) G_{s}^{s}(\tau,\tau_0,\kt)~\delta T^{\tau\tau}(\tau_0,\kt)\;, \nonumber 
\\
&& \delta T^{\tau i}(\tau,\kt)=-i \left( \frac{\TBg(\tau)}{\TBg(\tau_0)} 
\right) \frac{\kt^{i}}{\ktt} G_{s}^{v}(\tau,\tau_0,\kt)~\delta 
T^{\tau\tau}(\tau_0,\kt)\;, \nonumber \\
&& \delta T^{ij}(\tau,\kt)=\quad\left( \frac{\TBg(\tau)}{\TBg(\tau_0)} \right) 
\Big( G_{s}^{t,\delta}(\tau,\tau_0,\kt) \delta^{ij} + 
G_{s}^{t,k}(\tau,\tau_0,\kt) \frac{\kt^{i}\kt^{j}}{\ktt^2} \Big)~\delta 
T^{\tau\tau}(\tau_0,\kt) \\
&&
\text{Momentum perturbations:}   \nonumber \\
&& \delta T^{\tau\tau}(\tau,\kt)=-i \left( \frac{\TBg(\tau)}{\TBg(\tau_0)} 
\right)~\frac{\kt^{i}}{\ktt}~G_{v}^{s}(\tau,\tau_0,\kt)~\delta T^{\tau 
i}(\tau_0,\kt)\;, \nonumber \\
&& \delta T^{\tau i}(\tau,\kt)=\quad\left( \frac{\TBg(\tau)}{\TBg(\tau_0)} 
\right)~ \Big( G_{v}^{v,\delta}(\tau,\tau_0,\kt) \delta^{ij} + 
G_{v}^{v,k}(\tau,\tau_0,\kt) \frac{\kt^{i}\kt^{j}}{\ktt^2} \Big)~\delta T^{\tau 
j}(\tau_0,\kt)\;, \nonumber \\
&& \delta T^{ij}(\tau,\kt)= -i\left( \frac{\TBg(\tau)}{\TBg(\tau_0)} \right) 
\Big( G_{v}^{t,\delta}(\tau,\tau_0,\kt) \delta^{ij} \frac{\kt^{k}}{\ktt} + 
G_{v}^{t,m}(\tau,\tau_0,\kt) \frac{\delta^{ik} \kt^{j} + \delta^{jk} 
\kt^{i}}{2\ktt} + G_{v}^{t,k}(\tau,\tau_0,\kt) 
\frac{\kt^{i}\kt^{j}\kt^{k}}{\ktt^3} \Big)~\delta T^{\tau k}(\tau_0,\kt)  
\nonumber\\
\end{eqnarray}
It is then straightforward to derive the following relations between the 
Green functions
\begin{eqnarray}
&&
\text{Energy perturbations:}   \nonumber \\
&&
G_{s}^{t,\delta}(\tau,\tau_0,\kt) =
\Big( c_s^2+ \frac{2}{3}\frac{\eta/s}{\tau 
	T}+ \frac{4(1-c_s^2)}{9}\frac{\tau_\pi \eta-\lambda_1}{sT\tau^2} 
-\frac{2}{3} c_s^2 k^2 \frac{\tau_\pi \eta}{sT} 
\Big)~G_{s}^{s}(\tau,\tau_0,\kt)+\nonumber\\
&&\qquad -\frac{2}{3}\Big[-\frac{\eta/s}{\tau T} + \eta\tau_\pi 
\frac{c_s^2+\frac{2}{3}}{\tau s 
T}-\frac{8}{3}\frac{\lambda_1}{sT}\Big]~\frac{\ktt
 \tau sT}{\bar 
 T^{\tau\tau}+\frac{1}{2}\bar 
 T^{k}_{~k}}~G_{s}^{v}(\tau,\tau_0,\kt)\;, \nonumber \\
&&
G_{s}^{t,k}(\tau,\tau_0,\kt) = 
2(-\frac{\eta/s}{\tau T} + \eta\tau_\pi 
\frac{c_s^2+\frac{1}{3}}{\tau 
sT}-\frac{4}{3\tau}\frac{\lambda_1}{sT})~\frac{\ktt
\tau 
sT}{\TBg^{\tau\tau}+\frac{1}{2}\TBg^{k}_{~k}}~G_{s}^{v}(\tau,\tau_0,\kt)+2 
c_s^2 \ktt^2\tau^2 \frac{\tau_\pi \eta}{\tau^2 sT}~G_{s}^{s}(\tau,\tau_0,\kt)
 \;,    \\
&&
\text{Momentum perturbations:} \nonumber \\
&&G_{v}^{t,\delta}(\tau,\tau_0,\kt)=\Big( c_s^2+ \frac{2}{3}\frac{\eta/s}{\tau 
	T}+ \frac{4(1-c_s^2)}{9}\frac{\tau_\pi \eta-\lambda_1}{sT\tau^2} 
-\frac{2}{3} c_s^2 k^2 \frac{\tau_\pi \eta}{sT} 
\Big)~G_{v}^{s}(\tau,\tau_0,\kt) +\nonumber\\
&&\qquad\qquad +\frac{2}{3}\Big[-\frac{\eta/s}{\tau T} 
+ \eta\tau_\pi 
\frac{c_s^2+\frac{2}{3}}{\tau s 
	T}-\frac{8}{3}\frac{\lambda_1}{sT}\Big]\frac{sT\ktt
	\tau }{\TBg^{\tau\tau}+\frac{1}{2}\TBg^{k}_{~k}}~\Big( 
	G_{v}^{v,\delta}(\tau,\tau_0,\kt) + G_{v}^{v,k}(\tau,\tau_0,\kt)  
\Big) \nonumber \\
&&G_{v}^{t,m}(\tau,\tau_0,\kt)= 
2(-\frac{\eta/s}{\tau T} + \eta\tau_\pi 
\frac{c_s^2+\frac{1}{3}}{\tau 
	sT}-\frac{4}{3\tau}\frac{\lambda_1}{sT})~\frac{\ktt
	\tau 
	sT}{\TBg^{\tau\tau}+\frac{1}{2}\TBg^{k}_{~k}}~G_{v}^{v,\delta}(\tau,\tau_0,\kt)
	 \nonumber \\
&&G_{v}^{t,k}(\tau,\tau_0,\kt)=  
2(-\frac{\eta/s}{\tau T} + \eta\tau_\pi 
\frac{c_s^2+\frac{1}{3}}{\tau 
	sT}-\frac{4}{3\tau}\frac{\lambda_1}{sT})~\frac{\ktt
	\tau 
	sT}{\TBg^{\tau\tau}+\frac{1}{2}\TBg^{k}_{~k}}~G_{v}^{v,k}(\tau,\tau_0,\kt)+2
	 c_s^2 \ktt^2\tau^2 \frac{\tau_\pi \eta}{\tau^2 
	 sT}~G_{v}^{s}(\tau,\tau_0,\kt)\label{eq:G_momentum_const}
\end{eqnarray}
\end{widetext}
where to this order of accuracy in the hydrodynamic expansion one can
approximate
\begin{equation}
 \frac{sT}{\bar T^{\tau\tau}+\frac{1}{2}\bar T^{k}_{~k}}\approx 
1+\frac{2}{3}\frac{\eta/s}{\tau T}
\end{equation}
\subsection{Hydrodynamic response functions in the long wave-length limit\label{sec:hydrolimit}}

In the limit of $\ktt \to 
0$, one can derive the exact expressions for energy and momentum perturbations from
the conservation laws of energy and momentum~\cite{Keegan:2016cpi}. 
 Based on our symmetry assumptions the relevant evolution equations 
\begin{eqnarray}
\nabla_{\mu} T^{\mu\nu}=0\;,
\end{eqnarray}
take the explicit form
\begin{eqnarray}
\partial_{\tau} T^{\tau\tau}+\partial_{i}T^{i \tau} &=& -\frac{T^{\tau\tau}+T^{\eta}_{~\eta}}{\tau}\;, \\
\partial_{\tau} T^{\tau j} + \partial_{i} T^{ij}  &=& -\frac{T^{\tau j}}{\tau}\;.
\end{eqnarray}

Evolution equations for linearized energy perturbations then take the form
\begin{eqnarray}
\partial_{\tau}~\frac{\delta T^{\tau\tau}}{\TBg^{\tau\tau}}+\partial_{i}T^{i \tau} = - \frac{ \delta T^{\tau\tau} + \delta T^{\eta}_{~\eta}}{\tau \TBg^{\tau\tau}} +  \frac{\delta T^{\tau\tau}}{\TBg^{\tau\tau}} \frac{ T^{\tau\tau} +  T^{\eta}_{~\eta}}{\tau \TBg^{\tau\tau}}\;. \nonumber \\
\end{eqnarray}
which can be re-written as
\begin{eqnarray}
\tau \partial_{\tau}~\frac{\delta T^{\tau\tau}}{\TBg^{\tau\tau}}+\partial_{i}T^{i \tau} = -  \frac{\delta T^{\tau\tau}}{\TBg^{\tau\tau}} \Big( \frac{\delta T^{\eta}_{~\eta}}{\delta T^{\tau\tau}} - \frac{\TBg^{\eta}_{~\eta}}{\TBg^{\tau\tau}}\Big)
\end{eqnarray}
In this section we limit ourselves only to the first order terms in the constitutive equations described in \Sec{sec:hydroresp}. 
We will also assume in the following that the background solution for the longitudinal pressure $\TBg^{\eta}_{~\eta}(\tau)$ can be expressed in the following form
\begin{eqnarray}
\label{eq:HydroScalingAssumption}
\TBg^{\eta}_{~\eta}(\tau)=\TBg^{\tau\tau}(\tau)~f\Big( \TBg^{\tau\tau}(\tau)^{1/4} \tau \Big)
\end{eqnarray}
where the energy dependence of the function $f$ accounts for the change in 
evolution speed under variations of the energy scale. Based on this relation, 
we can  find the ratio $\frac{\delta T^{\eta}_{~\eta}}{\delta T^{\tau\tau}}$ at 
$k=0$ by varying according \Eq{eq:HydroScalingAssumption} with respect 
to $\TBg^{\tau\tau}$ to
\begin{eqnarray}
\left.\frac{\delta T^{\eta}_{~\eta}}{\delta T^{\tau\tau}}\right|_{k=0}&=& f\Big( \TBg^{\tau\tau}(\tau)^{1/4} \tau \Big)  \\
&&+ \frac{1}{4} \Big(\TBg^{\tau\tau}(\tau)\Big)^{1/4} \tau~f'\Big( \TBg^{\tau\tau}(\tau)^{1/4} \tau \Big) \;,  \nonumber
\end{eqnarray}
while for non-vanishing $k$ there is an additional contribution related to the momentum flow
\begin{eqnarray}
\frac{\delta T^{\tau\tau}}{\TBg^{\tau\tau}} \left(\frac{\delta T^{\eta}_{~\eta}}{\delta T^{\tau\tau}}-\left.\frac{\delta T^{\eta}_{~\eta}}{\delta T^{\tau\tau}}\right|_{k=0}\right)=\frac{8}{9} \xScInv \frac{\TBg^{\tau\tau}}{T^{\tau\tau}+\frac{1}{2}T^{k}_{k}}~\tau \partial_{i} \frac{\delta T^{i\tau}}{\TBg^{\tau\tau}} \nonumber \\
\end{eqnarray}
Based on Eq.~(\ref{eq:HydroScalingAssumption}) we obtain the relation
\begin{eqnarray}
\tau \partial_{\tau} \frac{\TBg^{\eta}_{~\eta}}{\TBg^{\tau}_{\tau}} = \Big( \frac{3}{4} - \frac{1}{4} \frac{\TBg^{\eta}_{~\eta}}{\TBg^{\tau}_{\tau}} \Big) \Big(\TBg^{\tau\tau}(\tau)\Big)^{1/4} \tau~f'\Big( \TBg^{\tau\tau}(\tau)^{1/4} \tau \Big)\;, \nonumber \\
\end{eqnarray}
which can be used to re-express the equation of motion as
\begin{align}
&\left( \tau\partial_{\tau} + 
\frac{\TBg^{\tau\tau}}{3\TBg^{\tau\tau}-\TBg^{\eta}_{~\eta}} \Big(\tau 
\partial_{\tau} \frac{\TBg^{\eta}_{~\eta}}{\TBg^{\tau}_{\tau}}\Big) \right) 
\frac{\delta T^{\tau\tau}}{\TBg^{\tau\tau}}\nonumber\\
&\quad + \left(1+\frac{8}{9} \xScInv 
\frac{\TBg^{\tau\tau}}{\TBg^{\tau\tau}+\frac{1}{2}\TBg^{k}_{k}}\right)~\tau 
\partial_{i} \frac{\delta T^{i\tau}}{\TBg^{\tau\tau}}=0\;.
\end{align}
Since the term associated with the momentum flow vanishes in the $k\to0$ limit we can obtain the solution in this case directly by noting that
\begin{eqnarray}
&&\left( \tau\partial_{\tau} + \frac{\TBg^{\tau\tau}}{3\TBg^{\tau\tau}-\TBg^{\eta}_{~\eta}} \Big(\tau \partial_{\tau} \frac{\TBg^{\eta}_{~\eta}}{\TBg^{\tau}_{\tau}}\Big) \right) \frac{\delta T^{\tau\tau}}{\TBg^{\tau\tau}}= \\
&& \qquad \qquad \qquad \frac{3\TBg^{\tau\tau}-\TBg^{\eta}_{~\eta}}{\TBg^{\tau\tau}} \tau \partial_{\tau} \left( \frac{\TBg^{\tau\tau}}{3\TBg^{\tau\tau}-\TBg^{\eta}_{~\eta}} \frac{\delta T^{\tau\tau}}{\TBg^{\tau\tau}} \right) \nonumber 
\end{eqnarray}
is proportional to a total time derivative. Hence the response function 
\begin{eqnarray}
G^{0}_{s}(\tau,\tau_0)= \left.\left.\frac{\delta T^{\tau\tau}(\tau)}{\TBg^{\tau\tau}(\tau)} \right/ \frac{\delta T^{\tau\tau}(\tau_0)}{\TBg^{\tau\tau}(\tau_0)}\right|_{k=0}\;,
\end{eqnarray}
can be found by direct inegration of the equation of motion yielding
\begin{eqnarray}
G^{0}_{s}(\tau,\tau_0)=\frac{3-\frac{\TBg^{\eta}_{~\eta}(\tau)}{\TBg^{\tau\tau}(\tau)}}{3-\frac{\TBg^{\eta}_{~\eta}(\tau_0)}{\TBg^{\tau\tau}(\tau_0)}}\;.
\end{eqnarray}
Neglecting $\TBg^{\eta}_{~\eta}(\tau_0)\approx0$ and using first order 
constitutive equations for $\TBg^{\eta}_{\eta}(\tau)$ the late time behavior 
is given by
\begin{align}
	\tilde{G}_{s}^{0}(\tau,\tau_0)&=\frac{8}{9} \Big(1+ 
	\frac{2}{3}\xScInv\Big)\;.\label{eq:Gs0visc}
\end{align}

Similarly for momentum perturbations the relevant equation of motion takes the form
\begin{eqnarray}
\left( \tau \partial_{\tau} +1 - \frac{\TBg^{\tau\tau}+\TBg^{\eta}_{~\eta}}{\TBg^{\tau\tau}}\right) \frac{\delta T^{\tau j}}{\TBg^{\tau\tau}} + \tau \partial_{i}T^{ij}=0\;, \nonumber \\
\end{eqnarray}
such that in the $k\to0$ limit 
\begin{eqnarray}
G^{0}_{v}(\tau,\tau_0)= \left.\left. \frac{\delta T^{\tau i}(\tau)}{\TBg^{\tau\tau}(\tau)} \right/ \frac{\delta T^{\tau i}(\tau_0)}{\TBg^{\tau\tau}(\tau_0)}\right|_{k=0}\;,
\end{eqnarray}
can be immediately obtained as
\begin{eqnarray}
G^{0}_{v}(\tau,\tau_0)=\left(\frac{\tau_0}{\tau}\right)~\frac{\TBg^{\tau\tau}(\tau_0)}{\TBg^{\tau\tau}(\tau)}
\end{eqnarray}
Beyond the $k=0$ limit the evolution equations are coupled and by use of the first order 
constitutive relations take the following form
\begin{widetext}
\begin{eqnarray}
&&\text{Energy perturbations:} \nonumber \\
&&\left(\tau \partial_{\tau} + 
\frac{\TBg^{\tau\tau}}{3\TBg^{\tau\tau}-\TBg^{\eta}_{~\eta}} \Big(\tau 
\partial_{\tau} \frac{\TBg^{\eta}_{~\eta}}{\TBg^{\tau\tau}}\Big)\right) 
G_s^{s}(\tau,\tau_0,\ktt) + \left(1+\frac{8}{9} \xScInv 
\frac{\TBg^{\tau\tau}}{\TBg^{\tau\tau}+\frac{1}{2}\TBg^{k}_{k}}\right) \ktt 
\tau~G_{s}^{v}(\tau,\tau_0,\ktt) =0\;, \nonumber \\
&&\left( \tau\partial_{\tau} +1 
-\frac{\TBg^{\tau\tau}+\TBg^{\eta}_{~\eta}}{\TBg^{\tau\tau}} \right)  
G_{s}^{v}(\tau,\tau_0,\ktt) - \frac{1}{3} \Big(1+2 \xScInv \Big) 
\ktt\tau~G_{s}^{s}(\tau,\tau_0,\ktt) + \frac{16}{9} \xScInv 
\frac{\TBg^{\tau\tau}}{\TBg^{\tau\tau}+\frac{1}{2}\TBg^{k}_{~k}}~\ktt^2 
\tau^2~G_{s}^{v}(\tau,\tau_0,\ktt) =0\;. \nonumber \\
&&\text{Momentum perturbations:} \nonumber \\
&&\left(\tau \partial_{\tau} + 
\frac{\TBg^{\tau\tau}}{3\TBg^{\tau\tau}-\TBg^{\eta}_{~\eta}} \Big(\tau 
\partial_{\tau} \frac{\TBg^{\eta}_{~\eta}}{\TBg^{\tau\tau}}\Big)\right) 
G_{v}^{s}(\tau,\tau_0,\ktt) - \left(1+\frac{8}{9} \xScInv 
\frac{\TBg^{\tau\tau}}{\TBg^{\tau\tau}+\frac{1}{2}\TBg^{k}_{k}}\right) \ktt 
\tau \Big( G_{v}^{v,\delta}(\tau,\tau_0,\kt) + G_{v}^{v,k}(\tau,\tau_0,\kt) 
\Big)=0\;, \nonumber \\
&& \left( \tau\partial_{\tau} +1 
-\frac{\TBg^{\tau\tau}+\TBg^{\eta}_{~\eta}}{\TBg^{\tau\tau}} \right) 
G_{v}^{v,\delta}(\tau,\tau_0,\kt) + 
\frac{4}{3}\xScInv~\frac{\TBg^{\tau\tau}}{\TBg^{\tau\tau}+\frac{1}{2}\TBg^{k}_{~k}}~\ktt^2
 \tau^2~G_{v}^{v,\delta}(\tau,\tau_0,\kt) =0  \nonumber\\ 
&& \left( \tau\partial_{\tau} +1 
-\frac{\TBg^{\tau\tau}+\TBg^{\eta}_{~\eta}}{\TBg^{\tau\tau}} \right) 
G_{v}^{v,k}(\tau,\tau_0,\kt) + \frac{1}{3} \Big( 1+ 2 \xScInv\Big)~\ktt 
\tau~G_{v}^{s}(\tau,\tau_0,\kt)  \nonumber \\
&& \qquad \qquad + \frac{4}{9}  
\xScInv~\frac{\TBg^{\tau\tau}}{\TBg^{\tau\tau}+\frac{1}{2}\TBg^{k}_{~k}}~\ktt^2 
\tau^2 \Big( G_{v}^{v,\delta}(\tau,\tau_0,\kt) + 4 
G_{v}^{v,k}(\tau,\tau_0,\kt)  \Big) =0 \nonumber\\ 
\end{eqnarray}
\end{widetext}
where we projected the equation of motions for the momentum perturbations in 
the directions parallel and orthogonal to the momentum. Based on the knowledge 
of the explicit solution at $\ktt=0$, we can then attempt to construct the 
solution at small $\ktt$ in terms of a series expansion

\begin{eqnarray}
&&\text{Energy perturbations:} \nonumber \\
&& G_s^{s}(\tau,\tau_0,\ktt)=G^{0}_{s}(\tau,\tau_0)~\Big( 1 - \frac{1}{2} 
\ktt^2 \tau^2 \tilde{s}^{(2)}_{s} + \cdots \Big)\;, \nonumber \\
&& G_s^{v}(\tau,\tau_0,\ktt)=G^{0}_{s}(\tau,\tau_0)~\Big( \ktt \tau 
\tilde{s}^{(1)}_{v}+ \cdots \Big)\;,  \\
&&\text{Momentum perturbations:} \nonumber \\
&& G_{v}^{s}(\tau,\tau_0,\ktt)=G^{0}_{v}(\tau,\tau_0)~\Big( \ktt \tau 
\tilde{v}^{(1)}_{s}+ \cdots \Big) \nonumber \\
&& G_{v}^{v,\delta}(\tau,\tau_0,\ktt)=G^{0}_{v}(\tau,\tau_0)~\Big( 1 - 
\frac{1}{2} \ktt^2 \tau^2 \tilde{v}^{(2)}_{v,\delta}+ \cdots \Big) \nonumber \\
&& G_{v}^{v,k}(\tau,\tau_0,\ktt)=G^{0}_{v}(\tau,\tau_0)~\Big(  - \frac{1}{2} 
\ktt^2 \tau^2 \tilde{v}^{(2)}_{v,k}+ \cdots \Big) \nonumber \\
\end{eqnarray}
By inserting this ansatz into the evolution equations, and using the fact that $G^{0}_{s}(\tau,\tau_0)$ is an explicit solution of the equation of motion at $\ktt=0$, we obtain evolution equations for the coefficient functions $s^{(n)}_{s/v}(\tau,\tau_0)$ at each order in $\ktt \tau$. By keeping only the lowest order terms in $\ktt\tau$  we obtain the following set of evolution equations
\begin{eqnarray}
&&\text{Energy perturbations:} \nonumber \\
&&\frac{1}{2}\tau\partial_{\tau} \tilde{s}^{(2)}_{s} + \tilde{s}^{(2)}_{s}=\left(1+\frac{8}{9} \xScInv \frac{\TBg^{\tau\tau}}{\TBg^{\tau\tau}+\frac{1}{2}\TBg^{k}_{k}}\right) \tilde{s}^{(1)}_{v}\;, \\
&&\tau\partial_{\tau}\tilde{s}^{(1)}_{v} -  \frac{1}{3} \Big(1+2 \xScInv \Big)=  \\
&& \qquad -\left( 2 -\frac{\TBg^{\tau\tau}+\TBg^{\eta}_{~\eta}}{\TBg^{\tau\tau}} -  \frac{\TBg^{\tau\tau}}{3\TBg^{\tau\tau}-\TBg^{\eta}_{~\eta}} \Big(\tau \partial_{\tau} \frac{\TBg^{\eta}_{~\eta}}{\TBg^{\tau\tau}} \Big) \right)~\tilde{s}^{(1)}_{v}\;, \nonumber \\
&&\text{Momentum perturbations:} \nonumber \\
&&\tau\partial_{\tau}\tilde{v}^{(1)}_{s}-\left(1+\frac{8}{9} \xScInv \frac{\TBg^{\tau\tau}}{\TBg^{\tau\tau}+\frac{1}{2}\TBg^{k}_{k}}\right)= \\
&&\qquad -\left(\frac{\TBg^{\tau\tau}+\TBg^{\eta}_{~\eta}}{\TBg^{\tau\tau}} + \frac{\TBg^{\tau\tau}}{3\TBg^{\tau\tau}-\TBg^{\eta}_{~\eta}} \Big(\tau \partial_{\tau} \frac{\TBg^{\eta}_{~\eta}}{\TBg^{\tau\tau}}\Big)\right)\tilde{v}^{(1)}_{s}\;, \nonumber \\
&&\frac{1}{2}\tau\partial_{\tau}\tilde{v}^{(2)}_{v,\delta}+\tilde{v}^{(2)}_{v,\delta}=\frac{4}{3}\xScInv~\frac{\TBg^{\tau\tau}}{\TBg^{\tau\tau}+\frac{1}{2}\TBg^{k}_{~k}}\;, \\
&&\frac{1}{2}\tau\partial_{\tau}\tilde{v}^{(2)}_{v,k}+\tilde{v}^{(2)}_{v,k}=\frac{1}{3} \Big(1+2 \xScInv \Big) v_{s}^{(1)} + \frac{4}{9} \xScInv \frac{\TBg^{\tau\tau}}{\TBg^{\tau\tau}+\frac{1}{2}\TBg^{k}_{~k}}\;. \nonumber \\
\end{eqnarray}
Evaluating the background coefficients to lowest order in the viscous corrections $\xScInv$, we have
\begin{eqnarray}
\frac{\TBg^{\tau\tau}+\TBg^{\eta}_{~\eta}}{\TBg^{\tau\tau}}\simeq\frac{4}{3}-\frac{16}{9} \xScInv\;, \nonumber \\ 
\frac{\TBg^{\tau\tau}}{3\TBg^{\tau\tau}-\TBg^{\eta}_{~\eta}} \Big(\tau \partial_{\tau} \frac{\TBg^{\eta}_{~\eta}}{\TBg^{\tau\tau}}\Big)\simeq\frac{4}{9} \xScInv\;, \nonumber \\ 
\xScInv \frac{\TBg^{\tau\tau}}{\TBg^{\tau\tau}+\frac{1}{2}\TBg^{k}_{k}} \simeq 
\frac{3}{4} \xScInv\;.
\end{eqnarray}
such that the evolution equations can be re-written as
\begin{eqnarray}
&&\text{Energy perturbations:} \nonumber \\
&&\frac{1}{2}\tau\partial_{\tau} \tilde{s}^{(2)}_{s} + 
\tilde{s}^{(2)}_{s}=\left(1+\frac{2}{3} \xScInv\right) \tilde{s}^{(1)}_{v}\;, 
\nonumber\\
&&\tau\partial_{\tau}\tilde{s}^{(1)}_{v}+  \left( \frac{2}{3} + \frac{4}{3} 
\xScInv \right)~\tilde{s}^{(1)}_{v}= \Big(\frac{1}{3}+\frac{2}{3} \xScInv 
\Big)\;,  \nonumber  \\
&&\text{Momentum perturbations:} \label{eq:ds2} \\
&&\tau\partial_{\tau}\tilde{v}^{(1)}_{s}+\left(\frac{4}{3}-\frac{4}{3}  \xScInv 
\right)\tilde{v}^{(1)}_{s}=\left(1+\frac{2}{3} \xScInv \right)\;,\nonumber \\
&&\frac{1}{2}\tau\partial_{\tau}\tilde{v}^{(2)}_{v,\delta}+\tilde{v}^{(2)}_{v,\delta}=\xScInv\;,
 \nonumber\\
&&\frac{1}{2}\tau\partial_{\tau}\tilde{v}^{(2)}_{v,k}+\tilde{v}^{(2)}_{v,k}=\Big(\frac{1}{3}
 +\frac{2}{3}  \xScInv \Big) v_{s}^{(1)} + \frac{1}{3} \xScInv\;.\nonumber\\
\end{eqnarray}

To specify the initial conditions we use the late time expansion of free 
streaming given by \Eqs{eq:freestrhighorder} and \Eq{eq:freestrhighordermom}. Namely
\begin{eqnarray}
	&& \qquad\tilde{s}^{(2)}_{s}(\tau_0)=\frac{1}{2} \;,  \quad
	\tilde{s}^{(1)}_{v}(\tau_0)=\frac{1}{2} \;,   \\
	&&\tilde{v}^{(1)}_{s}(\tau_0)=1\;, \quad
	\tilde{v}^{(2)}_{v,\delta}(\tau_0)=\frac{1}{4}\;,  \quad
	\tilde{v}^{(2)}_{v,k}(\tau_0)=\frac{1}{2}\;.  
\end{eqnarray}
noting that $\tilde{s}^{(1)}_{v}(\tau_0)=\frac{1}{2}$ is a stationary solution 
of 
\Eq{eq:ds2}, the leading order solution in viscous correction is
\begin{eqnarray}
	&&\text{Energy perturbations:} \nonumber \\
	&&\tilde{s}^{(2)}_{s}=\frac{1}{2} + \frac{1}{2}\xScInv \left[1- 
	\Big(\frac{\tau_0}{\tau}\Big)^{4/3} \right]\;, \\
	&&\tilde{s}^{(1)}_{v}=\frac{1}{2} \\
	&&\text{Momentum perturbations:} \nonumber \\
	&&\tilde{v}^{(1)}_{s}=\frac{3}{4} \left[1+ \frac{1}{3}
	\Big(\frac{\tau_0}{\tau}\Big)^{4/3} \right] \\
	&& \qquad \qquad  + \frac{5}{2} \xScInv \left[ 1 - \frac{1}{5} 
	\Big(\frac{\tau_0}{\tau}\Big)^{2/3} 
	-\frac{4}{5}\Big(\frac{\tau_0}{\tau}\Big)^{4/3} \right]\;, \nonumber \\
	&&\tilde{v}^{(2)}_{v,\delta}=\frac{1}{4}\Big(\frac{\tau_0}{\tau}\Big)^{2}+\frac{3}{2}
	 \xScInv \left[1- 
	\Big(\frac{\tau_0}{\tau}\Big)^{4/3} \right]\;, \\
	&&\tilde{v}^{(2)}_{v,k}=\frac{1}{4}\left[1+
	\Big(\frac{\tau_0}{\tau}\Big)^{4/3}\right] \\
	&& \qquad \qquad + \frac{5}{2}   \xScInv    \left[ 1 - \frac{4}{5} 
	\Big(\frac{\tau_0}{\tau}\Big)^{2/3} - 
	\frac{1}{5}\Big(\frac{\tau_0}{\tau}\Big)^{4/3} \right]\;, \nonumber
\end{eqnarray}
such that in the limit $\tau \gg \tau_0$ one has
\begin{eqnarray}
	&& \qquad\tilde{s}^{(2)}_{s}=\frac{1}{2}+ \frac{1}{2}\xScInv \;,  \quad
	\tilde{s}^{(1)}_{v}=\frac{1}{2}  \;,   \\
	&&\tilde{v}^{(1)}_{s}=\frac{3}{4} + \frac{5}{2} \xScInv\;, \quad
	\tilde{v}^{(2)}_{v,\delta}=\frac{3}{2} \xScInv\;,  \quad
	\tilde{v}^{(2)}_{v,k}=\frac{1}{4}+ \frac{5}{2}   \xScInv\;.    \nonumber 
\end{eqnarray}
which is the main result.

Note that for initial energy perturbations the leading Taylor expansion terms in energy and momentum response agrees between free streaming and hydrodynamic response. However, at higher orders
in $\ktt(\tau-\tau_0)$ this agreements disappears, e.g.\ at cubic order one finds
\begin{align}
		\text{ideal hydro} \quad
		\tilde 
		G_s^v&=\frac{1}{2}\ktt(\tau-\tau_0)-\frac{1}{32}(\ktt(\tau-\tau_0))^3\nonumber\\
	\text{free streaming}\quad	
	\tilde G_s^v&=\frac{1}{2}\ktt(\tau-\tau_0)-\frac{1}{16}(\ktt(\tau-\tau_0))^3\label{eq:cubic}
\end{align}
Consequently we find that in practice the extracted coefficients from kinetic evolution (see \Fig{fig:coefde}(a)) remain approximately constant for initial energy perturbations, while for initial momentum perturbations (see \Fig{fig:coefde}(b)) the approach to the hydrodynamic limit is well described by viscous asymptotics.

Once the response to conserved quantities is calculated, the shear response, i.e. $\delta T^{ij}$ is completely determined by the 
constitutive relations. So for the low $\ktt$ expansion
\begin{eqnarray}
	G_{s}^{t,\delta}(\tau,\tau_0,\ktt)&=&\tilde{G}_{s}^{0}(\tau,\tau_0)\Big(\tilde{s}^{(0)}_{t,\delta}-\frac{1}{2}\ktt^2(\tau-\tau_0)^2\tilde{s}^{(2)}_{t,\delta}+
	\cdots \Big)\;, \nonumber \\
	G_{s}^{t,k}(\tau,\tau_0,\ktt)&=&\tilde{G}_{s}^{0}(\tau,\tau_0)\Big(-\frac{1}{2}\ktt^2(\tau-\tau_0)^2\tilde{s}^{(2)}_{t,k}+
	\cdots \Big)\;, \nonumber \\
\end{eqnarray}
we get the response coefficients
\begin{align}
	\tilde{s}^{(0)}_{t,\delta}&=\frac{1}{3}\Big(1+ 2 \xScInv 
	\Big)\;, \nonumber\\
	\tilde{s}^{(2)}_{t,\delta}&=\frac{1}{3}\Big(1+2 \xScInv 
	\Big)~\tilde{s}^{(2)}_{s} -\frac{4}{3}\xScInv~\tilde{s}^{(1)}_{v} \;, 
	\nonumber 
	\\
	\tilde{s}^{(2)}_{t,k}&=4~\xScInv~\tilde{s}^{(1)}_{v}\;, 
\end{align}
and similarly for momentum perturbations
\begin{eqnarray}
G_{v}^{t,\delta/m/k}(\tau,\tau_0,\ktt)&=&\tilde{G}_{v}^{0}(\tau,\tau_0)\Big(\ktt
 (\tau-\tau_0)~\tilde{v}^{(1)}_{t,\delta/m/k} + \cdots \Big)\;, \nonumber \\
\end{eqnarray}
with the response coefficients
\begin{eqnarray}
&&\tilde{v}^{(1)}_{t,\delta}=\frac{1}{3}\tilde{v}^{(1)}_{s}\Big(1+2 
\xScInv\Big)-\frac{2}{3} \xScInv \;,  \nonumber \\
&& \qquad \tilde{v}^{(1)}_{t,m}=2 \xScInv\;, \qquad \tilde{v}^{(1)}_{t,k}=0\;. 
\end{eqnarray}

One important feature of the hydrodynamic Green functions is the fact that 
the solutions can be expressed entirely in terms of the two scaling variables 
$\TId\tau/(\eta/s)$ and $\ktt (\tau-\tau_0)$. The scaling of hydrodynamic response functions 
motivates the use of scaled variables for the non-equilibrium 
response functions, see \Sec{sec:response}.

\section{Derivation of long-wavelength limit of kinetic theory response\label{app:lowk}}
Below we provide the details of the derivation of the  long-wavelength limit of kinetic theory response, including the response in coordinate space as well as some additional discussion on the regulator dependence.
\subsection{Coordinate space implementation}
Starting point of our discussion is \Eq{eq:EvolutionLowKmain}, which we re-write as
\begin{eqnarray}
	\label{eq:EvolutionLowKapp}
	&&\bar{T}^{\tau\tau}_\xt(\tau_0)\frac{\delta T^{\mu\nu}(\tau,\xt)}{\bar{T}^{\tau\tau}_{\xt}(\tau)}=\\
	&& \qquad \int d^2\xt_0 G^{\mu\nu}_{\alpha\beta}\Big(\tau,\tau_0,\xt-\xt_0\Big)~(\bar{T}^{\alpha\beta}(\tau_0,\xt_0)-\bar{T}_\xt^{\alpha\beta}(\tau_0))\;, \nonumber
\end{eqnarray}
expressing the energy-momentum response at space-time point $(\tau,\xt)$ on the left hand side, in terms of the long wave-length components,
\begin{eqnarray}
\delta \mathcal{T}^{\alpha\beta}_{\xt}(\tau_0,\xt_0)=\bar{T}^{\alpha\beta}(\tau_0,\xt_0)-\bar{T}_\xt^{\alpha\beta}(\tau_0)\label{eq:curlyT}
\end{eqnarray}
where according to \Eq{eq:smoothTmunu} 
\begin{align}
\label{eq:dTlowKDefApp}
\bar{T}^{\mu\nu}(\tau_0,\x_0)= \int d^2\xt'~S_{\sigma}(\xt_0-\xt'_0)~T^{\mu\nu}(\tau_0,\xt_0')\;, 
\end{align}
denotes the coarse-grained energy momentum tensor, and $\bar{T}_\xt^{\mu\nu}(\tau_0)$ denotes the local average background. 

Explicitly the background $\bar{T}_\xt^{\mu\nu}(\tau_0)$ is constructed from the local average background energy $e_{\xt}(\tau_0)=\bar{T}^{\tau\tau}(\tau_0,\x)$ around the space-time point $(\tau_0,\xt)$ and defined to be diagonal 
\begin{align}
\bar{T}_\xt^{\mu\nu}(\tau_0)=\text{diag}\,\Big(e_{\xt}(\tau_0),P^{T}_{\xt}(\tau_0),P^{T}_{\xt}(\tau_0),P^{L}_{\xt}(\tau_0)/\tau_0^2\Big)\;. 
\end{align}
Since the background is taken to be diagonal, all off-diagonal components, including e.g. the initial momentum flow $\bar{T}^{\tau i}(\tau_0,\x_0)$, are treated explicitly as perturbations. Similarly the transverse and longitudinal pressure $P^{T/L}_{\xt}(\tau_0)$ in this expression correspond to values determined from the ``non-equilibrium constitutive relations'' determined by the universal background scaling curve (see \Sec{subsec:background}), noting that possible deviations such as e.g. $\delta T^{\eta\eta}_{\xt}(\tau_0,\xt_0)=P^{L}_{\xt}(\tau_0)-\bar{T}^{\eta\eta}(\tau_0,\xt_0)$, could be treated as tensor perturbations (which we neglect throughout this work).

Expressing the Greens functions on the right hand side of Eq.~(\ref{eq:EvolutionLowKapp}) in Fourier ($\kt$) space as
\begin{eqnarray}
	&&\bar{T}^{\tau\tau}_\xt(\tau_0)\frac{\delta T^{\mu\nu}(\tau,\xt)}{\bar{T}^{\tau\tau}_{\xt}(\tau)}=\\
	&&\int d^2\xt_0 \int \frac{d^2\kt}{(2\pi)^2} \tilde{G}^{\mu\nu}_{\alpha\beta}\Big(\tau,\tau_0,\kt \Big)~e^{i\kt(\xt-\xt_0)}~\delta \mathcal{T}^{\alpha\beta}_{\xt}(\tau_0,\xt_0)\;. \nonumber
\end{eqnarray}
we can then take advantage of the usual tensor decomposition in \Eqs{eq:G_energy_decomp} and \eq{eq:G_momentum_decomp} and perform the low $\kt$ expansion of the response functions as in \Eq{eq:lowkcoef}. By focusing on the $T^{\tau\tau}$ and $T^{\tau i}$ components of the response, we obtain to following expressions
\begin{eqnarray}
\label{eq:lowKresponseDerivation}
&&\text{Energy perturbations:} \nonumber \\
&&\bar{T}^{\tau\tau}_\xt(\tau_0)\frac{\delta T^{\tau\tau}(\tau,\xt)}{\bar{T}^{\tau\tau}_{\xt}(\tau)} \approx \int d^2\xt_0 \int \frac{d^2\kt}{(2\pi)^2}\tilde G_{s}^{0}\Big(\tau,\tau_0,\bar{T}^{\tau\tau}_\xt(\tau_0)\Big) \nonumber \\
&& \qquad \Big( 1- \frac{\ktt^2}{2}  (\tau-\tau_0)^2 \tilde{s}^{(2)}_{s}\Big)~e^{i\kt(\xt-\xt_0)}~\delta \mathcal{T}^{\tau\tau}_{\xt}(\tau_0,\xt_0)\;, \nonumber \\
&&\bar{T}^{\tau\tau}_\xt(\tau_0)\frac{\delta T^{\tau i}(\tau,\xt)}{\bar{T}^{\tau\tau}_{\xt}(\tau)} \approx \int d^2\xt_0 \int \frac{d^2\kt}{(2\pi)^2}\tilde G_{s}^{0}\Big(\tau,\tau_0,\bar{T}^{\tau\tau}_\xt(\tau_0)\Big) \nonumber \\
&& \qquad \qquad \Big( -i \kt^{i} (\tau-\tau_0) \tilde{s}^{(1)}_{v} \Big)~e^{i\kt(\xt-\xt_0)}~\delta \mathcal{T}^{\tau\tau}_{\xt}(\tau_0,\xt_0)\;, \nonumber \\
&&\text{Momentum perturbations:} \nonumber \\
&&\bar{T}^{\tau\tau}_\xt(\tau_0)\frac{\delta T^{\tau\tau}(\tau,\xt)}{\bar{T}^{\tau\tau}_{\xt}(\tau)} \approx \int d^2\xt_0 \int \frac{d^2\kt}{(2\pi)^2}\tilde G_{v}^{0}\Big(\tau,\tau_0,\bar{T}^{\tau\tau}_\xt(\tau_0)\Big) \nonumber \\
&& \qquad \qquad \Big( -i \kt_{i} (\tau-\tau_0) \tilde{v}^{(1)}_{s}\Big)~e^{i\kt(\xt-\xt_0)}~\delta \mathcal{T}^{\tau i}_{\xt}(\tau_0,\xt_0)\;, \nonumber \\
&&\bar{T}^{\tau\tau}_\xt(\tau_0)\frac{\delta T^{\tau i}(\tau,\xt)}{\bar{T}^{\tau\tau}_{\xt}(\tau)} \approx \int d^2\xt_0 \int \frac{d^2\kt}{(2\pi)^2}\tilde G_{v}^{0}\Big(\tau,\tau_0,\bar{T}^{\tau\tau}_\xt(\tau_0)\Big) \nonumber \\
&& \Big( \delta^{i}_j - \frac{\kt^2}{2}  (\tau-\tau_0)^2 [ \delta^{i}_j \tilde{v}^{(2)}_{\delta} + \frac{\kt^{i}\kt_{j}}{\kt^2}  \tilde{v}^{(2)}_{k}  ]    \Big)~e^{i\kt(\xt-\xt_0)}~\delta \mathcal{T}^{\tau j}_{\xt}(\tau_0,\xt_0)\;. \nonumber \\
\end{eqnarray}
where we also spelled out explicitly the dependence of the response functions $G_{s/v}^{0}$ on the the background energy density, noting that the coefficients $\tilde{s}^{(n)}/\tilde{v}^{(n)}$ also depend on the same arguments. By expressing powers $\kt^{i}$ as derivatives acting on the exponential, i.e. $\kt^{i} e^{i\kt(\xt-\xt_0)} \to +i \partial^{i}_{\xt_0} e^{i\kt(\xt-\xt_0)}$, we can then perform the integrals via integration by parts w.r.t. $\xt_{0}$ with vanishing boundary terms to obtain
\begin{eqnarray}
&&\text{Energy perturbations:} \nonumber \\
&&\bar{T}^{\tau\tau}_\xt(\tau_0)\frac{\delta T^{\tau\tau}(\tau,\xt)}{\bar{T}^{\tau\tau}_{\xt}(\tau)} \approx \tilde G_{s}^{0}\Big(\tau,\tau_0,\bar{T}^{\tau\tau}_\xt(\tau_0)\Big) \nonumber \\
&& \qquad \left. \Big( 1+ \frac{(\tau-\tau_0)^2}{2} \tilde{s}^{(2)}_{s} \partial^{k}_{\xt_0}(\partial_{\xt_0})_k  \Big)~\delta \mathcal{T}^{\tau\tau}_{\xt}(\tau_0,\xt_0)\right|_{\xt_0=\xt}\;, \nonumber \\
&&\bar{T}^{\tau\tau}_\xt(\tau_0)\frac{\delta T^{\tau i}(\tau,\xt)}{\bar{T}^{\tau\tau}_{\xt}(\tau)} \approx\tilde G_{s}^{0}\Big(\tau,\tau_0,\bar{T}^{\tau\tau}_\xt(\tau_0)\Big) \nonumber \\
&& \qquad \qquad \left. \Big( -(\tau-\tau_0) \tilde{s}^{(1)}_{v} \partial^{i}_{\xt_0} \Big)~\delta \mathcal{T}^{\tau\tau}_{\xt}(\tau_0,\xt_0) \right|_{\xt_0=\xt}\;, \nonumber \\
&&\text{Momentum perturbations:} \nonumber \\
&&\bar{T}^{\tau\tau}_\xt(\tau_0)\frac{\delta T^{\tau\tau}(\tau,\xt)}{\bar{T}^{\tau\tau}_{\xt}(\tau)} \approx \tilde G_{v}^{0}\Big(\tau,\tau_0,\bar{T}^{\tau\tau}_\xt(\tau_0)\Big) \nonumber \\
&& \qquad \qquad\left. \Big( - (\tau-\tau_0) \tilde{v}^{(1)}_{s} (\partial_{\xt_0})_k \Big)~\delta \mathcal{T}^{\tau k}_{\xt}(\tau_0,\xt_0)  \right|_{\xt_0=\xt}\;, \nonumber \\
&&\bar{T}^{\tau\tau}_\xt(\tau_0)\frac{\delta T^{\tau i}(\tau,\xt)}{\bar{T}^{\tau\tau}_{\xt}(\tau)} \approx \tilde G_{v}^{0}\Big(\tau,\tau_0,\bar{T}^{\tau\tau}_\xt(\tau_0)\Big) \nonumber \\
&&  \Big( \delta^{i}_j + \frac{(\tau-\tau_0)^2}{2}   [\tilde{v}^{(2)}_{\delta} \delta^{i}_j \partial^{k}_{\xt_0}(\partial_{\xt_0})_k  +  \tilde{v}^{(2)}_{k} \partial^{i}_{\xt_0}(\partial_{\xt_0} )_j  ]    \Big)\nonumber\\
&&\hspace{4.5cm} \left. \delta \mathcal{T}^{\tau j}_{\xt}(\tau_0,\xt_0)  \right|_{\xt_0=\xt} \;. 
\end{eqnarray}
Evaluating the derivatives $\partial^{k}_{\xt_0}$ of the long-wavelength perturbations $\delta \mathcal{T}^{\tau\tau}_{\xt}(\tau_0,\xt_0)$ based on Eq.~(\ref {eq:dTlowKDefApp}), noting that according to our definition of the background $\left.\delta \mathcal{T}^{\tau\tau}_{\xt}(\tau_0,\xt_0)\right|_{\xt=\xt_0}=0$ whereas $\left.\delta \mathcal{T}^{\tau i}_{\xt}(\tau_0,\xt_0)\right|_{\xt=\xt_0}\neq 0$, we can then express the long wave-length response in terms of derivatives of the coarse-grained energy momentum tensor 
\begin{eqnarray}
&&\text{Energy perturbations:} \nonumber \\
&&\frac{\delta T^{\tau\tau}(\tau,\xt)}{\bar{T}^{\tau\tau}_{\xt}(\tau)} \approx \frac{\tilde{G}_{s}^{0}\Big(\tau,\tau_0,\bar{T}^{\tau\tau}_\xt(\tau_0)\Big)}{\bar{T}^{\tau\tau}_\xt(\tau_0)}  \nonumber \\
&& \qquad \qquad \Big[\frac{(\tau-\tau_0)^2}{2} \tilde{s}^{(2)}_{s} \partial^{k}\partial_{k}  \Big]~\bar{T}^{\tau\tau}(\tau_0,\xt)\;, \nonumber \\
&&\frac{\delta T^{\tau i}(\tau,\xt)}{\bar{T}^{\tau\tau}_{\xt}(\tau)} \approx \frac{\tilde{G}_{s}^{0}\Big(\tau,\tau_0,\bar{T}^{\tau\tau}_\xt(\tau_0)\Big)}{\bar{T}^{\tau\tau}_\xt(\tau_0)} \nonumber \\
&& \qquad \qquad \Big[ -(\tau-\tau_0) \tilde{s}^{(1)}_{v} \partial^{i} \Big]~\bar{T}^{\tau\tau}(\tau_0,\xt)\;, \nonumber \\
&&\text{Momentum perturbations:} \nonumber \\
&&\frac{\delta T^{\tau\tau}(\tau,\xt)}{\bar{T}^{\tau\tau}_{\xt}(\tau)} \approx \frac{\tilde{G}_{v}^{0}\Big(\tau,\tau_0,\bar{T}^{\tau\tau}_\xt(\tau_0)\Big)}{\bar{T}^{\tau\tau}_\xt(\tau_0)} \nonumber \\
&& \qquad \qquad\Big[ - (\tau-\tau_0) \tilde{v}^{(1)}_{s} \partial_{i} \Big]~\bar{T}^{\tau i}(\tau_0,\xt) \;, \nonumber \\
&&\frac{\delta T^{\tau i}(\tau,\xt)}{\bar{T}^{\tau\tau}_{\xt}(\tau)} \approx\frac{\tilde{G}_{v}^{0}\Big(\tau,\tau_0,\bar{T}^{\tau\tau}_\xt(\tau_0)\Big)}{\bar{T}^{\tau\tau}_\xt(\tau_0)} \nonumber \\
&&  \Big[ \delta^{i}_j + \frac{(\tau-\tau_0)^2}{2}   [\tilde{v}^{(2)}_{\delta} \delta^{i}_j \partial^{k}\partial_{k}  +  \tilde{v}^{(2)}_{k} \partial^{i}\partial_{j}  ]    \Big]~\bar{T}^{\tau j}(\tau_0,\xt) \;. \nonumber \\
\end{eqnarray}
which is the result given in the main text. We note that corresponding expressions for long wave-length response of the shear-stress ($T^{ij}$) components can be derived in a similar fashion. However, since in practice the shear-stress tensor at hydrodynamization time is typically close to the Navier-Stokes value, the result is of limited practical use and we therefore refrain from providing explicit expressions.
 
 \subsection{Regulator dependence and renormalization\label{app:regsigma}}
We will now discuss in more detail how the long wavelength response depends on the choice of the regulator ${S}_\sigma(\x_0-\x_0')$ in \Eq{eq:smoothTmunu}, and how part of this dependence can be absorbed into the renormalization of the long wave-length coefficients. The background energy momentum tensor in $\kt$ space is given by
\begin{eqnarray}
\bar{T}^{\tau\tau(\sigma)}_{\kt}(\tau_0)=\int d^2\xt_0~\bar{T}^{\tau\tau(\sigma)}_{\xt_0}(\tau_0)~e^{-i\kt\xt_0}=\tilde{S}_{\sigma}(\kt) T^{\tau\tau}(\tau_0,\kt)\;, \nonumber \\
\end{eqnarray}
where $T^{\tau\tau}(\tau_0,\kt)$ denotes the Fourier transform of the original energy momentum tensor and $\tilde{S}_{\sigma}(\kt)$ is the Fourier transform of the smearing kernel.
 Then the long-wavelength component of the  initial energy $\delta T^{\tau\tau}$ perturbations in \Eq{eq:curlyT}) can be expressed as
\begin{align}
\label{eq:jhgf}
\delta \mathcal{T}^{\tau\tau(\sigma)}_{\xt}(\tau_0,\xt_0)=\int \frac{d^2\kt}{(2\pi)^2}~\tilde{S}_{\sigma}(\kt) \Big(e^{i\kt \xt_0} - e^{i\kt \xt}\Big)~T^{\tau\tau}(\tau_0,\kt)\;,
\end{align}
which makes the regulator $(\sigma)$ dependence explicit. 
Changing $\sigma$ also alters the smooth background, \Eq{eq:avgBG}, around which the linearization of perturbations is performed, affecting the reconstruction of the energy momentum tensor at later times. Since we anticipate this difference of the background to be small and dominated by long wave-length modes, will can treat it as an additional source of long wave-length fluctuations
\begin{eqnarray}
\label{eq:qwgrh}
\delta \mathcal{T}^{\tau\tau(\sigma)}_{\xt,(BG)}(\tau_0,\xt_0)&=&\int \frac{d^2\kt}{(2\pi)^2}~\Big(\bar{T}^{\tau\tau(\sigma)}_{\kt}(\tau_0)-T^{\tau\tau}(\tau_0,\kt)\Big) e^{i \kt\xt}\;, \nonumber \\
&=&\int \frac{d^2\kt}{(2\pi)^2}\Big(\tilde{S}_{\sigma}(\kt)-1\Big) T^{\tau\tau}(\tau_0,\kt)~e^{i \kt\xt}\;, \nonumber \\
\end{eqnarray}
Collecting both contributions in Eq.(\ref{eq:jhgf}) and (\ref{eq:qwgrh}) and inserting the expression into one of the response formulae in Eq.~(\ref{eq:lowKresponseDerivation}), we schematically obtain expressions of the form\footnote{Note that in the following discussion, we will neglect for simplicity the (weak) dependence of the response functions on the background energy scale $\bar{T}^{\alpha\beta}_{\xt}(\tau_0)$.}
\begin{eqnarray}
&& \bar{T}^{\tau\tau}(\tau_0)\frac{\delta T^{\alpha\beta(\sigma)}(\tau,\xt)}{\bar{T}^{\tau\tau}(\tau)}  =\nonumber \\
&& \qquad \int d^2\xt_0 \int \frac{d^2\kt}{(2\pi)^2}\tilde G_{s}^{0}(\tau,\tau_0)~e^{i\kt(\xt-\xt_0)} \nonumber \\
&& \qquad \Big[ \tilde{s}^{(0)}_{\alpha\beta}-i \kt^{i}(\tau-\tau_0) \tilde{s}^{(1),i}_{\alpha\beta}- \frac{\kt^{i}\kt^{j}}{2}  (\tau-\tau_0)^2 s^{(2),ij}_{\alpha\beta}+\ldots\Big] \nonumber \\
&& \qquad \int \frac{d^2\kt'}{(2\pi)^2}~\Big(\tilde{S}_{\sigma}(\kt') e^{i\kt' \xt_0} - e^{i\kt' \xt}\Big)~T^{\tau\tau}(\tau_0,\kt') \nonumber \\
\end{eqnarray}
where the long wave-length coefficients $s^{(n),i_1 \cdots i_n}_{\alpha\beta}$ characterize the response of the different components. By performing the $\xt_0$ integral the expression simplifies to
\begin{eqnarray}
&&\bar{T}^{\tau\tau}(\tau_0)\frac{\delta T^{\alpha\beta(\sigma)}(\tau,\xt)}{\bar{T}^{\tau\tau}(\tau)} = \int \frac{d^2\kt}{(2\pi)^2}\tilde G_{s}^{0}(\tau,\tau_0)~e^{i\kt\xt}~T^{\tau\tau}(\tau_0,\kt)  \nonumber \\
&& \qquad \Big[ \tilde{s}^{(0)}_{\alpha\beta} (\tilde{S}_{\sigma}(\kt)-1) -i \kt^{i}(\tau-\tau_0) \tilde{s}^{(1),i}_{\alpha\beta} \tilde{S}_{\sigma}(\kt) \nonumber \\
&& \qquad \qquad \qquad - \frac{\kt^{i}\kt^{j}}{2}  (\tau-\tau_0)^2 \tilde{s}^{(2),ij}_{\alpha\beta} \tilde{S}_{\sigma}(\kt)+\ldots\Big] \;,
\end{eqnarray}
which can be used to compare the long wave-length response for different regulator choices $\sigma$. Specializing on the Gaussian regulator $\tilde{S}_{\sigma}(\kt)=e^{-\frac{1}{2}\sigma^2 \kt^2}$, and expanding the regulator into a Taylor series, one finds that the difference for two different regulators $\sigma,\sigma'$ is given by
\begin{eqnarray}
\label{eq:reg-dependence-app}
&&\bar{T}^{\tau\tau}_\xt(\tau_0)\frac{\delta T^{\alpha\beta(\sigma)}(\tau,\xt)-\delta T^{\alpha\beta(\sigma')}(\tau,\xt)}{\bar{T}^{\tau\tau}_{\xt}(\tau)} = \nonumber \\
&& \qquad \int \frac{d^2\kt}{(2\pi)^2}\tilde G_{s}^{0}(\tau,\tau_0)~e^{i\kt\xt}~T^{\tau\tau}(\tau_0,\kt)  \nonumber \\
&& \qquad \qquad\Big[   -\tilde{s}^{(0)}_{\alpha\beta} \frac{\kt^{i}\kt^{j}}{2} (\sigma^2-\sigma'^2) \delta_{ij}  + \cdots\Big] \nonumber \\
\end{eqnarray}
where $\cdots$ denote terms of $\mathcal{O}(\kt^3)$ and higher the $\kt$ expansion. It is important to note that the leading term in Eq.~(\ref{eq:reg-dependence-app}), is of quadratic order indicating that the constant and linear response (i.e. terms proportional to $T^{\mu\nu}$ and its first order derivatives $\partial_{i} T^{\mu\nu}$ in the coordinate space response formula) are independent of the regulator choice. While naively one may expect that the leading difference appears at order $\kt^2$, it turns out that the dependence at this order can in fact be absorbed into the re-normalization of the quadratic response coefficients, i.e. by redefining regulator dependent response coefficient
\begin{eqnarray}
\tilde{s}^{(2),ij }_{\alpha\beta}(\sigma) = \tilde{s}^{(2),ij}_{\alpha\beta}(\sigma') -\tilde{s}^{(0)}_{\alpha\beta}\delta^{ij}\frac{\sigma^2-\sigma'^2}{(\tau-\tau_0)^2}\;.
\end{eqnarray}
However at $\mathcal{O}(\kt^3)$ this is no longer possible, as the original truncation of the long-wavelength response does not contain third order derivatives terms, and the regulator dependence persists at this order. In particular, for the momentum response $\tilde G_s^v$, \Eq{eq:lowkcoef}, the generated cubic term is 
\begin{align}
&\tilde G_s^v(\sigma)=G_s^0  \left((\tau-\tau_0)\ktt \tilde{s}^{(1)}_{s}(1-\frac{1}{2}\sigma^2\ktt^2+\ldots)\right) \nonumber\\
&=G_s^0\left((\tau-\tau_0)\ktt \tilde{s}^{(1)}_{s}-\frac{1}{2}\tilde{s}^{(1)}_{s}\frac{\sigma^2}{(\tau-\tau_0)^2}\ktt^3(\tau-\tau_0)^3+\ldots\right)
\end{align}
For $\sigma=\Delta \tau/2$, $\tilde s_s^{(1)}=\frac{1}{2}$ and $G_s^0\approx 1$ this  approximately reproduces the cubic  term in the free-streaming expansion, see \Eq{eq:freestrhighorder}. This accidental agreements significantly improves the agreement between low-$k$ expansion and full kinetic result in \Fig{fig:lowk}. Increasing $\sigma$ value reveals the actual linear order accuracy for velocity response.

\end{document}